\documentclass{aa}  
\pdfoutput=1
\usepackage{graphicx}
\usepackage{txfonts}
\usepackage{textcomp}
\usepackage{lscape}
\usepackage{rotating}
\usepackage{array} 
\usepackage{varwidth} 

\DeclareUnicodeCharacter{2212}{-}

\newcolumntype{A}{>{\begin{varwidth}{0.5cm}\centering}c<{\end{varwidth}}} 
\newcolumntype{B}{>{\begin{varwidth}{0.8cm}\centering}c<{\end{varwidth}}} 
\newcolumntype{D}{>{\begin{varwidth}{0.9cm}\centering}c<{\end{varwidth}}} 
\newcolumntype{E}{>{\begin{varwidth}{1.0cm}\centering}c<{\end{varwidth}}} 
\newcolumntype{F}{>{\begin{varwidth}{1.3cm}\centering}c<{\end{varwidth}}} 

\newcommand{\um}{$\mu$m}
\newcommand{\kms}{km s$^{-1}$}
\newcommand{\cmdue}{cm$^{-2}$}
\newcommand{\cmtre}{cm$^{-3}$}
\newcommand{\lsun}{L$_{\odot}$}
\newcommand{\msun}{M$_{\odot}$}

\def\wise{{\it WISE}}
\def\her{{\it Herschel}}
\def\spit{{\it Spitzer}}
\def \twomass{2Mass}

\begin{document}

   \title{A catalogue of dense cores and young stellar objects in the Lupus complex based on {\it \bf Herschel}\thanks{{\it Herschel} is an ESA space observatory with science instruments provided by European-led Principal Investigator consortia and with important participation from NASA.} Gould Belt Survey observations}

   \author{M. Benedettini\inst{1}, S. Pezzuto\inst{1}, E. Schisano\inst{1}, P. Andr\'{e}\inst{2}, V. K\"{o}nyves\inst{3,2}, A. Men\textquoteright shchikov\inst{2}, B. Ladjelate\inst{4,2}, J. Di Francesco\inst{5}, D. Elia\inst{1}, D. Arzoumanian\inst{6,2}, F. Louvet\inst{2}, P. Palmeirim\inst{7}, K. L. J. Rygl\inst{8}, N. Schneider\inst{9}, L. Spinoglio\inst{1}, D. Ward-Thompson\inst{3}
          }

   \institute{
INAF--Istituto di Astrofisica e Planetologia Spaziali, via Fosso del Cavaliere 100, 00133 Roma, Italy,
\email{milena.benedettini@inaf.it}
\and Laboratoire AIM, CEA/DSM-CNRS-Universit\'{e} Paris Diderot, IRFU/Service d'Astrophysique, Saclay, 91191 Gif-sur-Yvette, France
\and Jeremiah Horrocks Institute, University of Central Lancashire, Preston PR1 2HE, UK  
\and Instituto Radioastronomía Milim\'{e}trica, Av. Divina Pastora 7, Nucleo Central, 18012 Granada, Spain
\and National Research Council Canada, 5071 West Saanich Road, Victoria, BC V9E 2E7, Canada
\and Department of Physics, Graduate School of Science, Nagoya University, Furo-cho, Chikusa-ku, Nagoya 464-8602, Japan
\and Instituto de Astrof\'isica e Ci{\^e}ncias do Espa\c{c}o, Universidade do Porto, CAUP, Rua das Estrelas, PT4150-762 Porto, Portugal
\and Italian ALMA Regional Centre, INAF--Istituto di Radioastronomia, via P. Gobetti 101, 40129, Bologna, Italy
\and I. Physik. Institut, University of Cologne, 50939 Cologne, Germany  
}
\titlerunning{A catalogue of dense cores and young stellar objects in the Lupus complex}
\authorrunning{Benedettini et al.}
  \date{Received ; accepted}

 
  \abstract
   {How the diffuse medium of molecular clouds condenses in dense cores and how many of these cores will evolve in protostars is still a poorly understood step of the star formation process. Much progress is being made in this field thanks to the extensive imaging of star-forming regions carried out with the \her\, Space Observatory.}
   {The \her\, Gould Belt Survey key project mapped the bulk of nearby star-forming molecular clouds in five far-infrared bands with the aim of compiling complete census of prestellar cores and young, embedded protostars. From the complete sample of prestellar cores, we aim at defining the core mass function and studying its relationship with the stellar initial mass function. Young stellar objects (YSOs) with a residual circumstellar envelope are also detected.}
   {In this paper, we present the catalogue of the dense cores and YSOs/protostars extracted from the \her\, maps of the Lupus I, III, and IV molecular clouds. The physical properties of the detected objects were derived by fitting their spectral energy distributions.
   }
   {A total of 532 dense cores, out of which 103 are presumably prestellar in nature, and 38 YSOs/protostars have been detected in the three clouds. Almost all the prestellar cores are associated with filaments against only about one third of the unbound cores and YSOs/protostars. Prestellar core candidates are found even in filaments that are on average thermally sub-critical and over a background column density lower than that measured in other star forming regions so far. The core mass function of the prestellar cores peaks between 0.2 \msun\, and 0.3 \msun, and it is compatible with the log-normal shape found in other regions. \her\, data reveal several, previously undetected, protostars and new candidates of Class 0 and Class II with transitional disks. We estimate the evolutionary status of the YSOs/protostars using two independent indicators: the $\alpha$ index and the fitting of the spectral energy distribution from near- to far-infrared wavelengths. For 70\% of the objects, the evolutionary stages derived with the two methods are in agreement. 
   }
   {Lupus is confirmed to be a very low-mass star-forming region, both in terms of the prestellar condensations and of the diffuse medium. Noticeably, in the Lupus clouds we have found star formation activity associated with interstellar medium at low column density, usually quiescent in other (more massive) star forming regions.   
   }


   \maketitle
%

\section{Introduction} 
\label{sect:intro}

Stars form in the denser filamentary regions of Giant Molecular Clouds (GMCs) but how  diffuse matter gathers in cores and how many of these dense cores will form stars is still a poorly understood process. {To provide insight into} the physical mechanisms responsible for the growth of structure in the cold interstellar medium (ISM), leading to the formation of prestellar cores and protostars is one of the main scientific goals of the \her\, Gould Belt Survey (HGBS) key project that mapped the main nearby (d $\lesssim$ 500 pc) star-forming regions in five far-infrared (FIR) bands with {\it Herschel's} photometric instruments \citep{andre10}. The main products of the HGBS project are comprehensive and homogeneous catalogues of compact sources for all the observed regions. In this paper, we present catalogues of dense cores and young stellar objects (YSOs) extracted from the \her\, observations of the star-forming regions in the Lupus complex and we derive their physical properties. 

The Lupus dark cloud complex is located in the Scorpius–Centaurus OB association and consists of several loosely connected dark clouds showing different levels of star formation activity. With \her, we mapped the three main sites of star formation within the complex, namely the \object{Lupus I}, \object{Lupus III}, and \object{Lupus IV} clouds. 
These three clouds have been intensively observed with other instruments at several wavelengths to study the star formation process in the low-mass regime. The YSOs population in these clouds was initially investigated at near-infrared (NIR) and mid-infrared (MIR) wavelengths as part of the \spit-c2d survey \citep{merin08}. The Lupus I cloud was observed also at 450 \um\, and 850 \um\, as part of the James Clerk Maxwell Telescope Gould Belt Survey \citep{mowat17}, revealing a number of cold and dense condensations. Fifteen of theses were identified as disks of YSOs and 12 as prestellar or protostellar cores. Emission line maps from high density molecular tracers at 3 mm and 12 mm with the Mopra telescope have allowed to be made a chemical classification of the brighter and denser cores from which indications of their evolutionary status were inferred \citep{benedettini12}. 

In Benedettini et al. (2015; hereafter paper I), we presented the analysis of the filamentary structure of the clouds extracted from the \her-based column density maps. This study has revealed that Lupus I, III, and IV clouds have an ISM characterised by a very low column density in both the diffuse regions and in the dense material arranged in filaments. Indeed, the probability distribution function of column density (PDF) in the three regions peaks between 5$\times$10$^{20}$ \cmdue\, and 10$^{21}$ \cmdue, and the average column density of filaments is only $\sim$1.5$\times$10$^{21}$ \cmdue\, \citep{benedettini15}. The absence of very high column density gas ($\geq$ 5$\times$10$^{22}$ \cmdue) and the predominance of low column density gas likely has an impact on the typical mass of stars formed in these regions. Lupus is indeed a low-mass star-forming complex with a stellar population dominated by mid M-type stars \citep{hughes94,mortier11,alcala17}. A preliminary visual version of a catalogue of dense cores based on \her\, data was presented in \citet{rygl13}.

The Lupus complex is one of the closer sites of low-mass star formation, but the precise distances of its sub-regions is still a matter of debate. Recently, \citet{galli13} investigated the kinematic properties of the Lupus moving group of young stars and derived the following distances: $d$ = 182$^{+7}_{-6}$ pc for Lupus I,  $d$ = 185$^{+11}_{-10}$ pc for Lupus III and $d$ = 204$^{+18}_{-15}$ pc for Lupus IV. We derived distances from the {\it Gaia} DR 2 paralaxes \citep{luri18} for a sample of know YSOs members of the Lupus complex and we found the following modal values: $d$ = 155$^{+7}_{-14}$ pc for Lupus I,  $d$ = 160$^{+45}_{-8}$ pc for Lupus III and $d$ = 155$^{+30}_{-18}$ pc for Lupus IV, where the negative and positive errors corresponds to the minimum 5\% and maximum 95\% quantile of all the quantiles of the objects in each of the three clouds. Previous distance estimates, based also on different methods, give values between 140 pc and 200 pc for the different members of the complex. In this work, we assume the values indicated by \citet{comeron08} who reviewed all the distance estimates in the previous literature, concluding that a distance of 150 pc seems adequate for Lupus I and IV while a value of 200 pc is more appropriate for Lupus III. The same distances were used also in the \spit-c2d catalogue of YSOs \citep{merin08} and in \citet{benedettini15} and they are within the errors associated to the values derived by {\it Gaia} DR 2 data.

In this paper we present the catalogue of the dense cores and YSOs/protostars extracted from the \her\, maps of Lupus I, III, and IV. In Sect. 2 we describe the observations and data products. The procedures for compact source and protostar extraction and selection are presented in Sect. 3. In Sects. 4 and 5, we present the catalogues of dense cores and YSOs/protostars, respectively, and how the physical properties of the objects in the catalogues are derived.
The analysis of the main properties of the objects of the two catalogues are discussed in Sect. 6. The main conclusions of the paper are summarised in Sect. 7.

\section{Observations and data reduction}
\label{sect:reduction}

As part of the HGBS\footnote{http://gouldbelt-herschel.cea.fr} \citep{andre10} the three sub-regions of the Lupus complex, Lupus I, III, and IV, were observed in five photometric bands between 70 \um\, and 500 \um\, with the Photodetector Array Camera and Spectrometer (PACS; \citealt{poglitsch10}) and the Spectral and Photometric Imaging Receiver (SPIRE; \citealt{griffin10}) on-board the {\it Herschel} Space Observatory \citep{pilbratt10}. The observations were carried out in the parallel observing mode with a scanning velocity of 60 \arcsec/sec and each cloud was observed twice along orthogonal scanning directions. 

The data reduction pipeline and the final maps were presented in \citet{benedettini15} to which we refer for details. Here we recall that the final maps, in unit of MJy/sr, have a pixel size 3\arcsec\, at 70 \um\, and 160 \um, 6\arcsec\, at 250 \um, 10\arcsec\, at 350 \um\, and 14\arcsec\, at 500 \um. The absolute flux calibration for point sources is $\leq$6\% for PACS \citep{nielbock13} and $\leq$5\% for SPIRE \citep{bendo13}. The extended sources calibration is more uncertain and we assume a conservative error of 20\%. For both PACS and SPIRE maps we added a zero-level offset derived by comparing the {\it Herschel} data with the {\it Planck} and {\it IRAS} data of the same area of the sky and adopting a dust model for extrapolating the flux at the {\it Herschel} wavelengths \citep{bernard10}.  The calibrated maps were used to produce an H$_2$ column density map at the higher resolution of the SPIRE 250 \um\, data of 18\farcs2, by applying a method based on a multi-scale decomposition of the data  \citep{palmeirim13} and a dust opacity law of $\kappa_{\lambda}=\kappa_{\rm 300} ~ (\lambda/300~\mu m)^{-\beta}$ with $\kappa_{\rm 300}$=0.1 cm$^2$g$^{-1}$ (already accounting for a gas-to-dust ratio of 100), a grain emissivity parameter $\beta$=2 \citep{hildebrand83}, and a mean molecular weight $\mu$=2.8, values adopted as the standard by the HGBS consortium. The five \her\, maps, as well as the high-resolution H$_2$ column density map, for the three clouds are available in the HGBS archive\footnote{http://gouldbelt-herschel.cea.fr/archives}.

Notably, the Lupus I maps were affected by stray Moonlight, visible as a bright vertical band in each image. This problem does not affect the estimates of the point sources fluxes since those are measured by applying a background subtraction, but it required correction before the derivation of the column density map. We removed the stray-light contamination in the four images at 160 \um, 250 \um, 350 \um\, and 500 \um, by evaluating its contribution from the difference between the observed map (affected by stray-light) and the model used for the flux calibration, producing a column density map corrected from the stray-light effect.

\section{Compact sources identification}

\subsection{Source classification}

The wavelength range covered by the {\it Herschel} photometric instruments is where the thermal emission of cold dust (T$\approx$10-30\,K) peaks. Therefore, the cold dust condensations which may potentially form stars are clearly visible as compact, roundish structures that emerge from the background in the \her\, maps and in the H$_2$ column density maps. Moreover, in the nearest molecular clouds as those of the Lupus complex, the extremely sensitive {\it Herschel} instruments were able to detect even the fainter cores. Most of them are not dense enough to undergo gravitational collapse and will dissipate. Finally, circumstellar envelope of YSOs are detectable in \her\, maps.

To classify the compact sources extracted from the \her\, maps we used the following scheme (see also \citealt{andre14}). The detection in the PACS 70 \um\, map was used to identify the YSOs/protostars since the presence inside the cores of an accreting object produces an internal heating that increments the emission at wavelengths $\lesssim$ 100 \um\, with respect to grey-body SED shape, typical of prestallar cores. In fact, the PACS 70 \um\, sensitivity is adequate to detect YSOs emission even in low luminosity sources but not that of grey-body emitters with temperature around 10 K. Sources without the 70 \um\, detection are considered starless dense cores and are divided into two groups: the gravitationally unbound cores, that are expected to dissipate in the future, and the gravitationally bound cores, that are considered good candidates prestellar cores. To establish if a starless core is gravitationally bound or unbound we used the ratio between its observed mass and the Bonnor-Ebert mass calculated by using the core measured radius and temperature. More details are given in Sect. \ref{sect:phys:gravity}.

\subsection{Sources extraction from Herschel maps}

To generate an extensive catalogue of dense condensations in the Lupus regions we used the strategy developed inside the HGBS consortium to ensure homogeneous data analysis among all the star-forming regions observed in the project \citep{konyves15}. SPIRE and PACS images were processed with {\it getsources} v1.140127 an algorithm that performs the extraction of compact sources and filamentary structures at a multi-scale and multi-wavelength level \citep{menshchikov12,menshchikov13}. The source extraction method is divided into two stages: a first `detection' stage and a second `measurement' stage. At the detection stage, {\it getsources} analyses `single-scale' images across a wide range of scales and across all observed wavebands, identifying the sources and their footprints. At the measurement stage, fluxes and sizes of detected sources are measured in the observed images at each wavelength, considering deblending of overlapping sources. Background is subtracted by linear interpolation under the source footprints found at the detection stage, constrained by different angular resolutions in each waveband. Aperture corrections are applied by {\it getsources} using tables of the encircled energy fraction values for the actual point spread functions provided by the PACS and SPIRE ICCs \citep{balog14,bendo13}.

Two different {\it getsources} extractions are performed, optimised for the detection of dense cores and protostars, respectively. In the first set, we used the Herschel maps at 160 \um, 250 \um, 350 \um, and 500 \um. In addition, we included the high-resolution column density image - as an additional `wavelength' - to ensure that detected sources correspond to genuine column density peaks. Furthermore, the 160 \um\, component at the detection stage is `temperature-corrected' to reduce the effects of anisotropic temperature gradients. The temperature-corrected 160 \um\, map is obtained by converting the original observed 160 \um\, map (13\farcs5 resolution) to an approximate column density image, using the colour-temperature map derived from the intensity ratio between 160 \um\, and 250 \um\, (at the 18\farcs2 resolution of the 250 \um\, map). The second set of {\it getsources} extractions is optimised to detect YSOs/protostars. In this case, at the detection stage we used only the 70 \um\, \her\, image. Indeed, the presence of point-like 70 \um\, emission is a strong indication of the presence of a protostar that is warming up its circumstellar envelope.

At the measurement stage of both sets of extractions, source properties are measured at the detected positions of either cores or YSOs/protostars, using the observed, background-subtracted, and deblended images at all five Herschel wavelengths, plus the high-resolution column density map. The advantage of this two-pronged extraction strategy is that it provides more reliable detections and measurements of column-density cores and 70 \um\, luminous YSOs/protostars, respectively.

\subsection{Selection criteria}
\label{sect:selection}

We filtered the raw source lists of the {\it getsources} extractions to select only reliable sources by applying the following criteria. These selection criteria are uniformly applied to all the catalogues produced in the HGBS consortium and are described in detail in \citet{konyves15}.

For candidate dense cores identified from the first set of extractions selection criteria are:

\begin{itemize}
 \item Column density detection significance greater than 5 in the high-resolution column density map;
 \item Global detection significance over all wavelengths greater than 10;
 \item Global goodness $\geqslant$ 1, where goodness is an output quality parameter of {\it getsources}, combining global signal to noise ratio (S/N) and source reliability;
 \item Column density measurement S/N \textgreater 1 in the high-resolution column density map;
 \item Monochromatic detection significance greater than 5 in at least two bands between 160 \um\, and 500 \um;
 \item Flux measurement with S/N \textgreater 1 in at least one band between 160 \um\, and 500 \um\, for which the monochromatic detection significance is simultaneously greater than 5.
\end{itemize}

For candidate YSOs and protostars identified from the second set of extractions selection criteria are:

 \begin{itemize}
  \item Monochromatic detection significance greater than 5 in the 70 \um\, band;
  \item Positive peak and integrated flux densities at 70 \um;
  \item Global goodness greater than or equal to 1;
  \item Flux measurement with S/N \textgreater 1.5 in the 70 \um\, band;
  \item Full Width Half Maximum (FWHM) source size at 70 \um\, smaller than 1.6 times the 70 \um\, beam size (i.e. \textless1.6 $\times$ 8\farcs4 = 13\farcs44);
  \item Estimated source elongation \textless1.30 at 70 \um, where source elongation is defined as the ratio of the major and minor FWHM sizes.
 \end{itemize}

We also include in our final catalogues three well known YSOs that were initially present in the raw list, but were successively excluded by the automatic selection based on the above criteria. In principle some of the sources of our catalogue could be galaxies. To reduce extragalatic contamination as much as possible, we removed any source located within 6\arcsec\, of known galaxies found by the NASA Extragalactic Database\footnote{https://ned.ipac.caltech.edu/forms/nearposn.html} (NED) and the SIMBAD\footnote{http://simbad.u-strasbg.fr/simbad/} database.

\begin{figure*}
\includegraphics[width=0.90\textwidth]{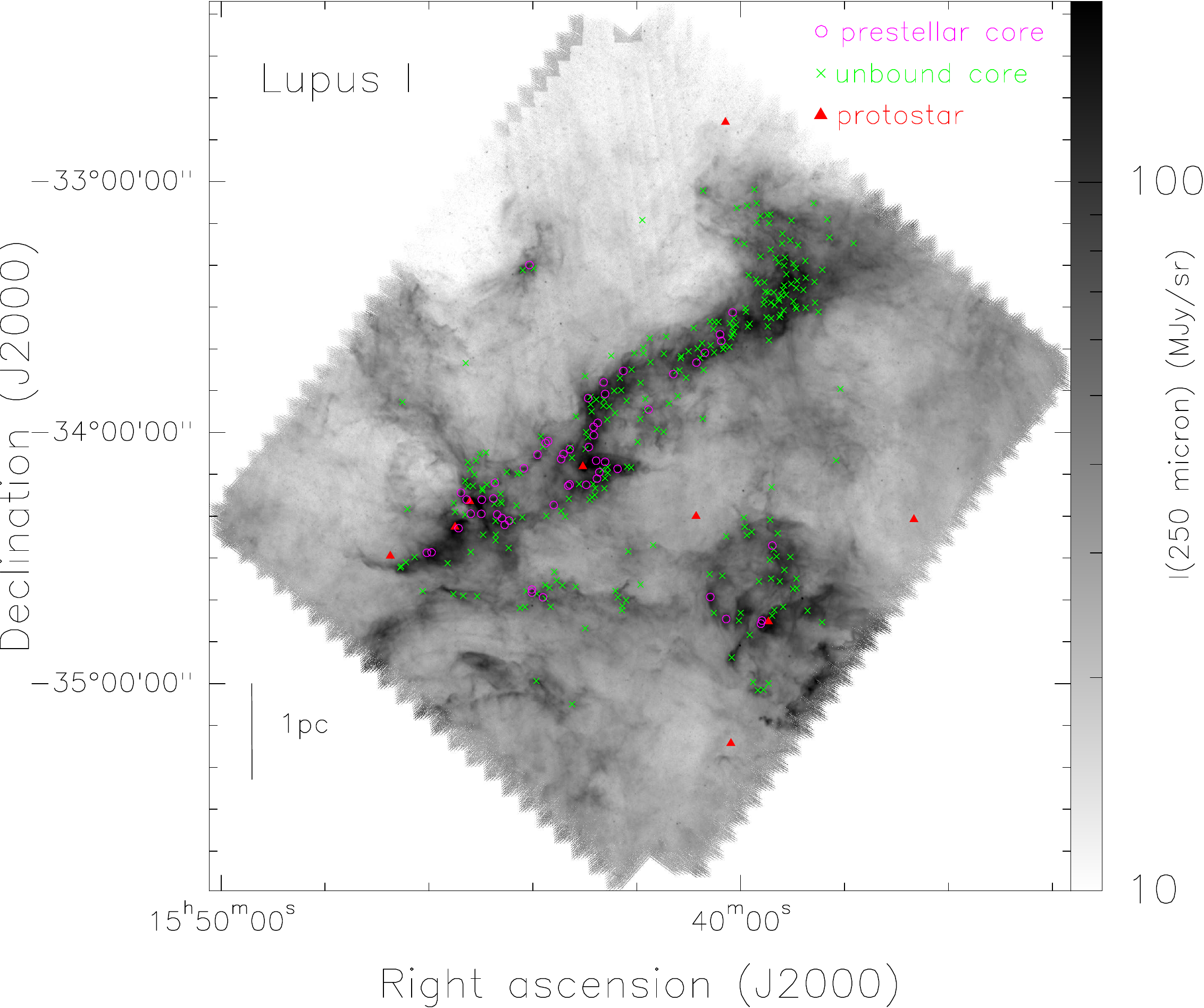}
\caption{{\it Herschel} 250 \um\, surface brightness image of the Lupus I region. The selected sample of starless unbound cores (crosses), prestellar cores (circle) and protostars (triangle) are indicated.}
\label{fig:lup1_sou}       
\end{figure*}

\begin{figure*}
\includegraphics[width=0.90\textwidth]{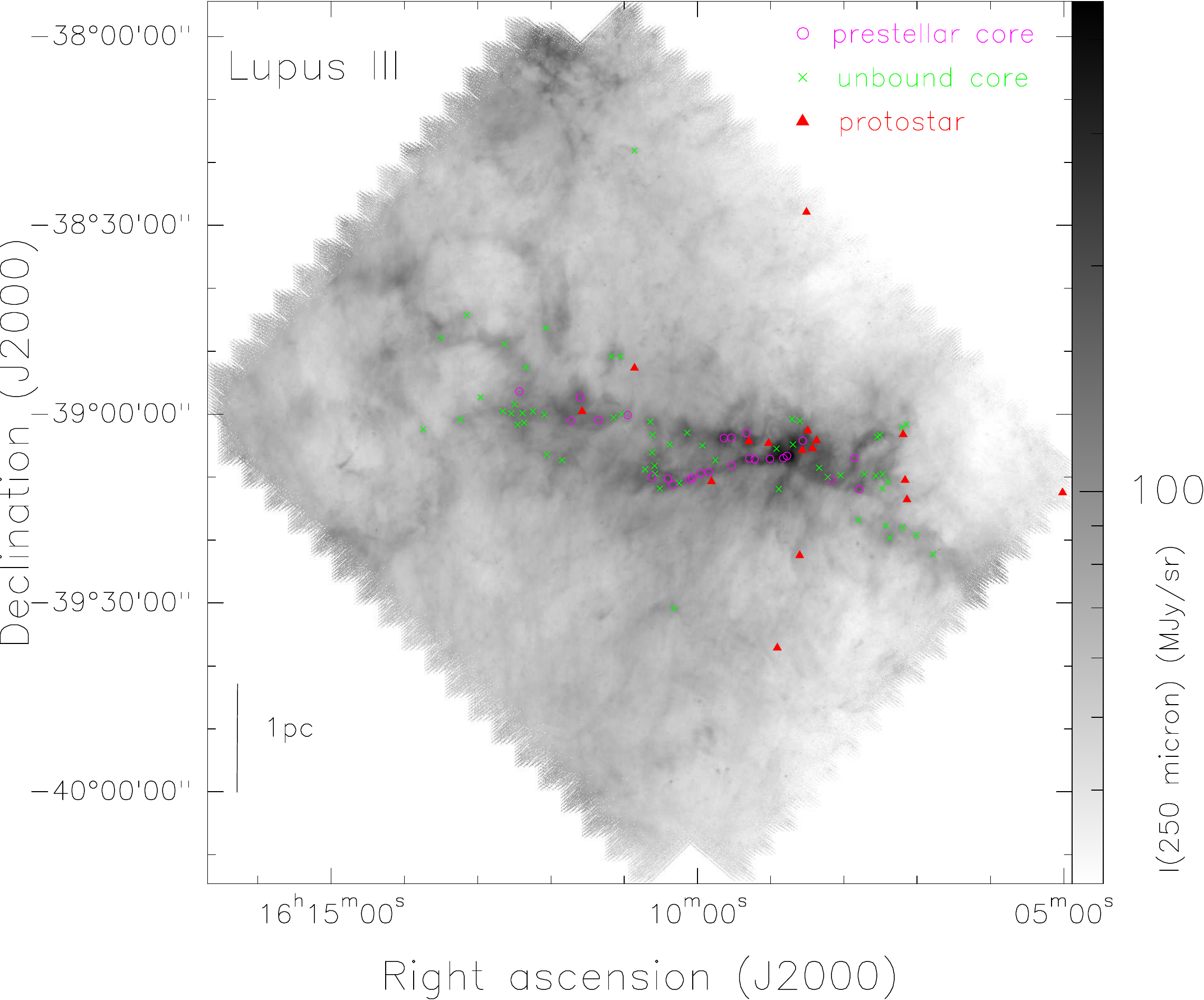}
\caption{{\it Herschel} 250 \um\, surface brightness image of the Lupus III region. The selected sample of starless unbound cores (crosses), prestellar cores (circle) and protostars (triangle) are indicated.}
\label{fig:lup3_sou}       
\end{figure*}

\begin{figure*}
\includegraphics[width=0.90\textwidth]{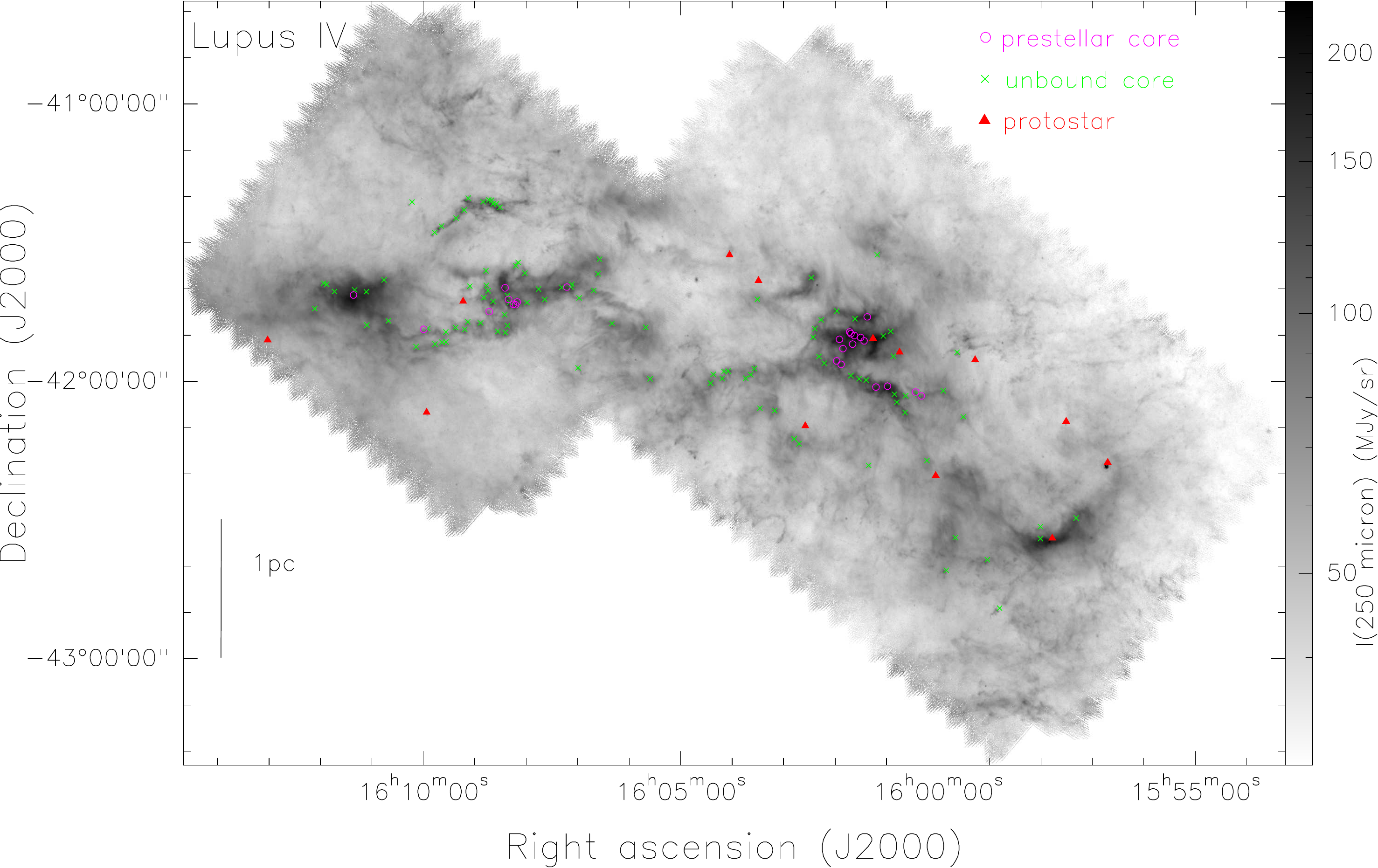}
\caption{{\it Herschel} 250 \um\, surface brightness image of the Lupus IV region. The selected sample of starless unbound cores (crosses), prestellar cores (circle) and protostars (triangle) are indicated.}
\label{fig:lup4_sou}       
\end{figure*}

We performed a final visual inspection of the selected sample to remove any dubious or not clearly visible source to produce the final catalogues of cores and YSOs/protostars. These  catalogues are available in the online material of this paper with few example lines listed in Table \ref{tab:cat}. In these tables, we report all photometric and geometrical quantities measured by {\it getsources} in the \her\, maps, that are: position, flux densities, and sizes with their respective errors and also associations with known sources. In Table \ref{tab:cat_sum}, we summarised the numbers of objects present in the catalogues for the three Lupus clouds and in Figs. \ref{fig:lup1_sou}, \ref{fig:lup3_sou} and \ref{fig:lup4_sou} we show the sources position on the 250 \um\, maps.

\begin{table*}
\caption{Summary of the number of entries in the catalogues divided for classification and regions.}             
\label{tab:cat_sum}      
\centering                          
\begin{tabular}{l c c c c}        
\hline\hline                 
region & total number of objects & unbound cores & prestellar cores (robust) & YSOs/protostars \\    
\hline                        
Lupus I   & 328  & 265 & 54 (24)  &  9 \\
Lupus III & 100  &  59 & 25 (19)  & 16 \\
Lupus IV  & 142  & 105 & 24 (16)  & 13 \\
Total     & 570  & 429 &103 (59)  & 38 \\
\hline                                   
\end{tabular}
\end{table*}

Disentangling the emission of a compact source from the emission of the underlying background in the {\it Herschel} maps is not a trivial task because of the usually bright and non-uniform background. Indeed, several algorithms, based on different approaches, have been developed to this purpose. Hence, to provide additional confidence of the reliability of our catalogues we used a second method to identify compact sources. Namely, we processed the {\it Herschel} maps with the {\it Curvature Thresholding EXtractor ({\it CuTEx})} algorithm \citep{molinari11}. {\it CuTEx} extracts compact sources independently on images at each wavelength band. Therefore, we searched for positional matches between our {\it getsources} sources and each single-band {\it CuTEx} catalogue within the ellipse that defines the size of the source at that band. For the cores catalogue, we consider confirmed matches those that coincide with a {\it CuTEx} detection in at least three distinct bands. With this approach, 43$\%$ of the {\it getsources} cores were also found by {\it CuTEx}. Very similar percentages of cross matches were found in Aquila and Taurus \citep{konyves15,marsh16} when comparing their {\it getsources} catalogues with the results from an alternative core extractor such as the {\it Cardiff Sourcefinding AlgoRithm (CSAR}, \citealt{kirk13}). For the YSOs/protostars catalogue, we consider a matched source one that has been detected with {\it CuTEx} in the 70 \um\, band. For this class of objects, the percentage of correspondence is much higher, between 85\% and 94\% for the three Lupus regions. This improvement occurs at 70 \um\, because at this wavelength in Lupus the background sky is almost dark while protostars are quite bright and well defined, making the detection of such sources easier.
In Table \ref{tab:cat}, we supply a robustness flag indicating if the {\it getsources} source was also found with the {\it CuTEx} algorithm.

\section{Dense cores catalogue}

\subsection{Physical properties}
\label{sect:phys}

For the starless dense cores, we estimate the mass and line-of-sight-averaged dust temperature by fitting a modified black-body function to the observed spectral energy distribution (SED) constructed from the set of measured flux densities in the wavelength range 160 \um\, -- 500 \um, weighted by 1/$\sigma^2\times$\verb+SIG_MON+, where $\sigma$ corresponds to the flux error and \verb+SIG_MON+ is the monochromatic detection significance parameter as estimated by {\it getsources}. The dust opacity law is the same as that used for the column density calculation (see Sect. \ref{sect:reduction}). We carried out SED fitting for those dense cores where: i) the monochromatic detection significance is greater than five and the S/N of the flux is greater than 0.1 in at least three wavelengths between 160 \um\, and 500 \um; and ii) the source has a larger integrated flux density at 350 \um\, than at 500 \um. 

In total 86 \% of starless cores fulfil the above criteria and we consider their SEDs fitting to be reliable. For the rest of the cores, the masses were directly estimated from the measured integrated flux density at the longest wavelength with a significant detection in each case, assuming optically thin dust emission and the median dust temperature found for starless cores with reliable SED fits (i.e. 11.5~K for Lupus I, 9.6~K for Lupus III, and 11.6~K for Lupus IV). The corresponding cores are marked as having `no SED fits' in the last column of Table \ref{tab:cores_phys} and they have more uncertain properties with respect to sources for which the fit was performed.
The derived physical parameters for each starless dense core are reported in the online material of this paper with few example lines listed in Table \ref{tab:cores_phys}.
In our catalogue, we provide the observed size of the core as the geometrical average between the major and minor FWHM sizes measured in the high-resolution column density map and we estimate the core outer radius as the observed size deconvolved with the HPBW of the map (18\farcs2). The peak (or central beam) column density, the average column density, the central-beam volume density, and the average volume density were also derived for each core based on its estimated mass and radius. 

The physical parameters derived from the SED fitting, namely the line-of-sight-averaged dust temperature and mass are affected by some sources of uncertainty: {\it i)} the error of the flux points due to the absolute flux calibration ($\leq$ 20\%), the irregular background that could be not well subtracted and the sources deblending that could be incorrect especially in crowed regions; {\it ii)} the uncertain of the distances that could be up to 30\% (see Sect. \ref{sect:intro}); {\it iii)} the uncertainty related to the assumption in the opacity law that is $<$ 50\% for cores in the H$_2$ column density range between
$\sim$3 $\times$ 10$^{21}$ \cmdue\, and $\sim$ 10$^{23}$ \cmdue\, \citep{roy14} but could be as high as a factor of two for lower H$_2$ column density \citep{benedettini15} as those we find in many of the Lupus cores; and {\it iv)} the assumption of single temperatures while prestellar cores have temperature profiles that drop towards their centres. In particular, the temperature profile of the starless dense cores is expected to have a minimum around 9--10 K at the core centre, where most of the core mass resides, and to increase up to 12--13 K at the core outer radius (e.g. \citealt{roy14,tafalla04}). The mean value of temperature derived from the SED fitting in our catalogue is $\sim$ 11.5 K. Therefore, the average SED dust temperature may overestimate the mass-averaged dust temperature within a dense core, leading to an underestimate of the core mass. This effect, however, decreases with the decreasing of the core mass, becoming less than 30\% for core masses less than 1 \msun\, \citep{menshchikov16}. To quantify the cumulative effect of all these factors on the final mass estimate is really complex and it is beyond the scope of the present paper. The simulation that we used for estimating the completeness of the prestellar cores sample (see Sect. \ref{sect:compl}), however, can help give an idea of the typical error on the mass derived by our sources extraction and SED fitting procedures. We found that the distribution of ratios between the mass derived by the SED fitting and the true core mass can be approximated by a Gaussian function centred around one and with $\sigma \sim$ 0.4. This implies that our strongest approximation of single temperature within the cores does not produce systematic underestimates of the masses, in the low mass regime of the Lupus prestellar cores.

\subsection{Selecting self-gravitating prestellar cores}
\label{sect:phys:gravity}

In the sample of starless cores, we are interested in identifying those cores that are dense enough to eventually form a star or a multiple system, namely the prestellar cores. These objects must be starless and gravitationally bound. For a self-gravitating core, the ratio between its virial mass and its measured mass is $M_{\rm vir}/M_{\rm obs}\leq$ 2. The virial mass can be derived by measuring the mean velocity dispersion of the gas from spectroscopic observations. Spectroscopic observations, at a spatial resolution similar to that of the \her\, maps, however, are usually unavailable for all the dense cores detected in the \her\, maps. Therefore, a simplified approach is adopted. We look at the ratio between the core mass and the thermal value of the critical Bonnor-Ebert (BE) mass, $\alpha_{\rm BE} = M_{\rm BE,crit}/M_{\rm obs}$, where the critical Bonnor-Ebert mass is given by $M_{\rm BE,crit} = 2.4 R_{\rm BE}c_{\rm s}^2/G$ with $R_{\rm BE}$ being the BE radius, $G$ being the gravitational constant, and $c_{\rm s} = \sqrt{kT/\mu m_{\rm H}}$ being the isothermal sound speed where the non-thermal component of the velocity dispersion is neglected, with $k$ being the Boltzmann's constant, $T$ being the temperature, $\mu$=2.33 being the molecular weigh per particle, and $m_{\rm H}$ being the Hydrogen mass. To get a rough estimate of the contribution of the non-thermal velocity dispersion to the total velocity dispersion that we are neglecting, we used the FWHM of the CS (1-0) line observed towards several dense cores in the Lupus clouds \citep{benedettini12}. These cores are larger than those in our catalogue and have a total velocity dispersion ranging from 0.25 \kms\, to 0.45 \kms, assuming a temperature of 10 K. Since the thermal component of the velocity dispersion at 10 K is 0.19 \kms, we expect that $M_{\rm BE,crit}$ underestimates the $M_{\rm vir}$ by at most a factor of about two. For each object, we estimated the thermal BE mass by using as gas temperature the dust temperature estimated by the SED fitting and $R_{\rm BE}$ as the deconvolved core radius given in Table \ref{tab:cores_phys}.

In line with the virial criterion, we classified a prestellar core as `robust' if $\alpha_{\rm BE} = M_{\rm BE,crit}/M_{\rm obs}\leq$ 2. However, previous papers of the HGBS consortium \citep{konyves15,marsh16} have shown that a $\alpha_{\rm BE}$ = 2 criterion is too restrictive for selecting gravitationally bound cores. It causes many marginally resolved sources to be misclassified since the critical BE mass depends linearly on radius. To take this effect into consideration, we assume a size-dependent limit $\alpha_{\rm BE}\leq 5 \times (HPBW_{\rm NH_2}/FWHM_{\rm NH_2})^{0.4}$, where $FWHM_{\rm NH_2}$ is the measured FWHM source diameter in the high-resolution column density map and $HPBW_{\rm NH_2}$ = 18\farcs2 is the HPBW resolution of the map. In this case, the limiting $\alpha_{\rm BE}$ ranges from $\sim$ 2 for well-resolved cores to $\sim$ 5 for unresolved cores. Monte-Carlo simulations performed on the \her\, data of the Aquila region have shown that the size-dependent limit allows about  95\% of the prestellar cores to be identified \citep{konyves15}.

\begin{figure}
\includegraphics[width=0.48\textwidth]{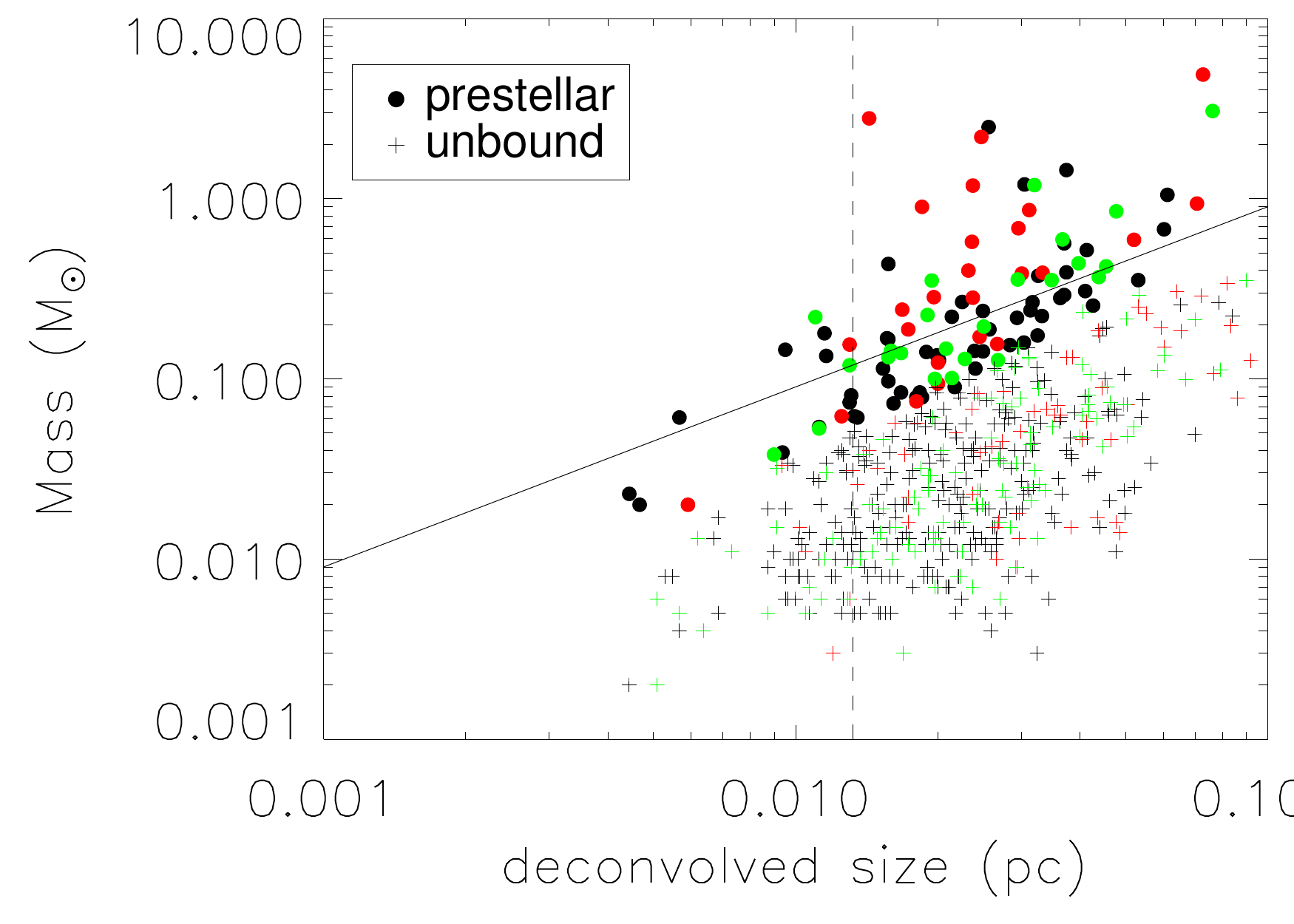}
\caption{Mass versus size diagram for the entire population of starless cores. The core sizes were estimated from the geometrical average between the major and minor FWHM measured by {\it getsources} in the high-resolution column density map and deconvolved from the 18\farcs2 HPBW resolution of the map. The 18\farcs2 resolution at the distance of 150 pc is indicated by the dashed line. Candidate prestellar cores are plotted with circular symbols and unbound cores with crosses. Colors indicate cores from the three regions: Lupus I (black), Lupus III (red), Lupus IV (green). The solid line indicates the virial criterion ($M_{\rm BE,crit}/M_{\rm obs}$ = 2), assuming $T$ = 11 K and $d$ = 150 pc; cores above this line are considered robust candidate prestellar cores.}
\label{fig:mass_radius}       
\end{figure}

Following the above criteria we identify a total of 103 candidate prestellar cores in the three Lupus regions out of which 59 cores can be considered {\it robust} candidates. All the rest of the starless cores are classified as unbound cores. The goodness of our classification can be easily assessed by looking at the mass versus radius plot, shown in Fig. \ref{fig:mass_radius}. As expected, the unbound cores populate the lower-right part of the plot, while the candidate prestellar cores occupy the upper-left part. In particular, the robust candidate prestellar cores are those above the solid line. We note that the cores of the three Lupus clouds share a very similar range of physical properties, indicating that they form a sample of similar objects. Therefore, we merged the cores catalogues of the three Lupus clouds and applied the following analysis to the merged catalogue. 

The percentage of prestellar cores with respect to the total number of starless cores is 19\%, in line with the that found in Taurus \citep{marsh16}, another nearby, low-mass star-forming region, but lower than that found in more distant and massive star-forming regions. Looking at some of the star-forming regions observed by \her\, in the HGBS and HOBYS (Herschel imaging survey of OB Young Stellar objects) key projects, we note that the percentage of prestellar cores with respect to the total number of detected starless cores increases with the distance of the region. For example, it is $\sim$ 60\% in Aquila \citep{konyves15} at a distance of $d$ = 260 pc \citep{cambresy99}, $\sim$ 84\% in Orion \citep{polychroni13} at $d$ = 414 pc \citep{menten07}, $\sim$ 94\% in Vela C \citep{giannini12} at $d$ = 700 pc \citep{liseau92}, and up to 100\% for regions more distant of 1 kpc in the Galactic Plane \citep{elia13}. This increasing percentage could be interpreted as a major efficiency in forming gravitationally bound cores in more massive clouds. In the adopted definition of bound cores, however, there is a distance bias since,  for a given SED, the derived mass varies as $d^2$, while $r_{\rm BE}$ varies as $d$, thus making it easier for the respective object to fulfil the BE criterion as $d$ increases \citep{elia13}. Moreover, there is an observational bias induced by the distance that prevents the detection of smaller and less massive unbound cores in more distant regions. As can be seen looking at the mass versus size plots of the starless cores populations in the mentioned regions \citep{marsh16,konyves15,polychroni13,giannini12}, the large number of small ($R\lesssim$ 0.05 pc) and low-mass ($M\lesssim$ 0.01 \msun) unbound cores present in Lupus and Taurus are missing in the more distant regions.

\subsection{Completeness of the prestellar core catalogue}
\label{sect:compl}

\begin{figure}
\includegraphics[width=0.5\textwidth]{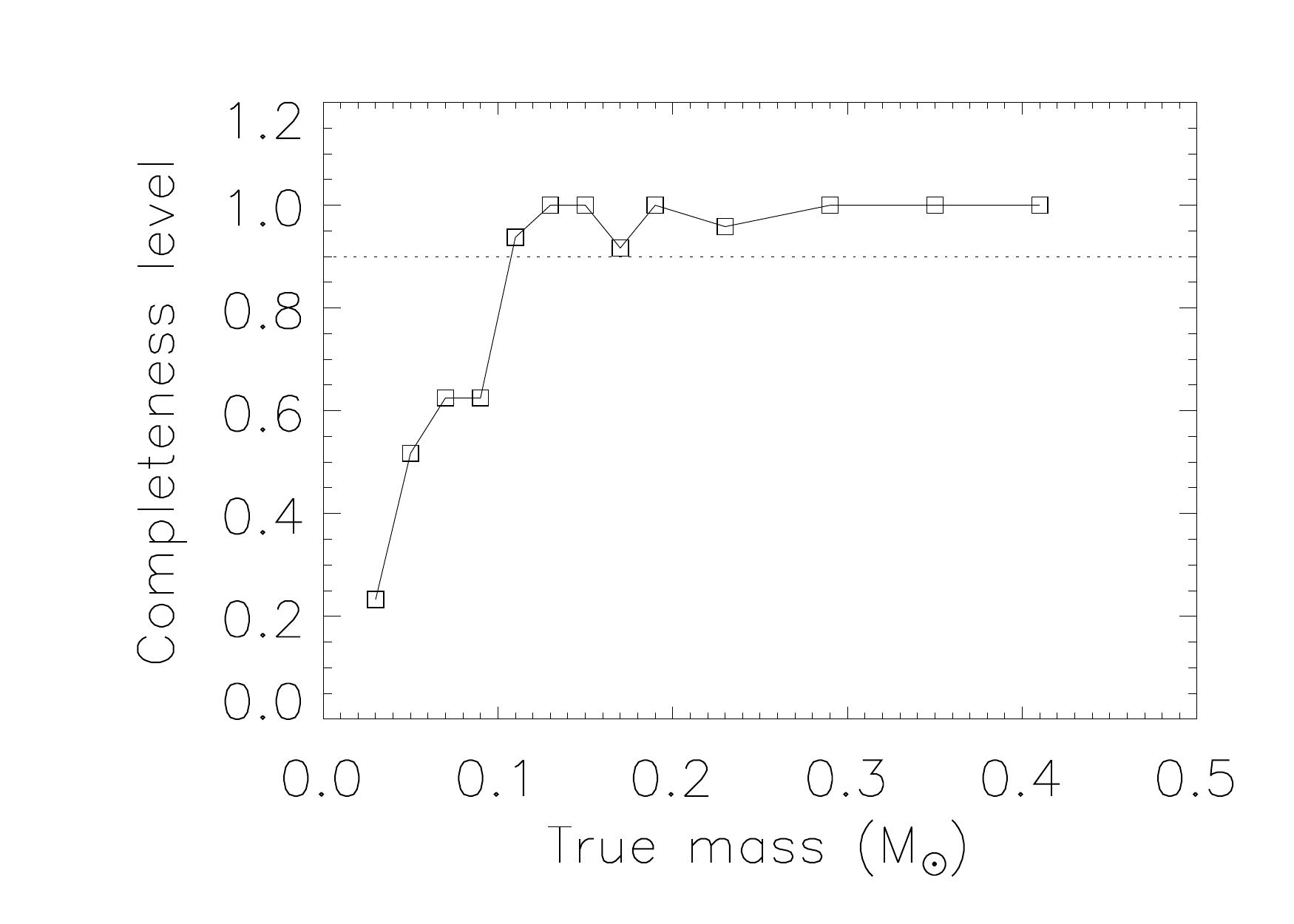}
\caption{Completeness level for the synthetic prestellar cores sample added to the \her\, maps as a function of the their mass. From this plot we can estimate that our sample of candidate prestellar cores is complete at 90\% level above a core mass of $\sim$ 0.1 \msun.}
\label{fig:compl}       
\end{figure}

We estimated the completeness of our sample of prestellar cores in the three Lupus clouds by using the following simulation. Briefly, we constructed simulated maps at all {\it Herschel} wavelengths (including the column density map) by adding synthetic sources, representative of a genuine population of prestellar cores, to realistic background emission. As background, we used the {\it Herschel} images where the emission of the compact sources identified with {\it getsources} was subtracted. We then inserted a population of 264 models of BE cores with masses between 0.039 \msun\, and 0.42 \msun, sampling the low-mass end of the observed prestellar core population. The synthetic sources were distributed across the three Lupus clouds to generate a full set of synthetic maps for the five Herschel bands plus the column density image. Compact sources extraction and classification was then performed with {\it getsources} for such synthetic skies in the same way as for the observed images.
In Fig. \ref{fig:compl} we show the completeness curve for the Lupus clouds as a function of the true core mass. Based on the results of these simulations, we estimate that our {\it Herschel} census of candidate prestellar cores is complete at 90\% level above a core mass of $\sim$ 0.1 \msun\, (see Fig.\ref{fig:compl}). A similar completeness level for the \her\, prestellar cores catalogues was found in Taurus \citep{marsh16} and in Corona Australis \citep{bresnahan18} which are at about the same distance of Lupus.

\begin{table*}
\caption{SED classification of the YSOs/protostars in our catalogue. Three parameters of the SED fit, disk mass ($M_{\rm disk}$), envelope density at the reference radius ($\rho_0^{\rm env}$) and cavity density at the reference radius ($\rho_0^{\rm cav}$), are listed to indicate the presence of a disk, envelope and cavity in the best fit model of the observed SED, respectively. Objects for which the SED fit fails are indicated with `no fit' in the eighth column.}             
\label{tab:par_yso}      
\centering                          
\begin{tabular}{c c c c c c c c c c}        
\hline\hline                 

name &  RA(J2000) & Dec(J200) & $L_{\rm bol}$ & $L_{\rm smm}$ & $\alpha$ & Class & $M_{\rm disk}$ \tablefootmark{(a)} & $\rho_0^{\rm env}$ \tablefootmark{(b)}  & $\rho_0^{\rm cav}$ \tablefootmark{(c)} \\    
 HGBS\_YSO-    & (h m s)    & (\degr\, \arcmin\, \arcsec\,) & (\lsun) & (\lsun) &  & & (\msun)  & (g cm$^{-3}$)  & (g cm$^{-3}$)\\
\hline   \\                     
LUPUS I \\
\hline    
J153640.0-342145 & 15 36 40.0  & -34 21 46.6  & 6.8e-01 & 1.7e-04 & -0.071 & Flat & 3.052e-04 & 1.377e-21 & 2.442e-21\\
J153927.9-344616 & 15 39 28.0  & -34 46 17.8  & 3.5e+00 & 9.1e-04 & -0.826 &  II  & 8.815e-04 & 4.996e-23 & 1.159e-23\\
J154011.3-351522 & 15 40 11.3  & -35 15 22.4  & 3.7e+02 & 0.0e+00 & -1.820 & III  & 3.318e-06 & ... & ... \\
J154017.6-324649 & 15 40 17.7  & -32 46 49.1  & 4.7e-03 & 5.6e-05 &  0.649 &  I/0 & 2.754e-07 & 2.765e-19 & 1.813e-21\\
J154051.6-342102 & 15 40 51.6  & -34 21 03.7  & 2.1e-02 & 1.9e-04 &  0.587 &   I & 2.137e-02 & 8.265e-22 & 7.478e-23\\
J154302.3-340908 & 15 43 02.4  & -34 09 08.2  & 2.9e-01 & 3.5e-02 &  1.469 &   0  & ...       & 3.017e-18 & 1.519e-23\\
J154512.8-341729 & 15 45 12.9  & -34 17 30.8  & 1.1e+01 & 7.5e-04 & -0.637 &  II  & 7.914e-05 & ... & ... \\	
J154529.8-342339 & 15 45 29.9  & -34 23 39.2  & 2.3e-01 & 5.7e-03 & -0.952 & II/0 & 9.364e-02 & 1.244e-23 & 3.096e-23\\
J154644.6-343034 & 15 46 44.7  & -34 30 35.9  & 2.6e+00 & 7.3e-04 & -0.835 &  II  & 5.221e-05 & 3.589e-24 & 4.890e-22 \\	
\hline \\
LUPUS III\\
\hline    
J160500.9-391301& 16  05 01.0  & -39 13 01.2 & 1.0e+03 & 0.0e+00 & -2.625 & III & 5.243e-06 & ... & ... \\
J160708.4-391407& 16  07 08.4  & -39 14 08.7 & 7.4e-02 & 9.1e-04 &  0.018 &Flat/0&7.356e-05& 9.630e-24 & ... \\
J160709.9-391102& 16  07 10.0  & -39 11 03.9 & 2.1e+00 & 3.2e-04 & -0.894 & II  & 2.619e-08 & 6.397e-20 & 2.907e-21 \\
J160711.5-390347& 16  07 11.5  & -39 03 47.3 & 1.1e+00 & 7.8e-04 & -1.910 & III & 1.406e-08 & 1.208e-21 & 2.167e-23 \\
J160822.4-390445& 16  08 22.4  & -39 04 45.2 & 4.6e+00 & 1.2e-03 & -0.464 & II  & 1.017e-03 & 4.002e-22 & 4.129e-22\\
J160825.7-390600& 16  08 25.7  & -39 06 01.7 & 1.1e+00 & 1.3e-04 & -0.627 & II  & 5.053e-03 & ... & ... \\
J160829.6-390309& 16  08 29.6  & -39 03 10.8 & 1.4e-02 & 0.0e+00 &  0.721 & I   & ... & 1.769e-21 & 1.358e-22 \\
J160830.7-382826& 16  08 30.7  & -38 28 26.5 & 3.8e+00 & 0.0e+00 & -0.832 & II  & 1.878e-02 & 1.952e-23 & 1.538e-22 \\
J160834.1-390617& 16  08 34.2  & -39 06 17.4 & 1.8e+02 & 4.6e-04 & -1.463 & II  & no fit \\	
J160836.1-392300& 16  08 36.2  & -39 23 01.9 & 3.3e+00 & 7.7e-03 & -1.225 & II  & 6.117e-02 & 3.771e-23 & 9.300e-23 \\
J160854.5-393743& 16  08 54.6  & -39 37 43.5 & 2.8e+00 & 2.1e-03 & -1.352 & II  & 5.094e-02 & ...  & ... \\
J160901.8-390511& 16  09 01.8  & -39 05 11.2 & 8.5e-01 & 0.0e+00 & -0.447 & II  & 6.486e-04 & 1.128e-23 & 5.924e-22 \\
J160917.9-390453& 16  09 17.9  & -39 04 53.0 & 9.9e-02 & 1.5e-02 &  1.235 & 0   & 4.764e-06 &  1.235e-19 & 1.081e-22 \\
J160948.5-391116& 16  09 48.5  & -39 11 16.4 & 1.8e+00 & 1.5e-04 & -0.863 & II  & 1.506e-04 & ... & ... \\
J161051.5-385314& 16  10 51.5  & -38 53 14.1 & 1.1e+00 & 2.2e-04 & -1.015 & II  & 2.380e-03 &  3.772e-23 & 2.516e-23 \\
J161134.4-390008& 16  11 34.4  & -39 00 08.2 & 3.9e+01 & 3.6e-04 & -1.572 & II  & no fit \\
\hline  \\
LUPUS IV\\
\hline    
J155641.9-421925 & 15 56 42.0 & -42 19 25.3 & 2.6e+01 & 1.9e-02 & -0.749 & II & 8.610e-03 & 2.970e-19 & 1.009e-23 \\
J155730.4-421032 & 15 57 30.5 & -42 10 32.3 & 8.1e-01 & 1.3e-04 & -0.549 & II & ... & 3.093e-20 & ... \\
J155746.6-423549 & 15 57 46.6 & -42 35 50.8 & 9.4e-01 & 1.4e-03 & -0.569 & II & 6.572e-03 & 4.594e-24 & 12.164e-22 \\	
J155916.5-415712 & 15 59 16.6 & -41 57 12.2 & 2.6e+00 & 6.5e-04 & -0.957 & II & 1.687e-02 & ... & ... \\
J160002.5-422216 & 16 00 02.5 & -42 22 16.3 & 7.2e-01 & 4.8e-04 & -1.124 & II & no fit \\
J160044.6-415530 & 16 00 44.6 & -41 55 31.9 & 2.8e+00 & 1.1e-03 & -0.502 & II & 1.368e-02 & 3.488e-24 & 3.113e-21 \\
J160115.5-415233 & 16 01 15.6 & -41 52 34.9 & 3.7e-02 & 3.0e-03 &  0.015 & Flat/0 & 4.258e-08 & 1.126e-20 & 2.154e-23 \\
J160234.6-421129 & 16 02 34.6 & -42 11 29.3 & 3.8e-02 & 5.4e-05 & -0.164 & Flat & 1.009e-03 & 1.153e-24 & 1.197e-22 \\
J160329.2-414001 & 16 03 29.2 & -41 40 01.2 & 4.1e-01 & 3.5e-04 & -0.965 & II & 3.946e-04 & 1.768e-24 & 7.543e-21 \\
J160403.0-413427 & 16 04 03.0 & -41 34 27.1 & 3.4e+02 & 0.0e+00 &  ...   & ... & 6.677e-04 & ... & ... \\
J160913.7-414430 & 16 09 13.7 & -41 44 30.5 & 2.5e+01 & 0.0e+00 & -1.170 &  II & 1.031e-05 & ... & ... \\
J160956.3-420834 & 16 09 56.3 & -42 08 34.2 & 5.7e+02 & 0.0e+00 & ...    & ... & 1.218e-04 & ... & ... \\
J161301.6-415255 & 16 13 01.7 & -41 52 55.5 & 2.8e+01 & 0.0e+00 & -2.772 & III & 5.518e-06 & ... & ... \\   
\hline                        
\end{tabular}
\tablefoot{
\tablefoottext{a}{$M_{\rm disk}$ is the mass of the disk.}
\tablefoottext{b}{$\rho_0^{\rm env}$ is the envelope density at the reference radius $r_0$ and it is serves to scale the envelope density.}
\tablefoottext{c}{$\rho_0^{\rm cav}$ is the constant density in the inner regions of the cavity where $\rho_0^{\rm cav}$ is lower than the envelope density, in the other regions, the cavity density is set to be equal to the envelope density.}
}
\end{table*}

\begin{figure*}[!ht]
\includegraphics[width=0.33\textwidth]{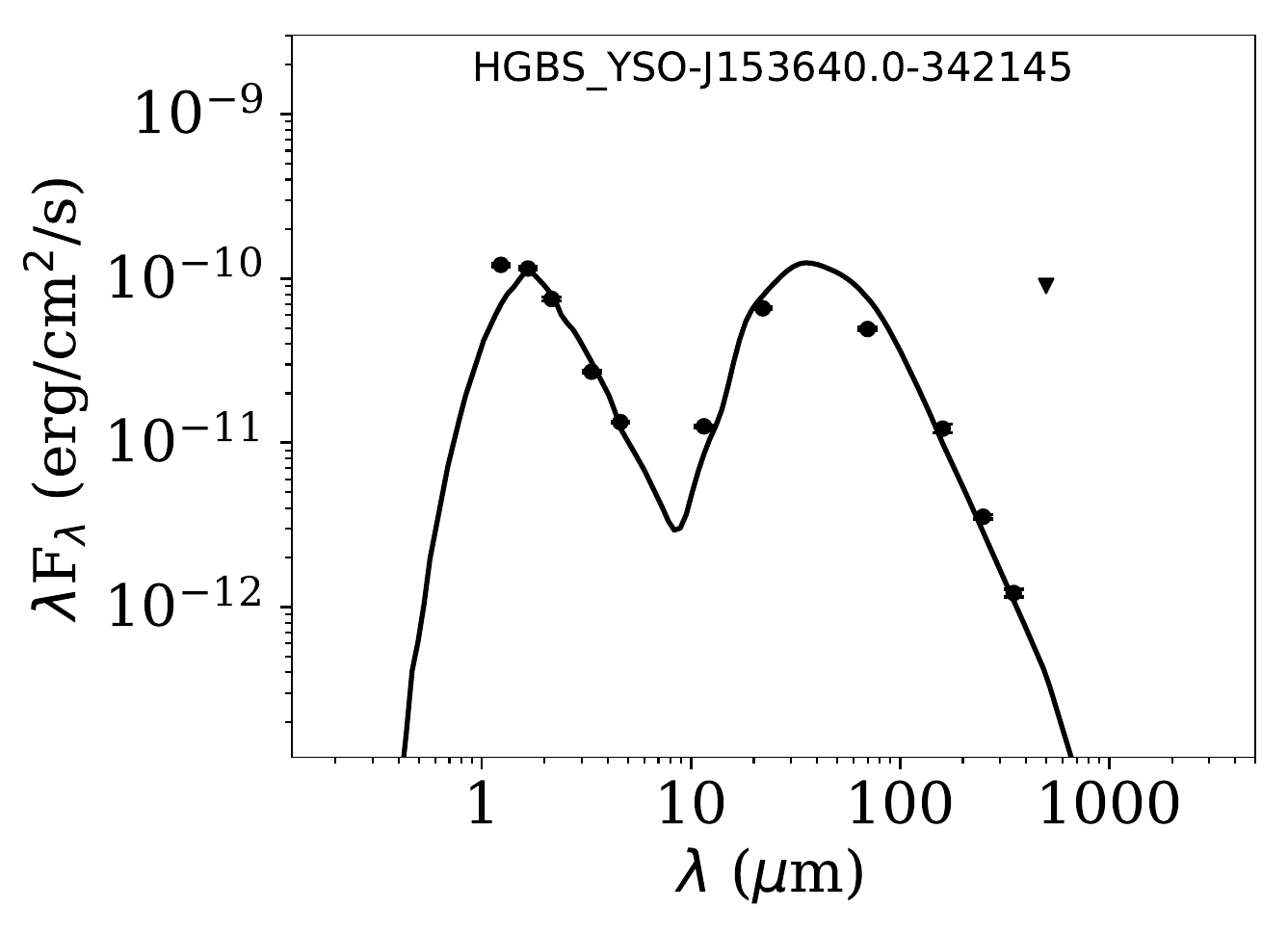} 
\includegraphics[width=0.33\textwidth]{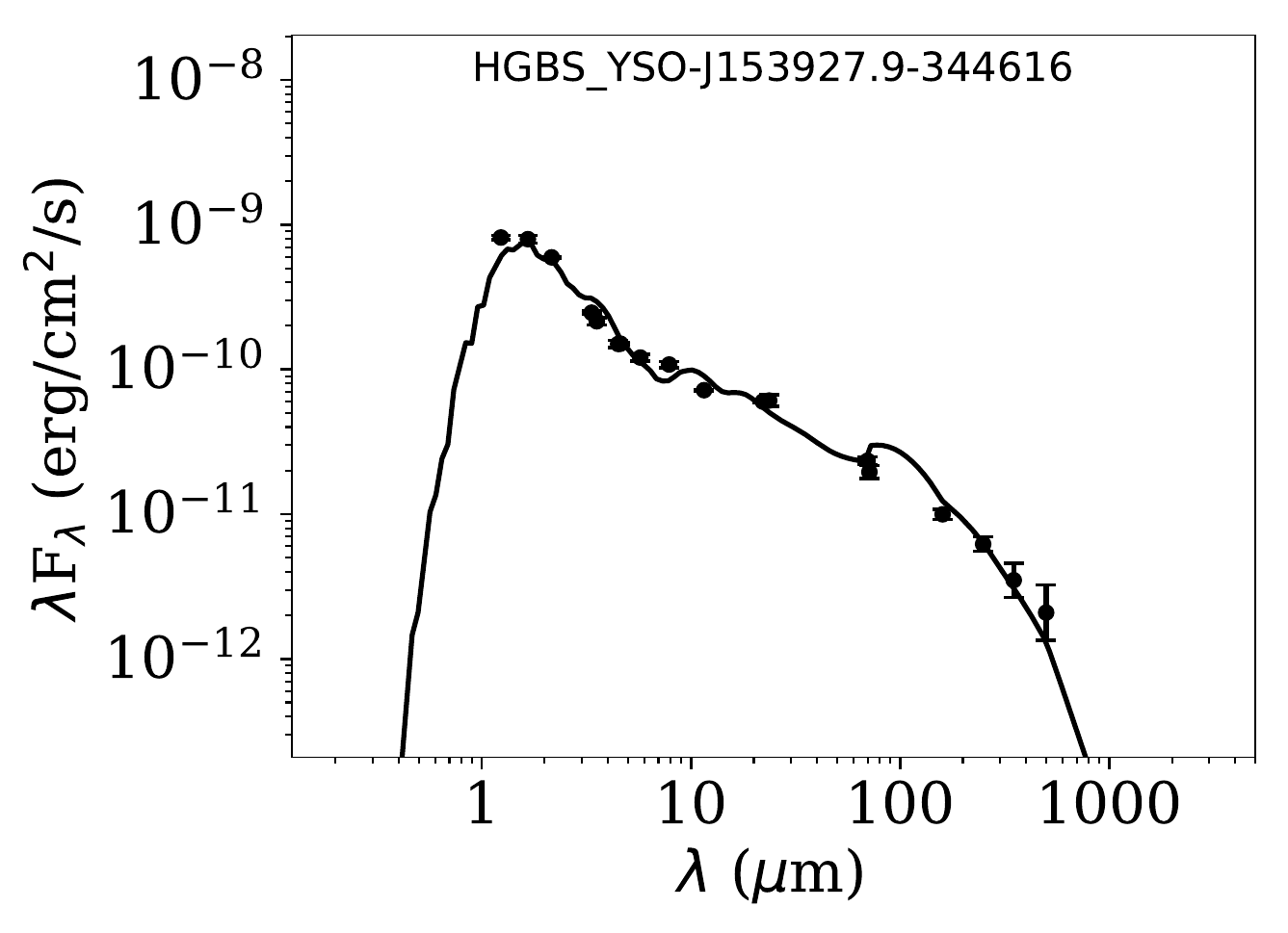}
\includegraphics[width=0.33\textwidth]{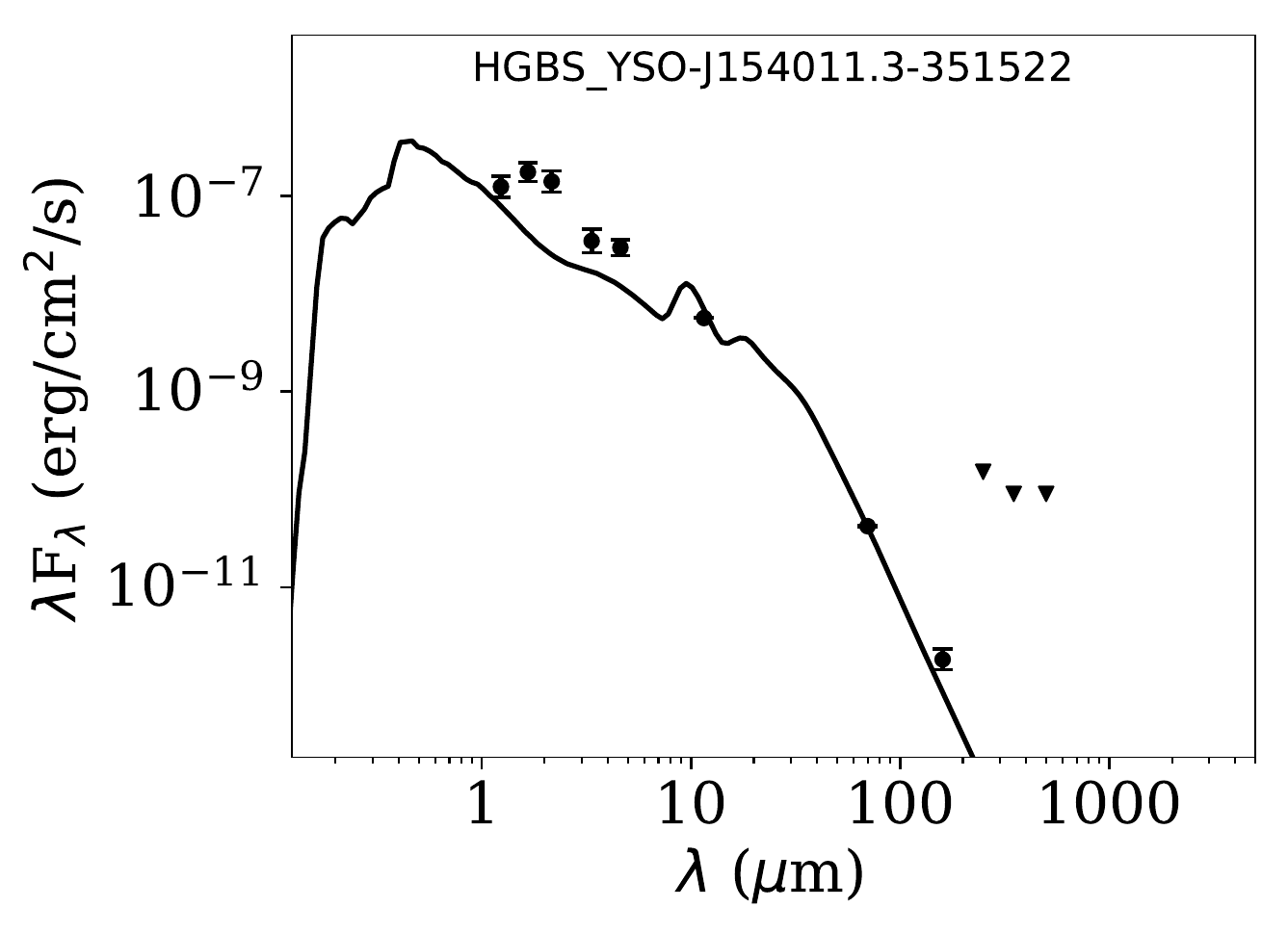}

\includegraphics[width=0.33\textwidth]{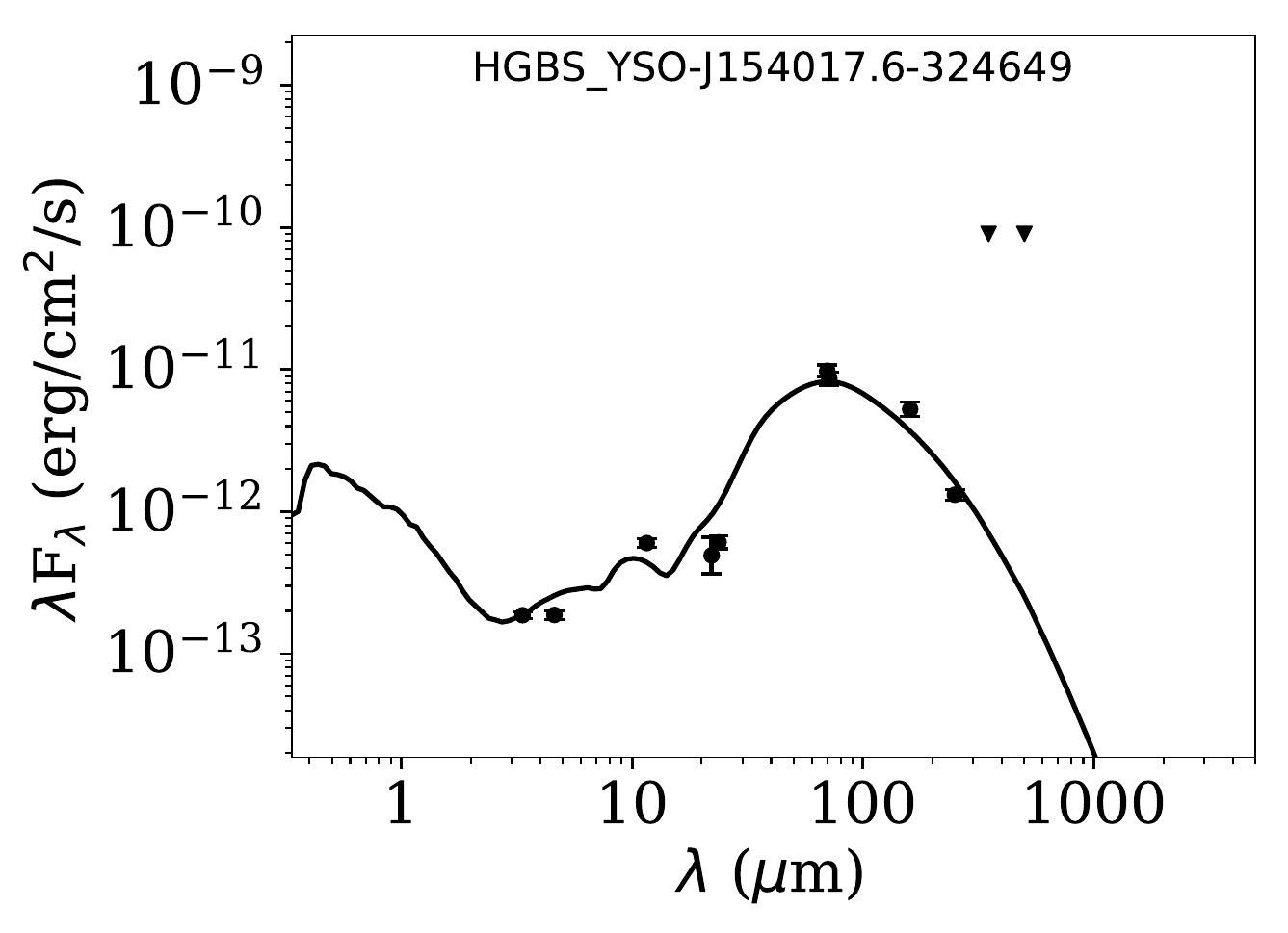}  
\includegraphics[width=0.33\textwidth]{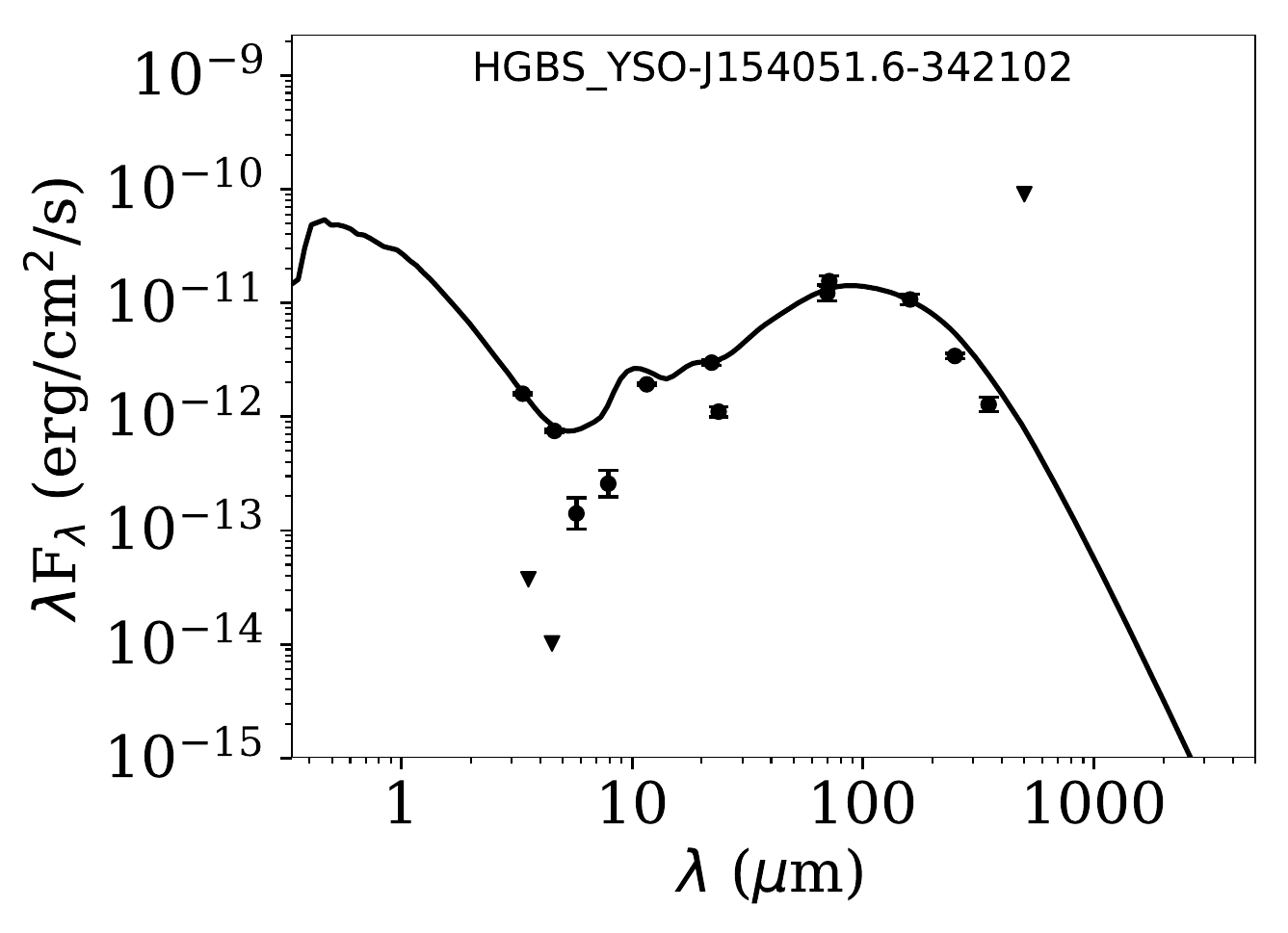} 
\includegraphics[width=0.33\textwidth]{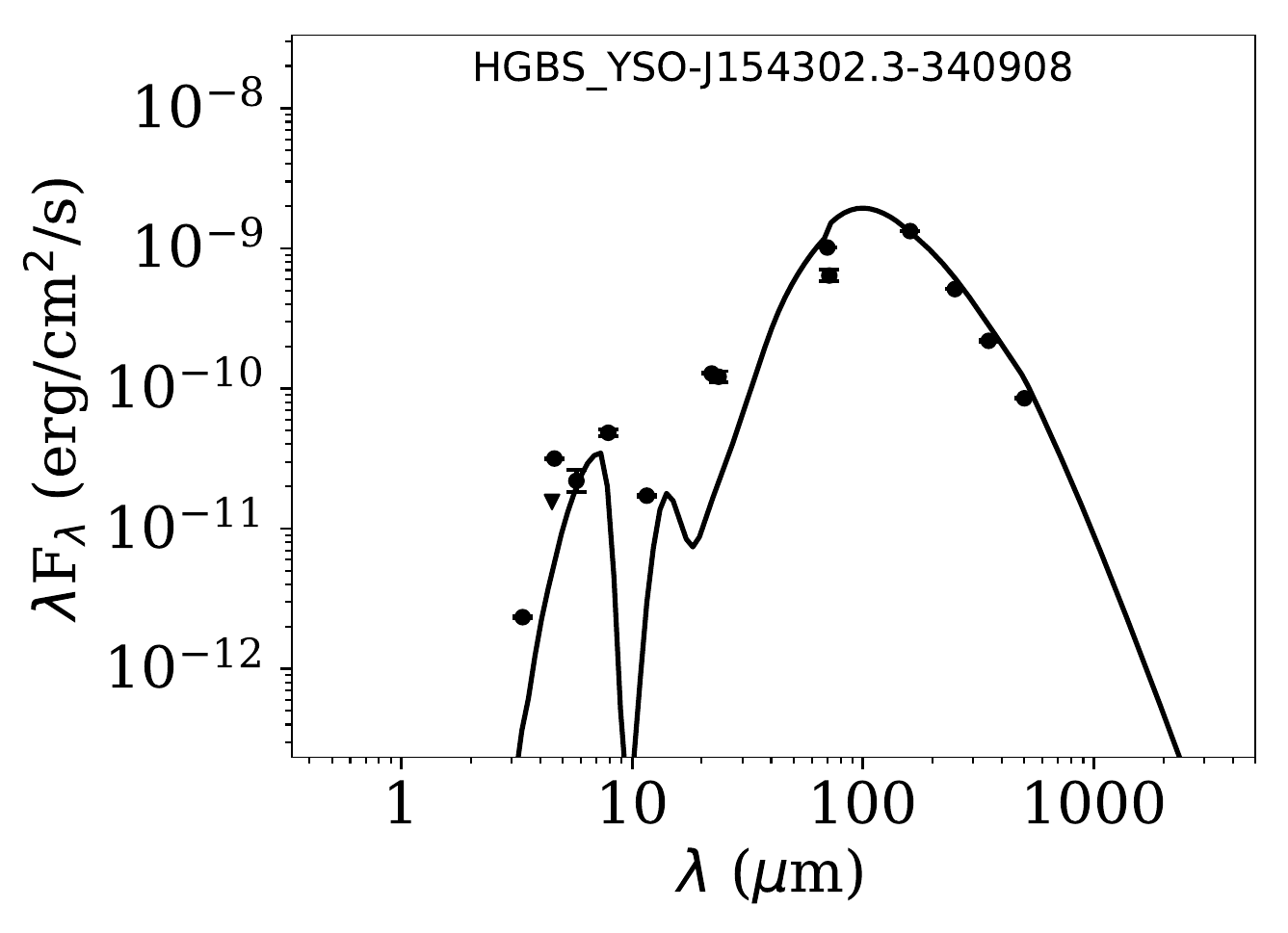} 

\includegraphics[width=0.33\textwidth]{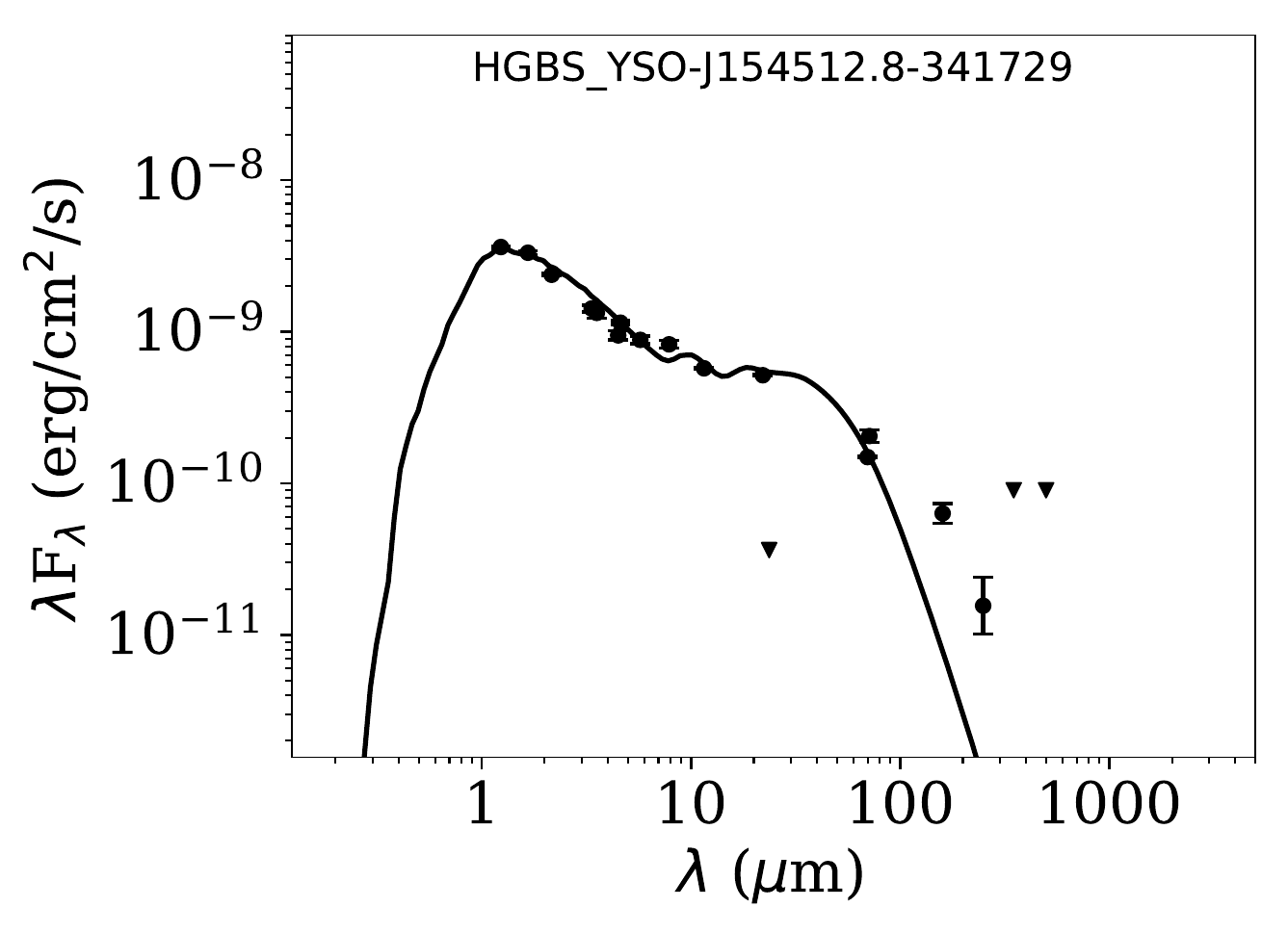} 
\includegraphics[width=0.33\textwidth]{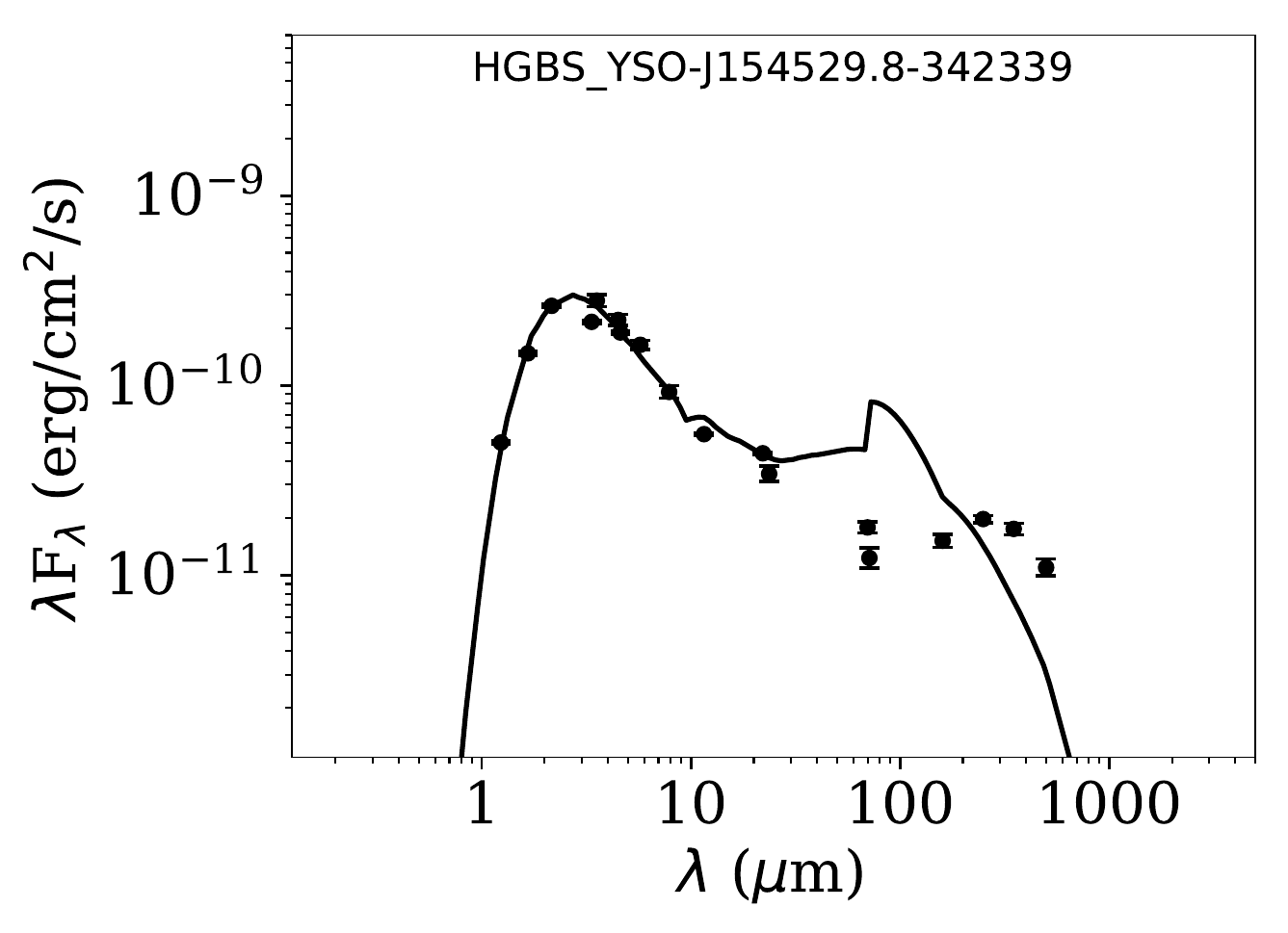} 
\includegraphics[width=0.33\textwidth]{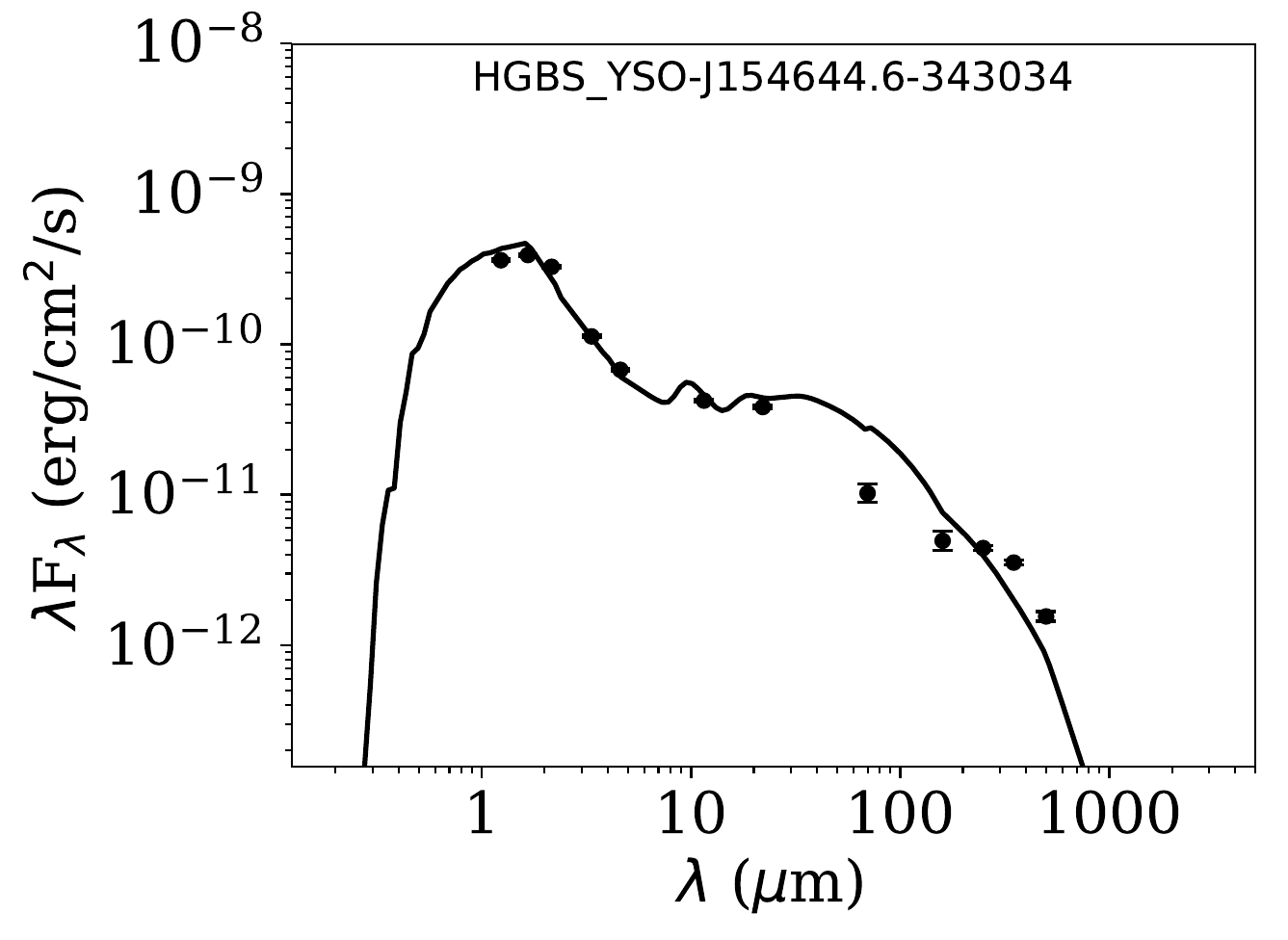} 

\includegraphics[width=0.33\textwidth]{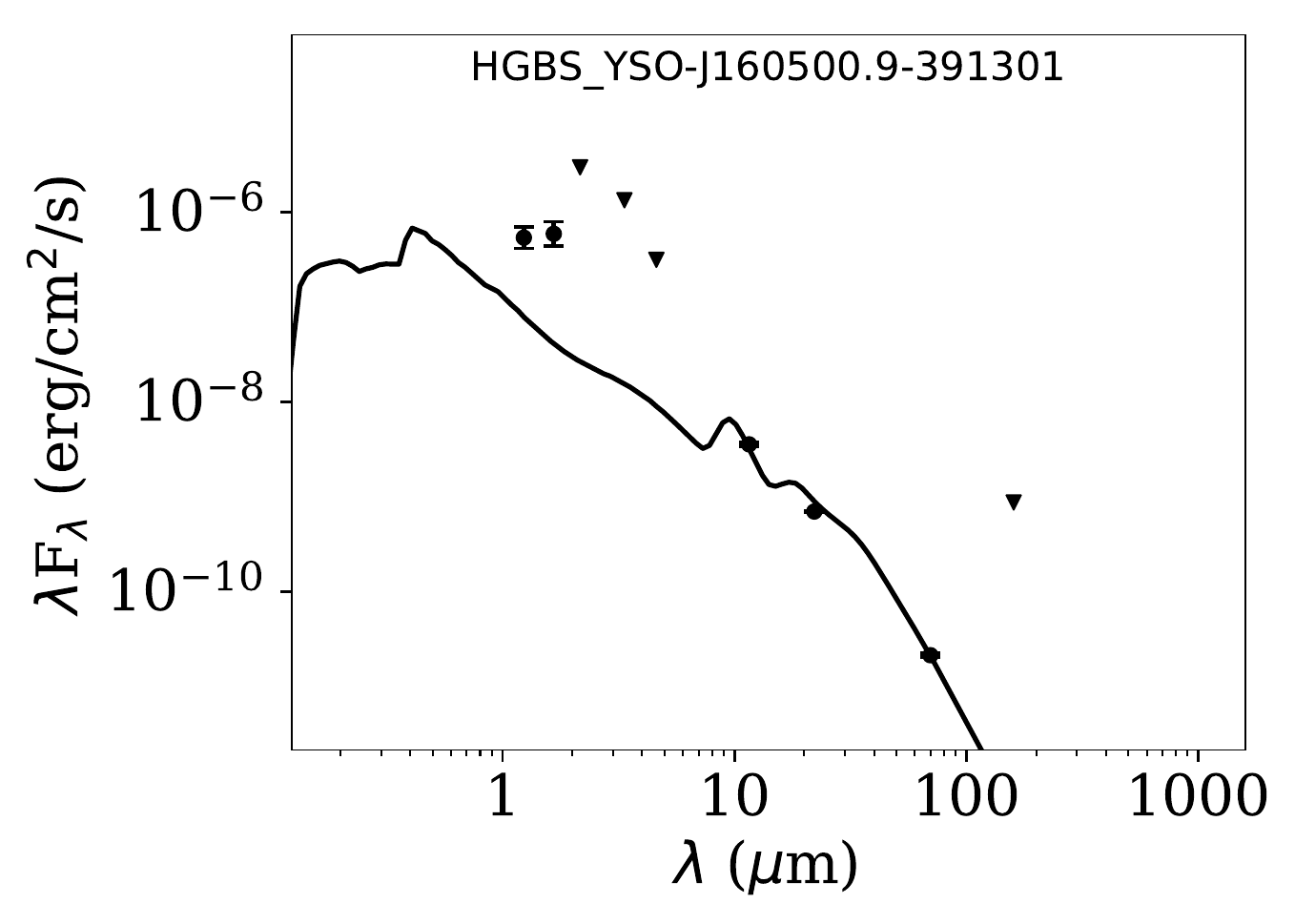} 
\includegraphics[width=0.33\textwidth]{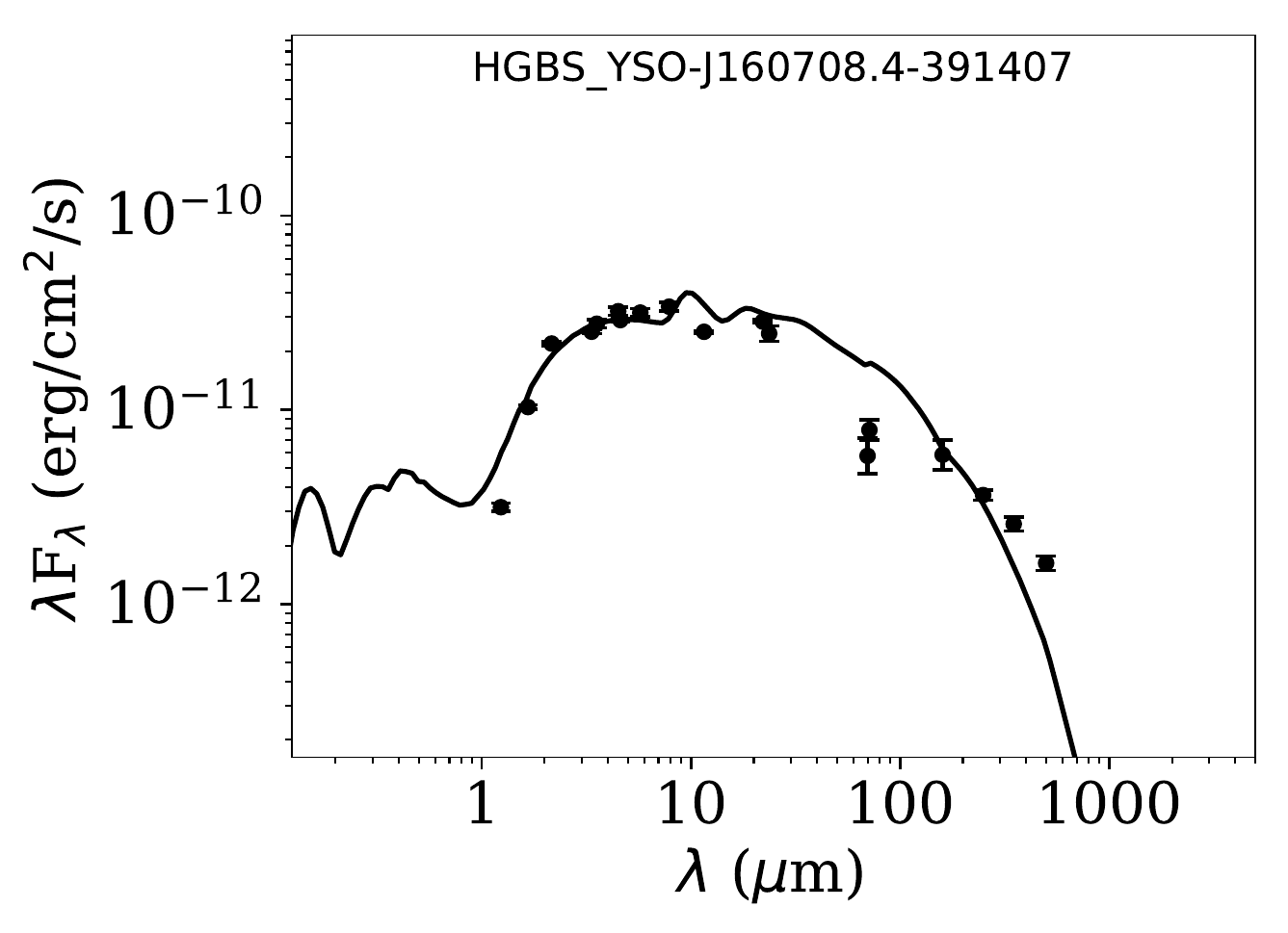}
\includegraphics[width=0.33\textwidth]{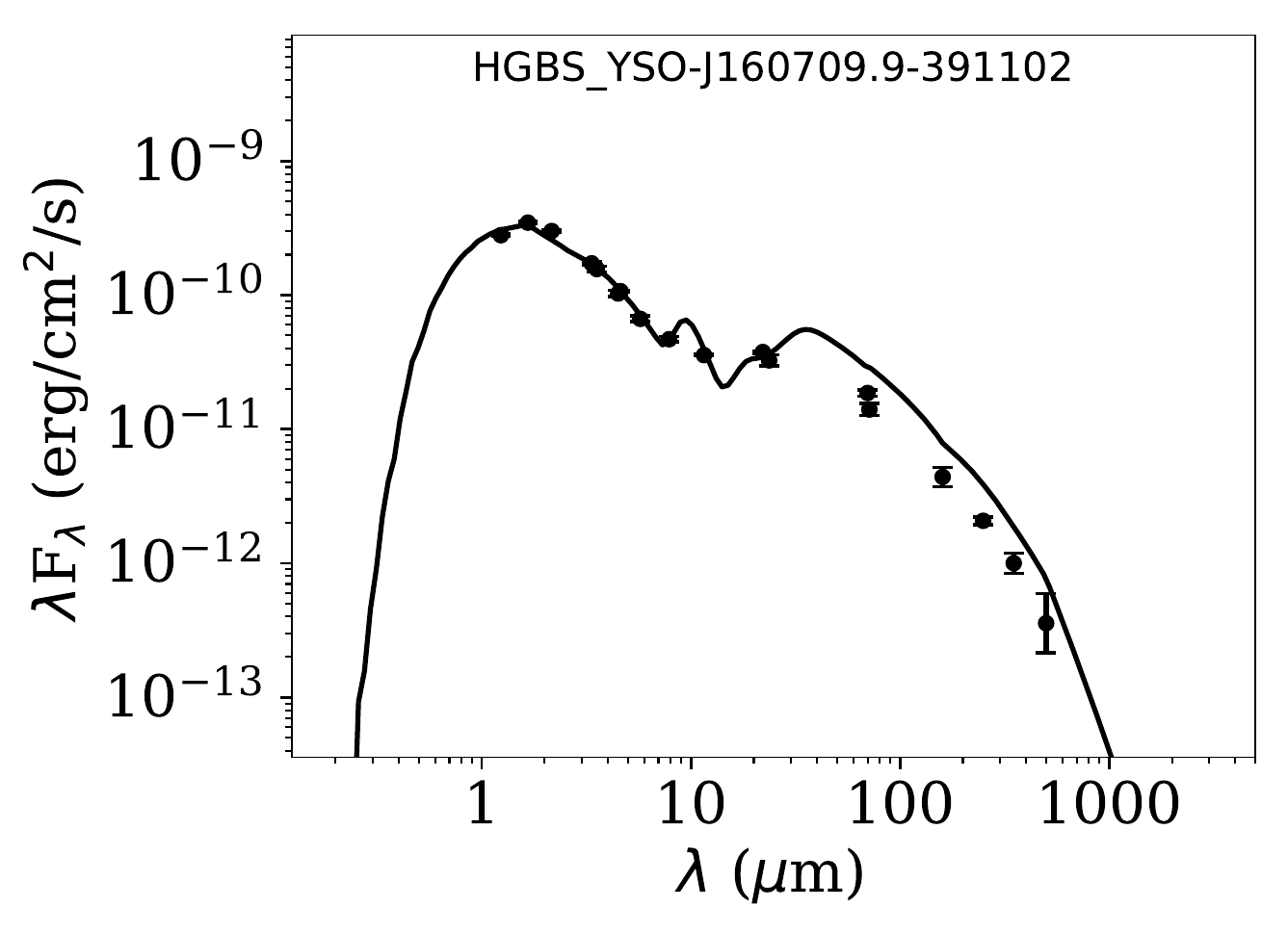}

\includegraphics[width=0.33\textwidth]{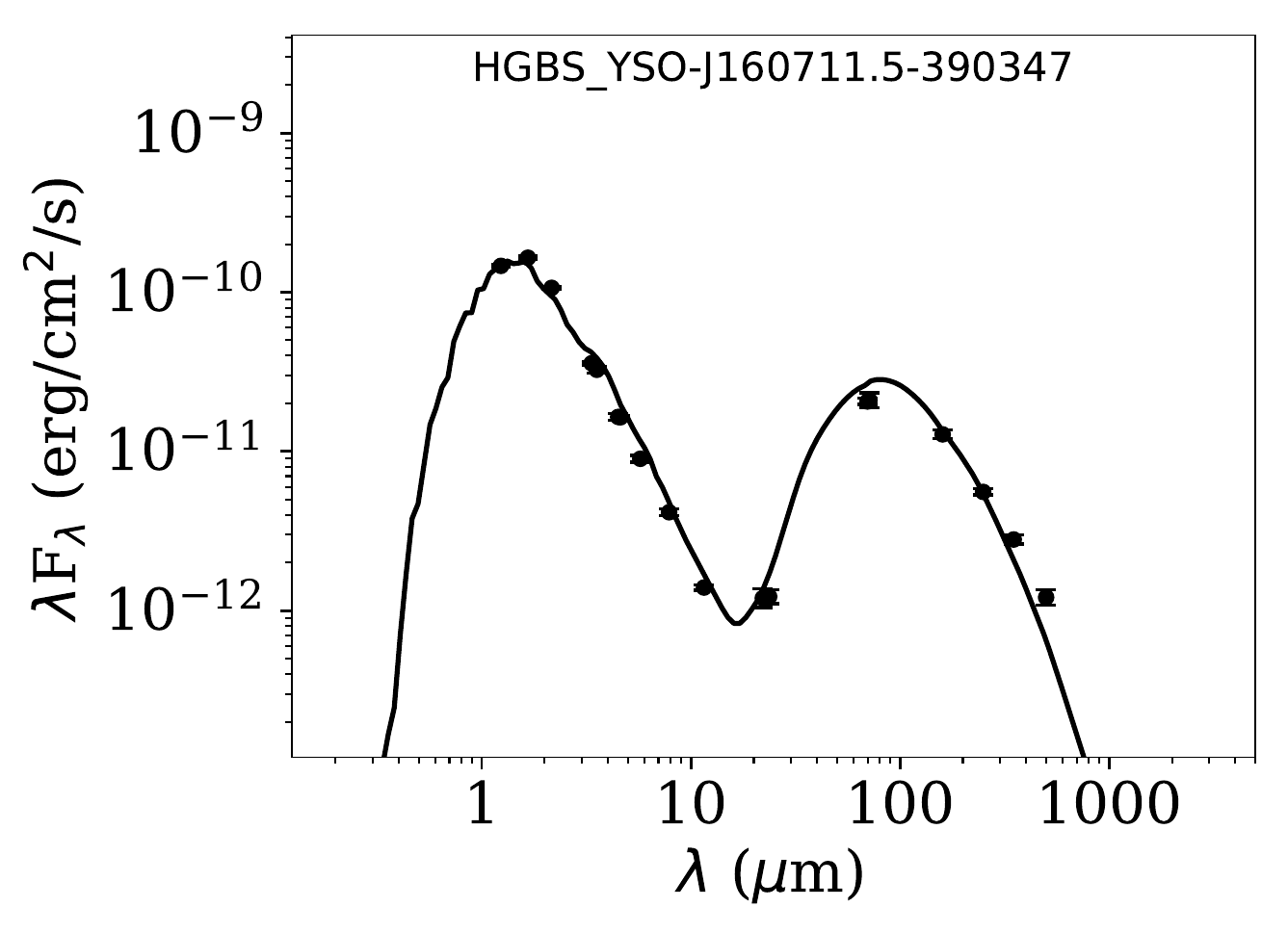} 
\includegraphics[width=0.33\textwidth]{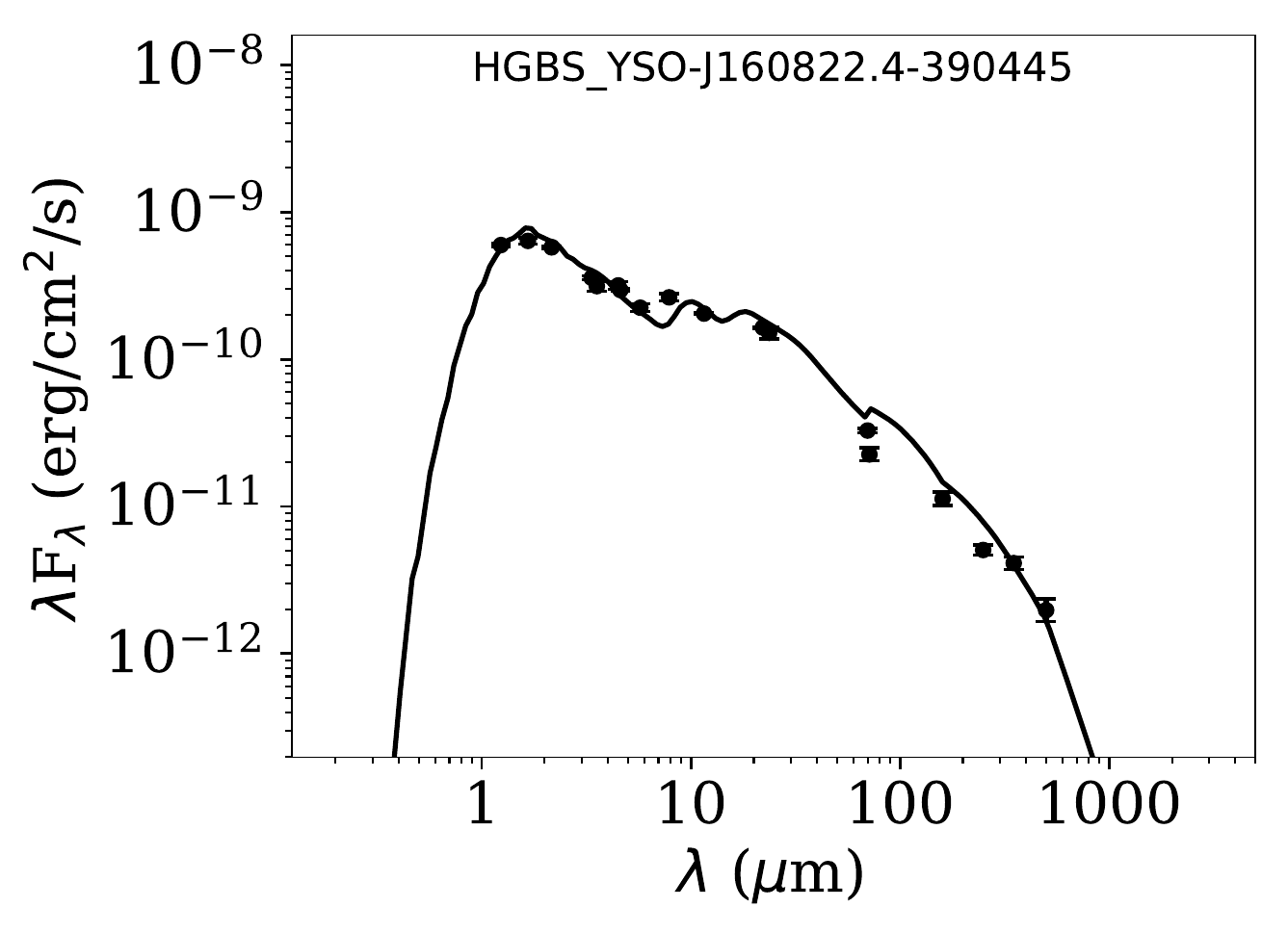} 
\includegraphics[width=0.33\textwidth]{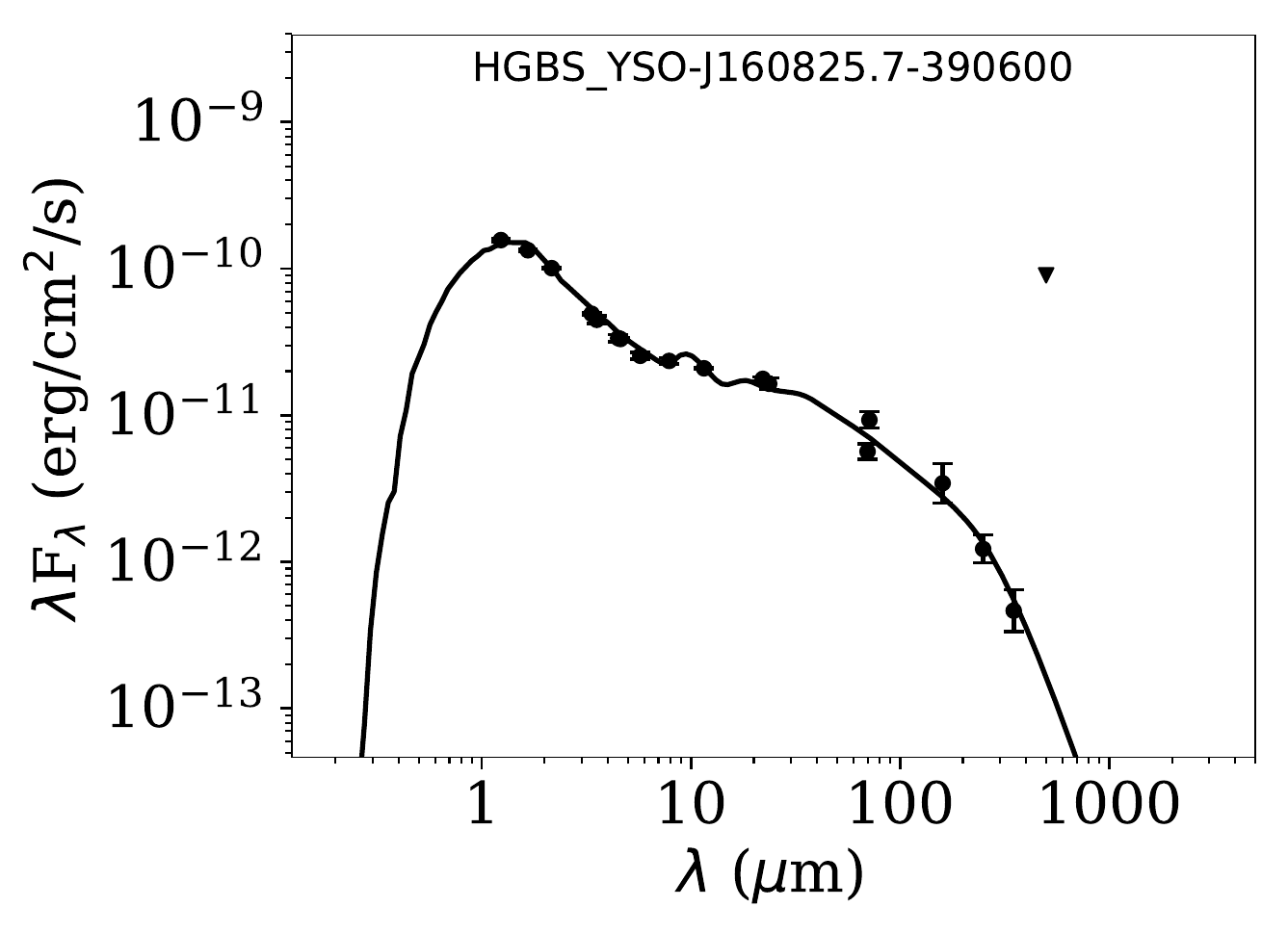}

\caption{SED of the YSOs/protostars of the \her\, catalogue from NIR to FIR wavelengths. Triangles are upper limits. The best fit model of the SED fitting with the \citet{robitaille17} synthetic SED is shown.}
\label{fig:sed}       
\end{figure*}

\begin{figure*}[!ht]
\renewcommand{\thefigure}{\arabic{figure} (Cont.)}
\addtocounter{figure}{-1}

\includegraphics[width=0.33\textwidth]{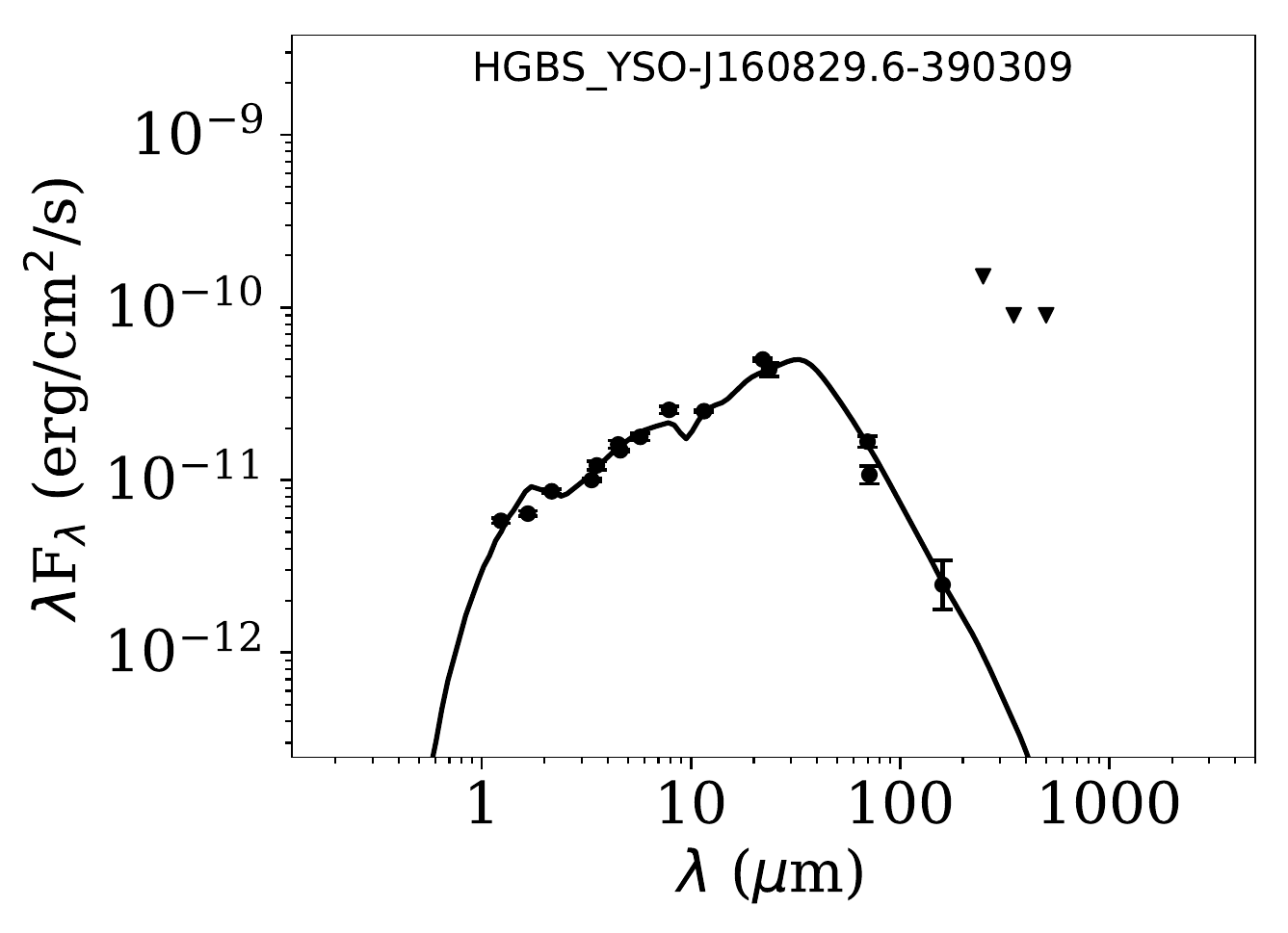} 
\includegraphics[width=0.33\textwidth]{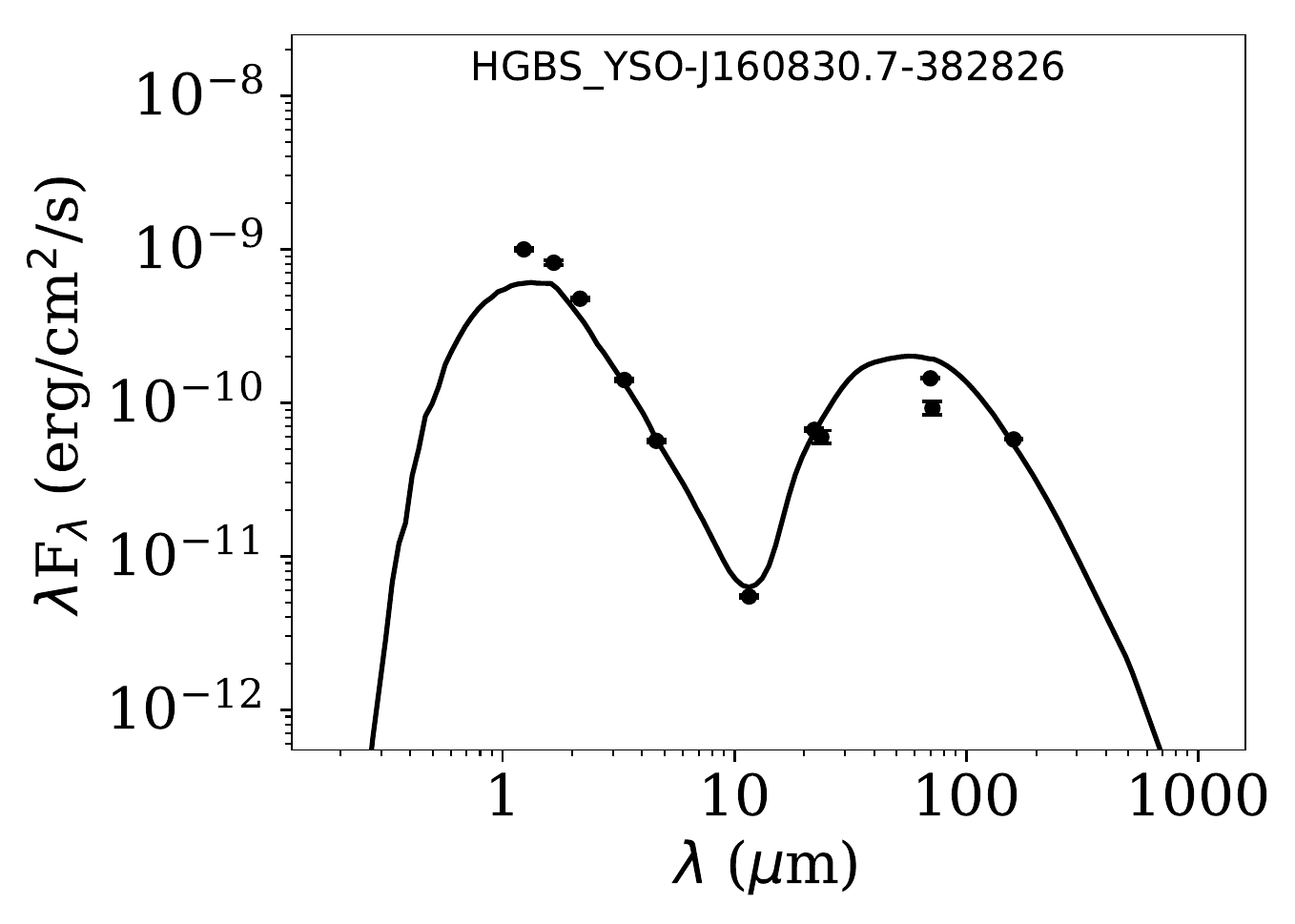} 
\includegraphics[width=0.33\textwidth]{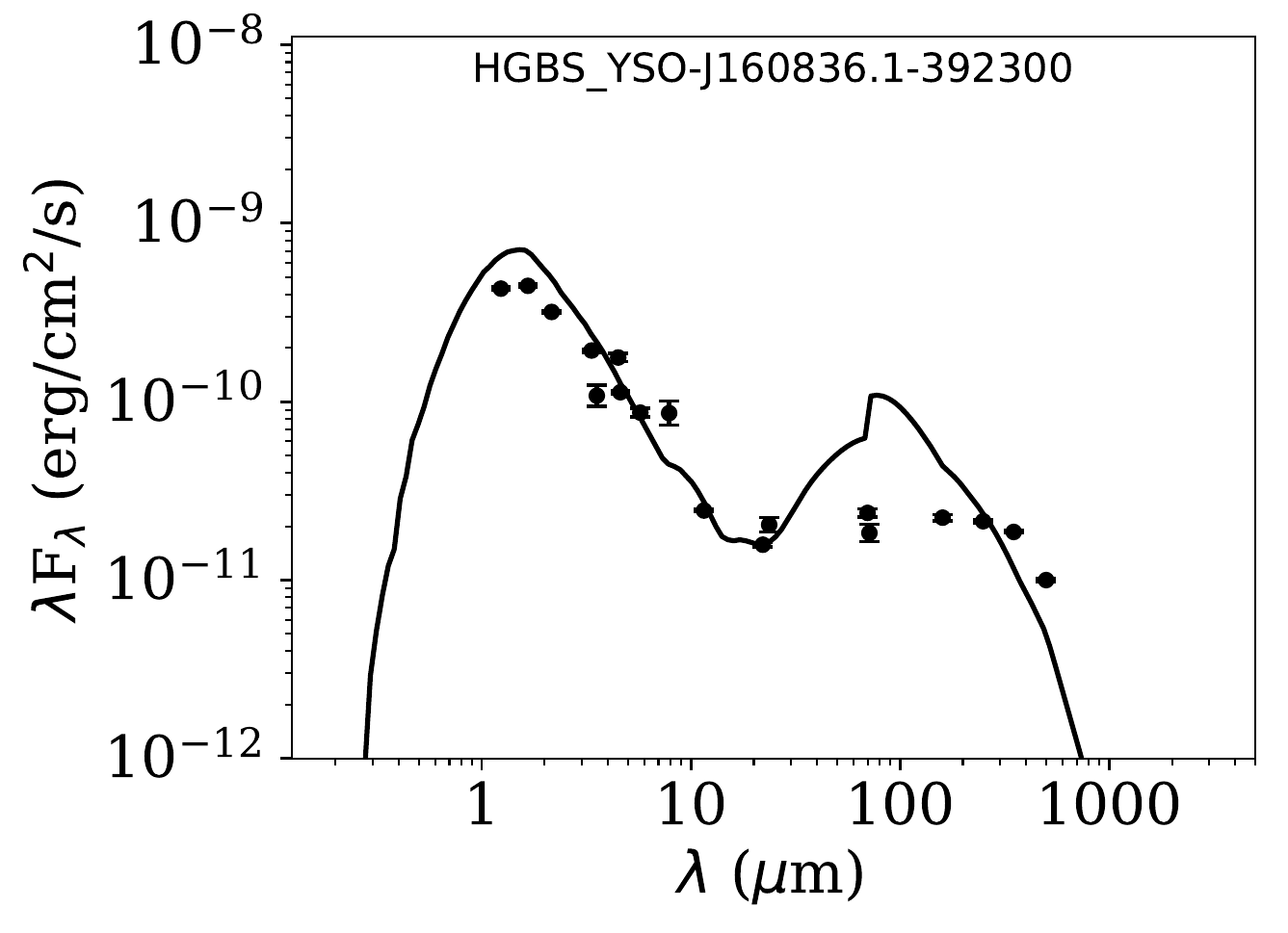} 

\includegraphics[width=0.33\textwidth]{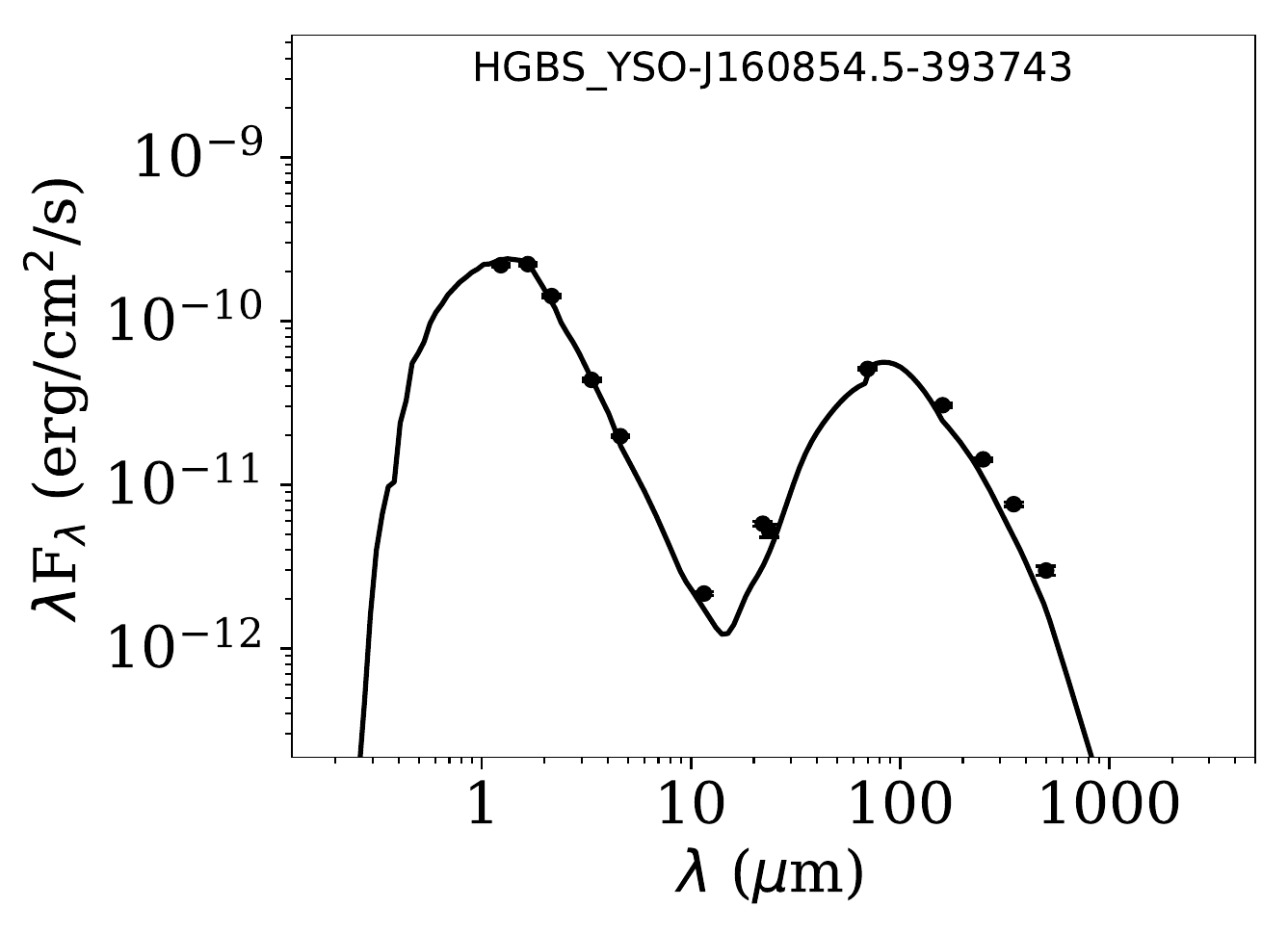} 
\includegraphics[width=0.33\textwidth]{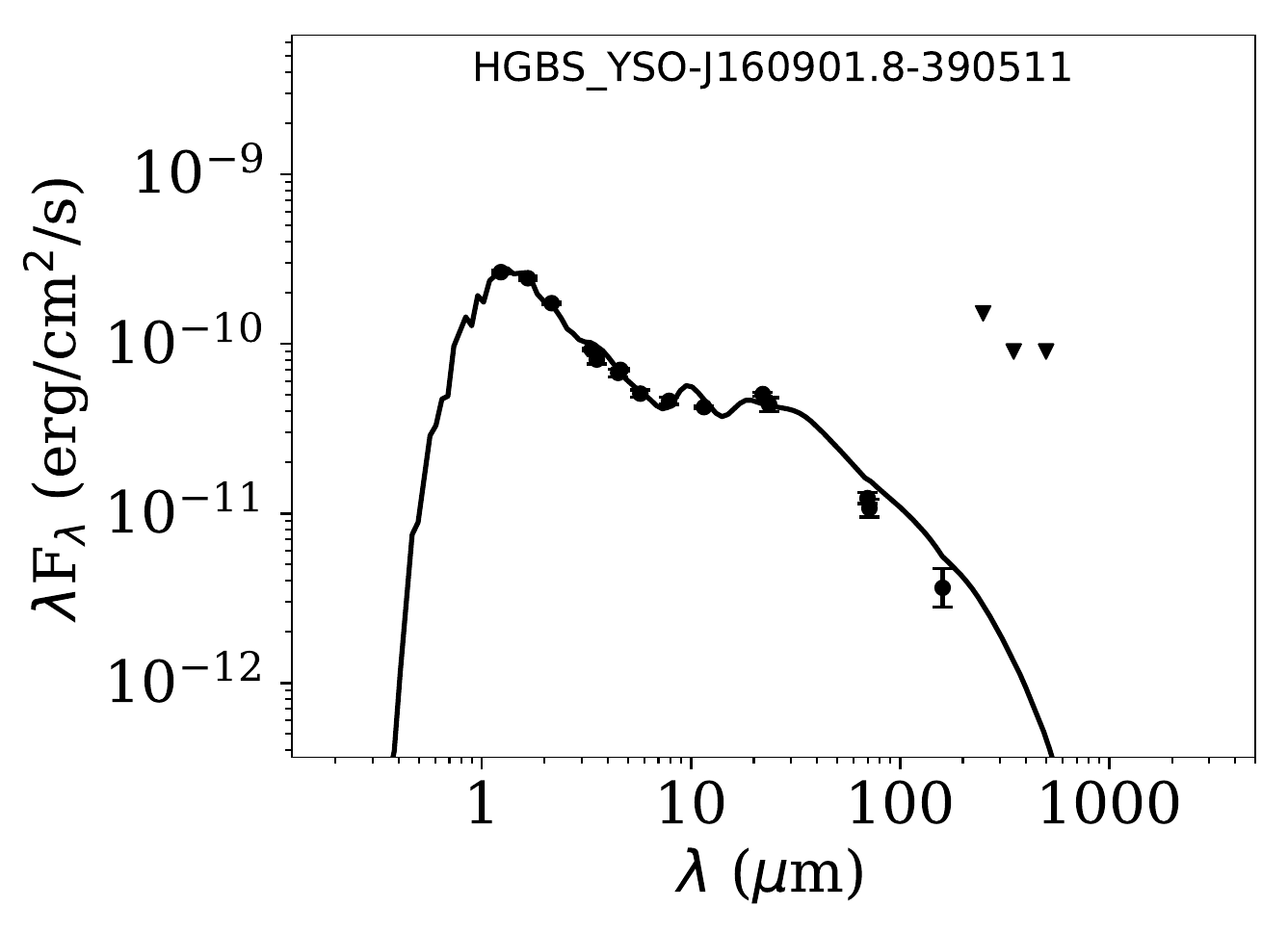} 
\includegraphics[width=0.33\textwidth]{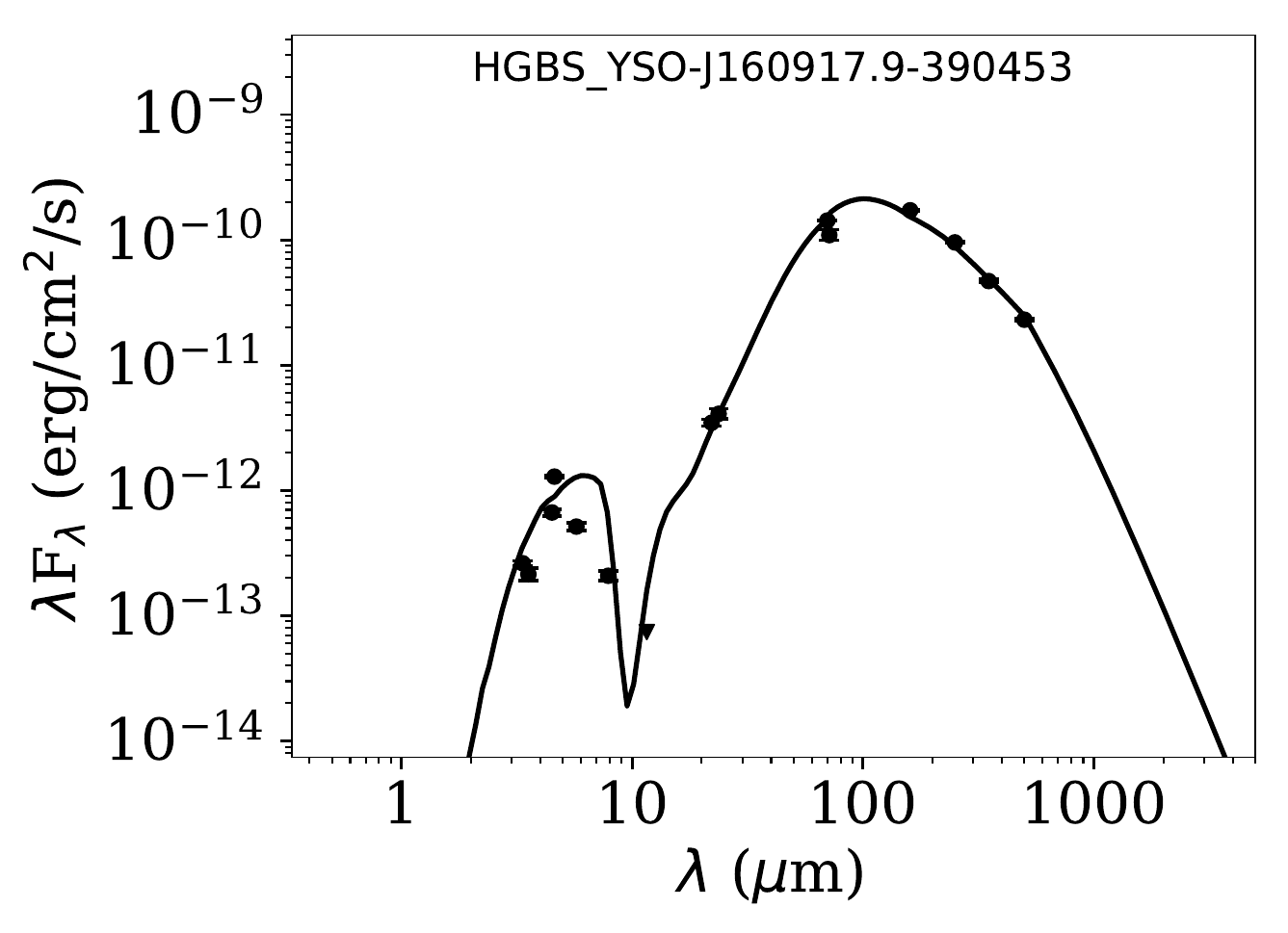}

\includegraphics[width=0.33\textwidth]{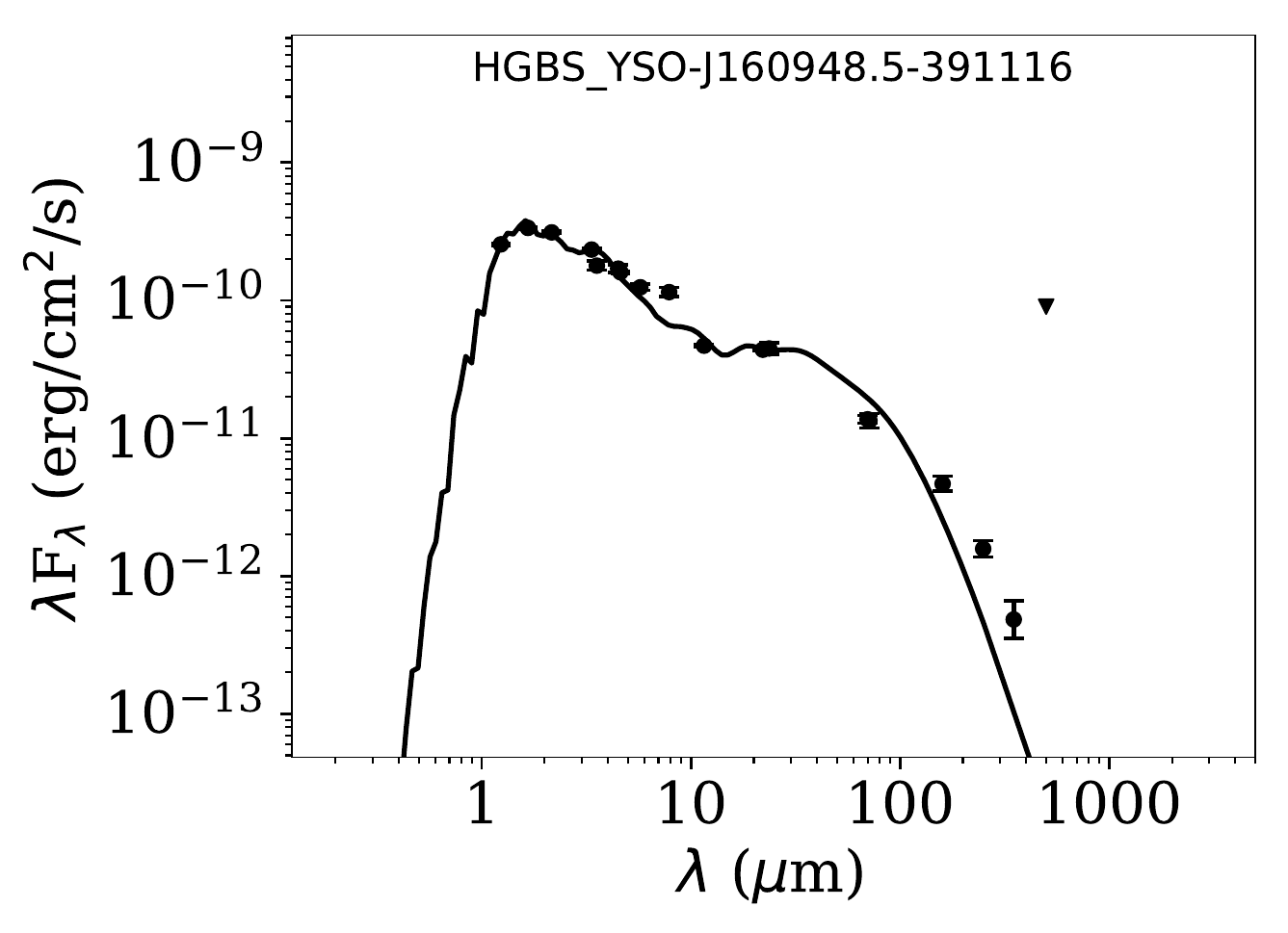}
\includegraphics[width=0.33\textwidth]{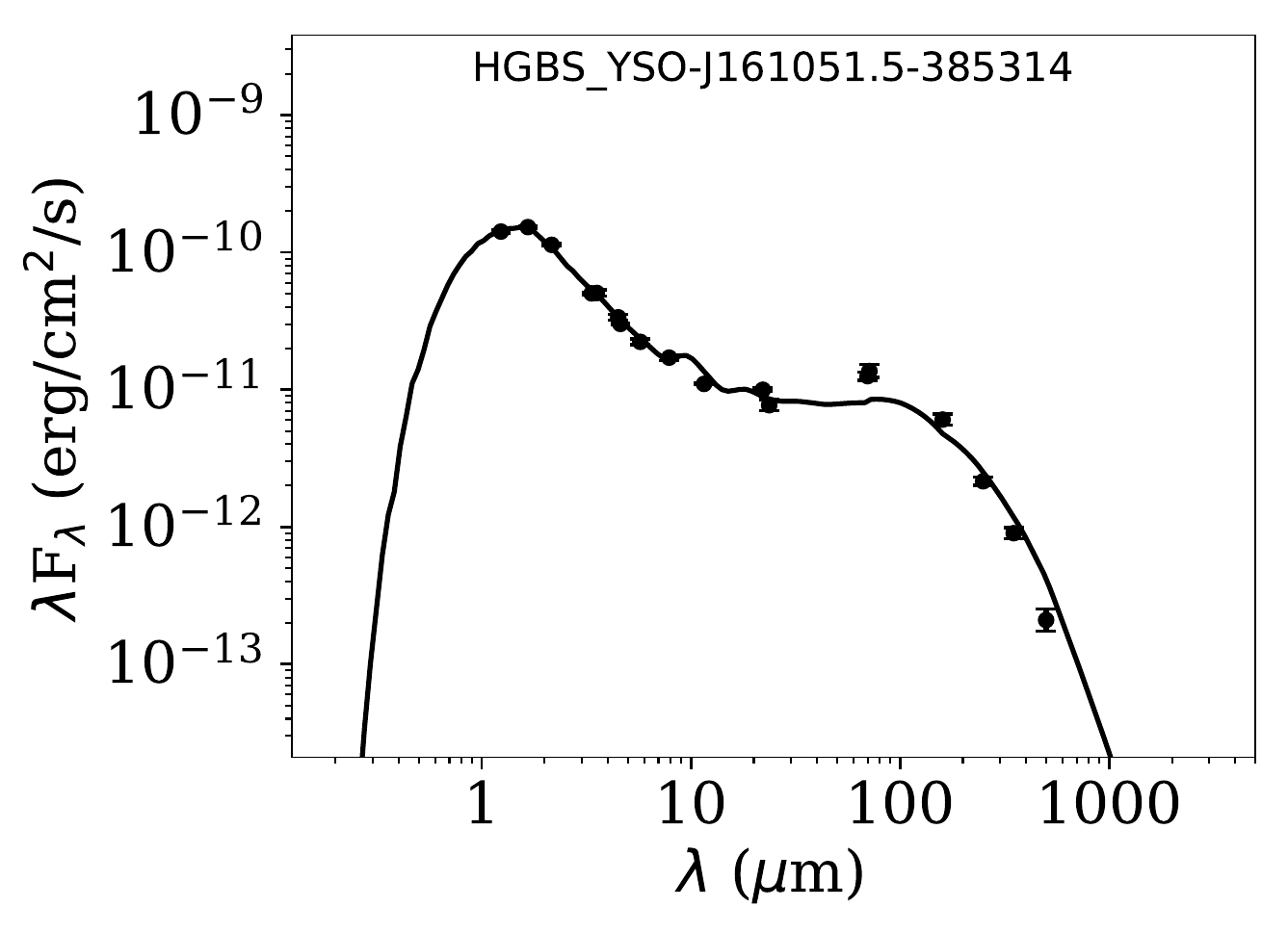}
\includegraphics[width=0.33\textwidth]{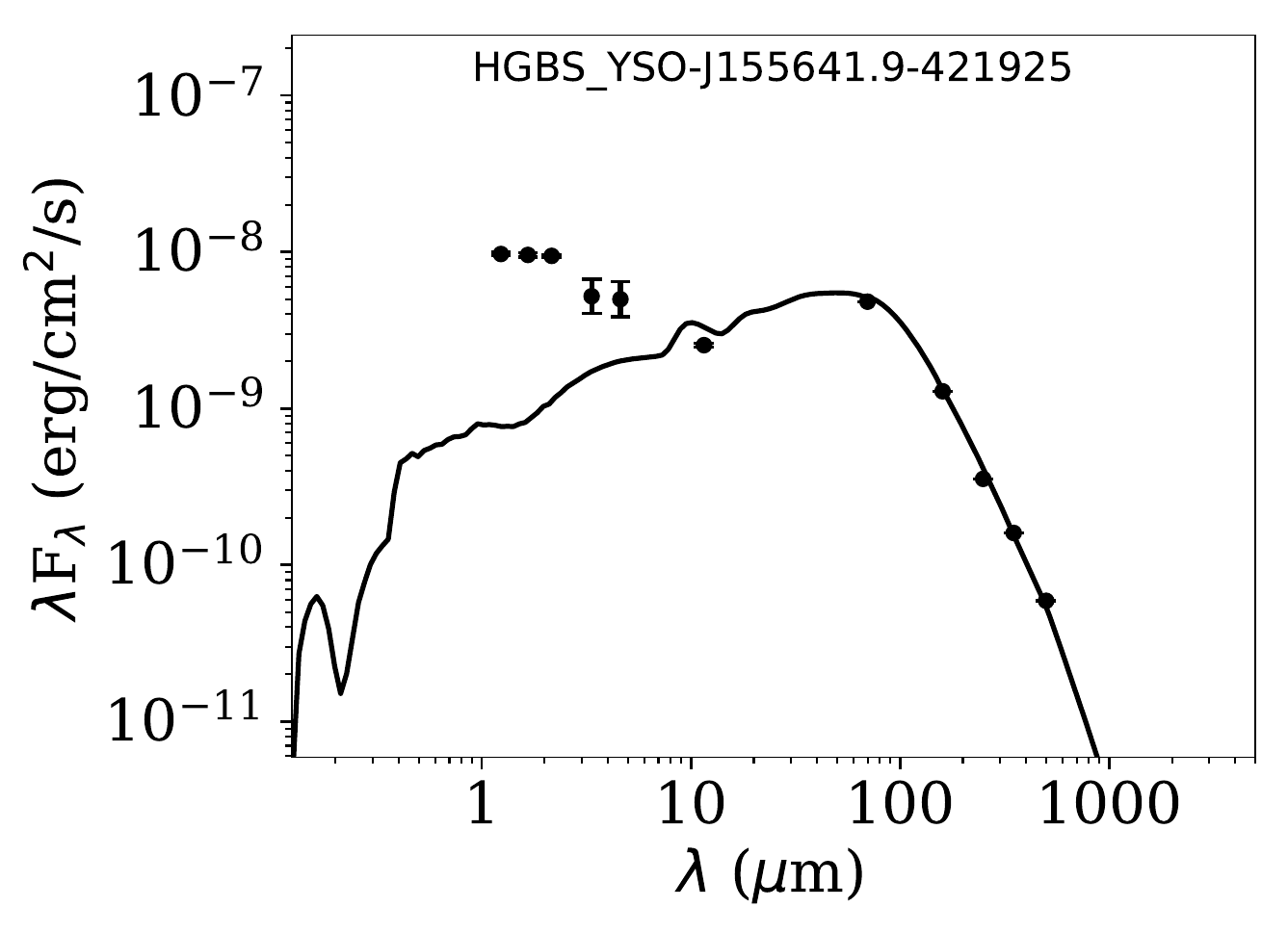}

\includegraphics[width=0.33\textwidth]{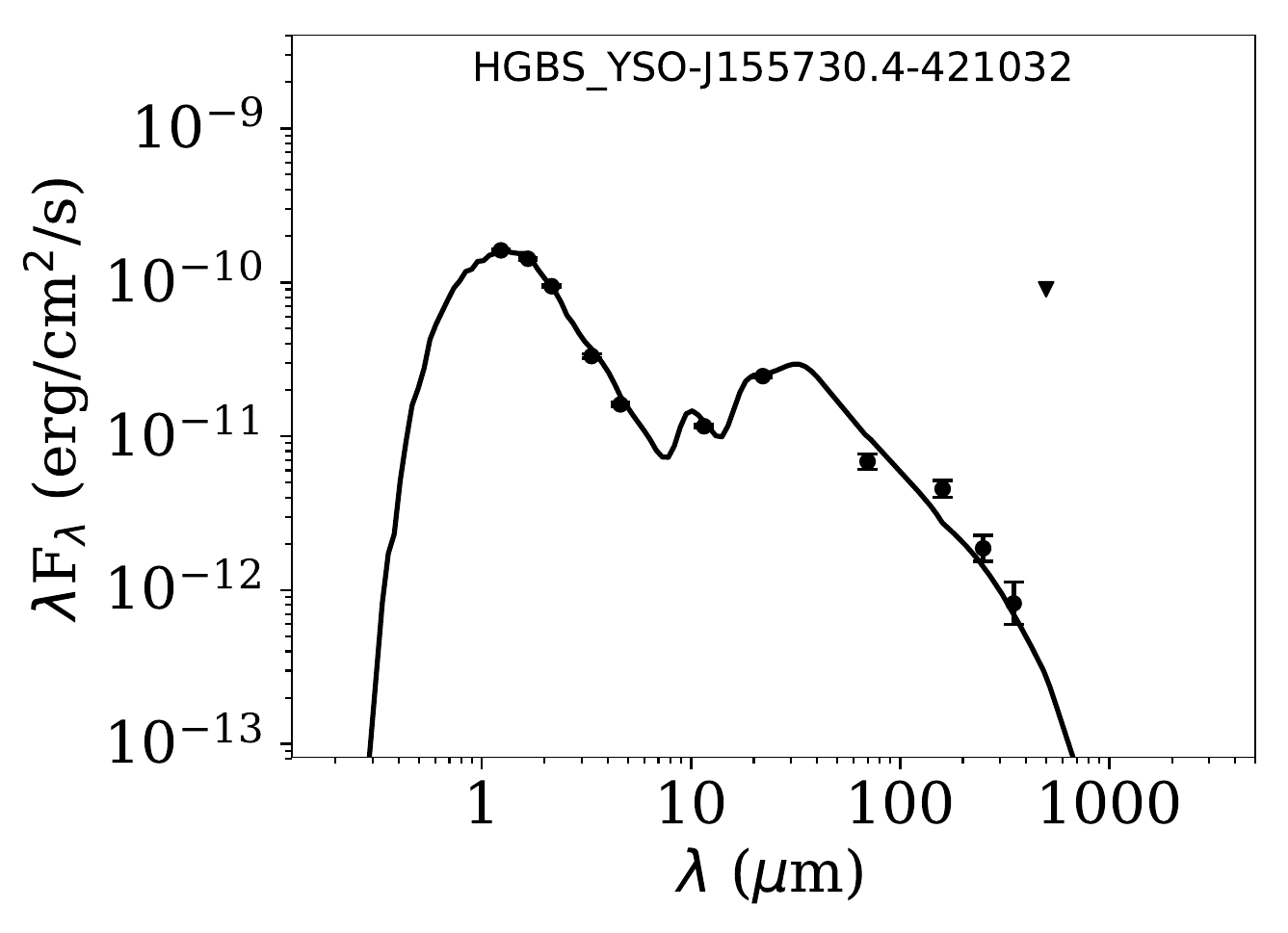}
\includegraphics[width=0.33\textwidth]{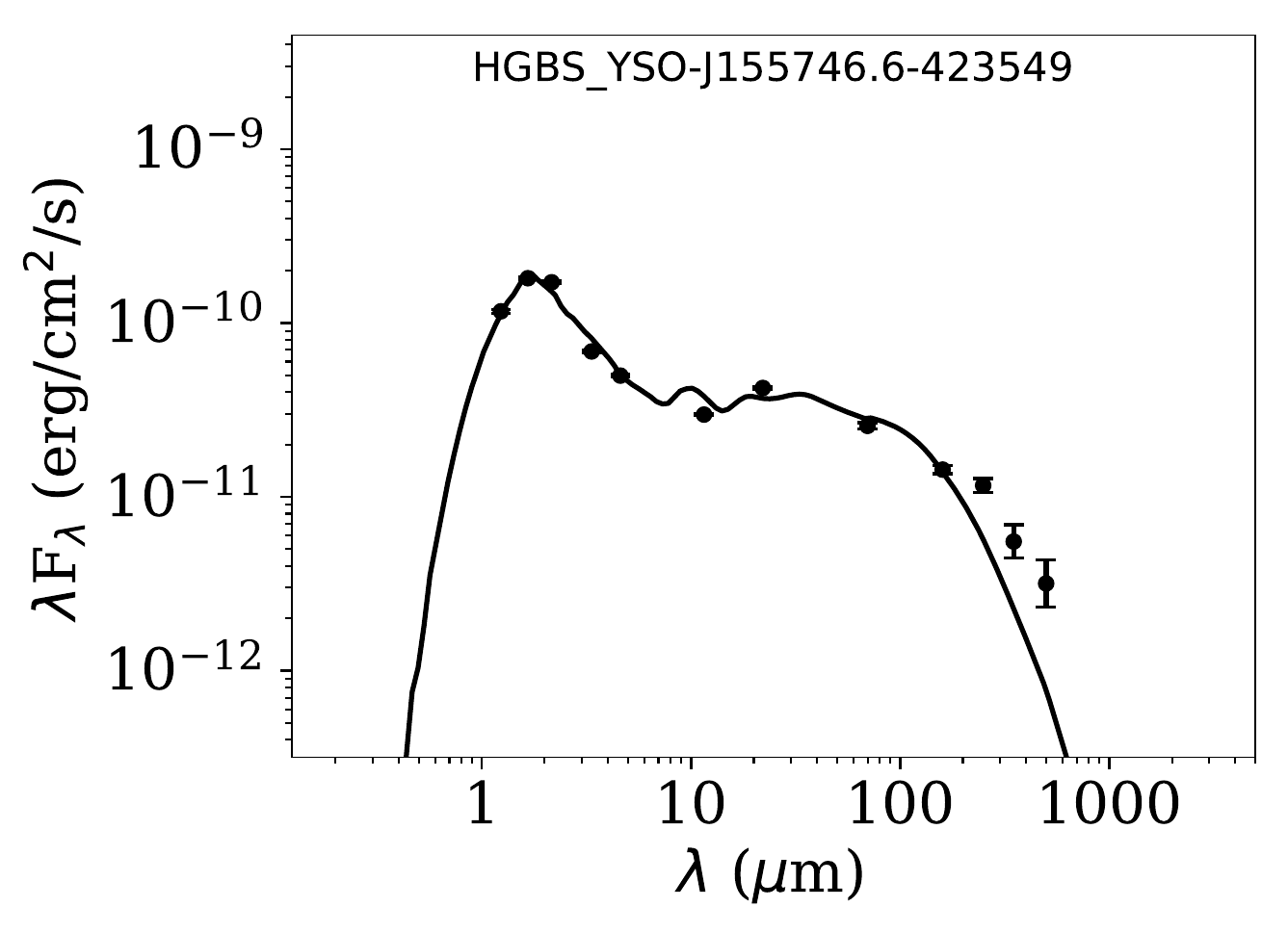}
\includegraphics[width=0.33\textwidth]{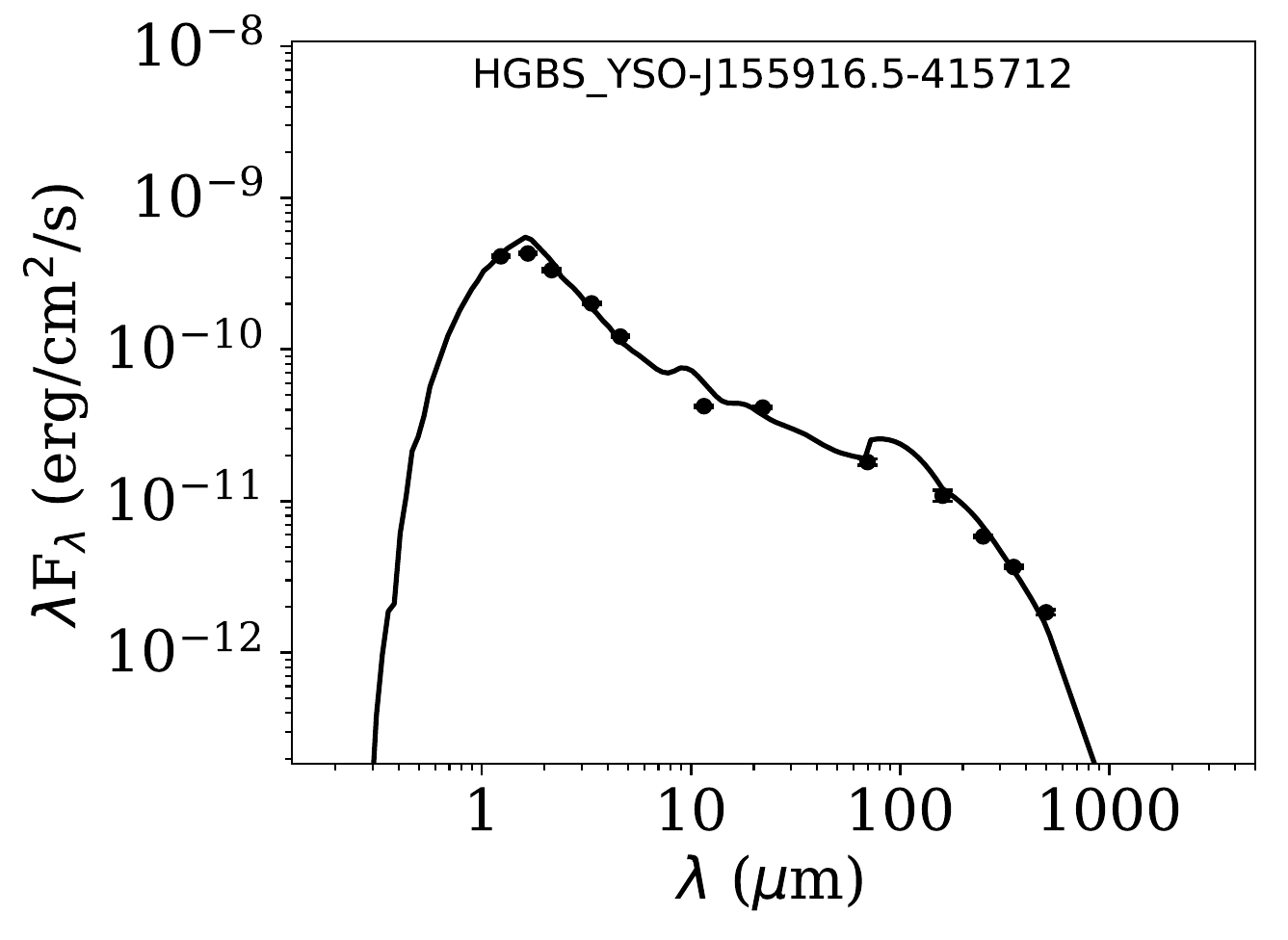}

\includegraphics[width=0.33\textwidth]{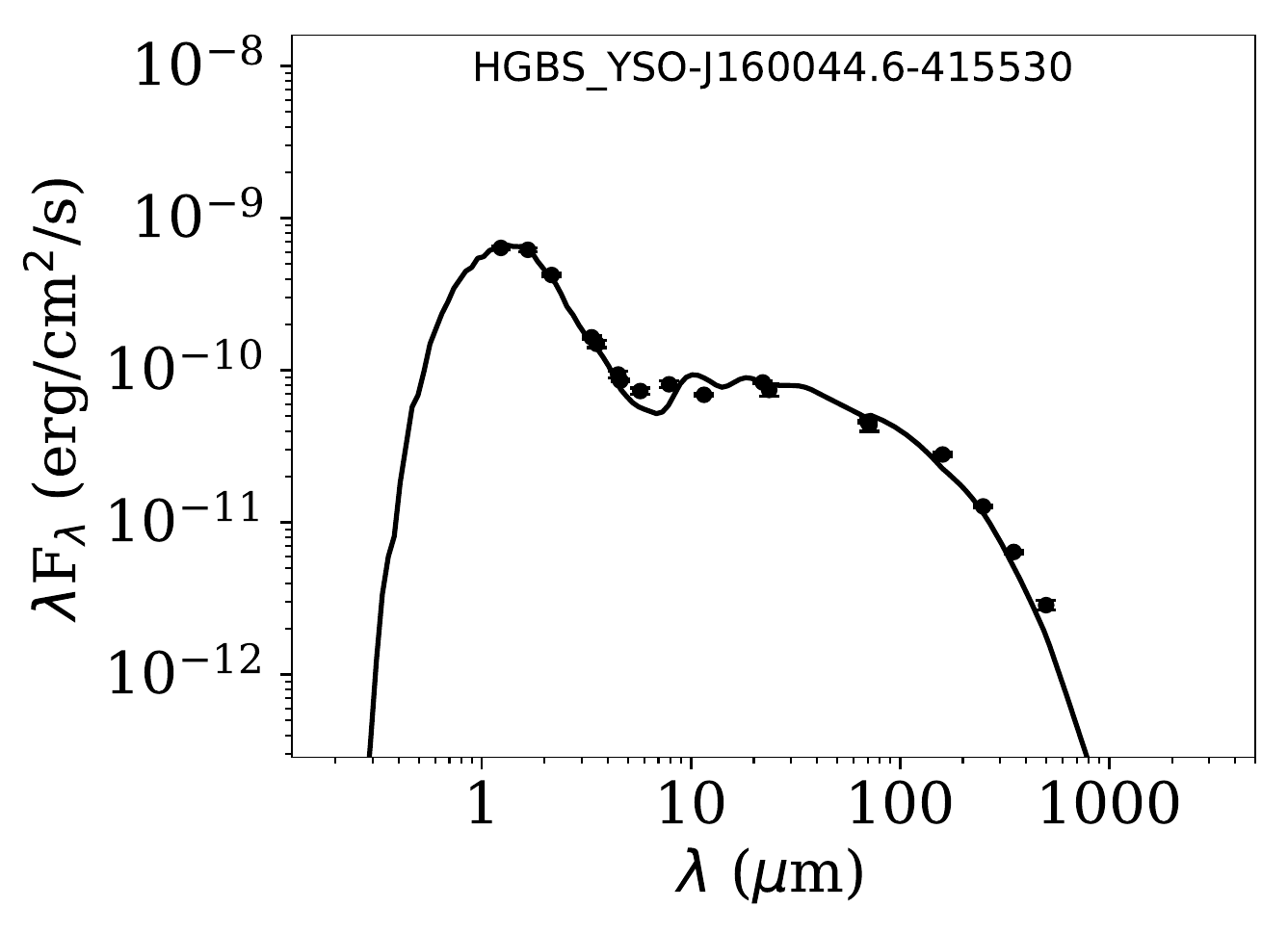}
\includegraphics[width=0.33\textwidth]{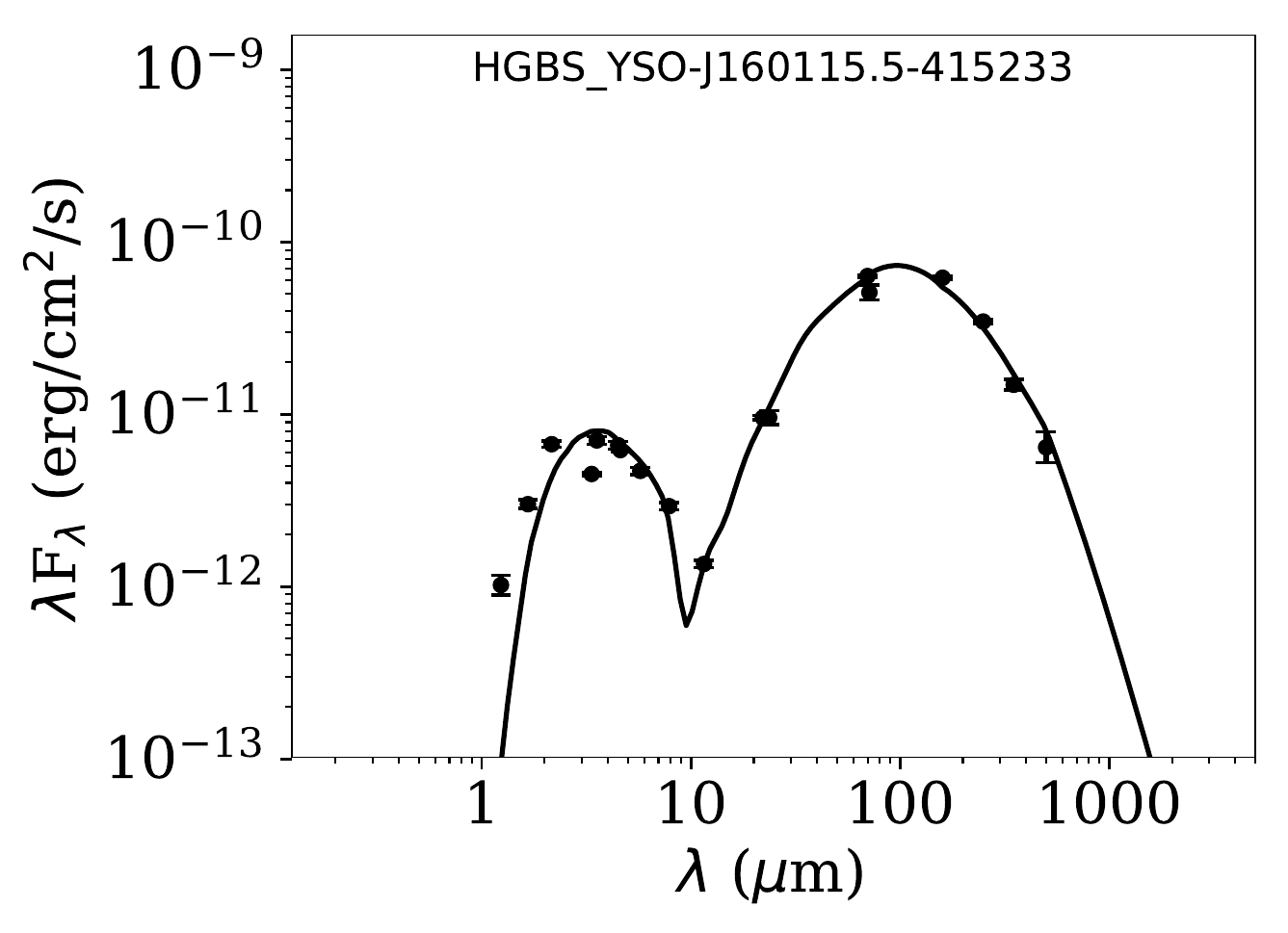}
\includegraphics[width=0.33\textwidth]{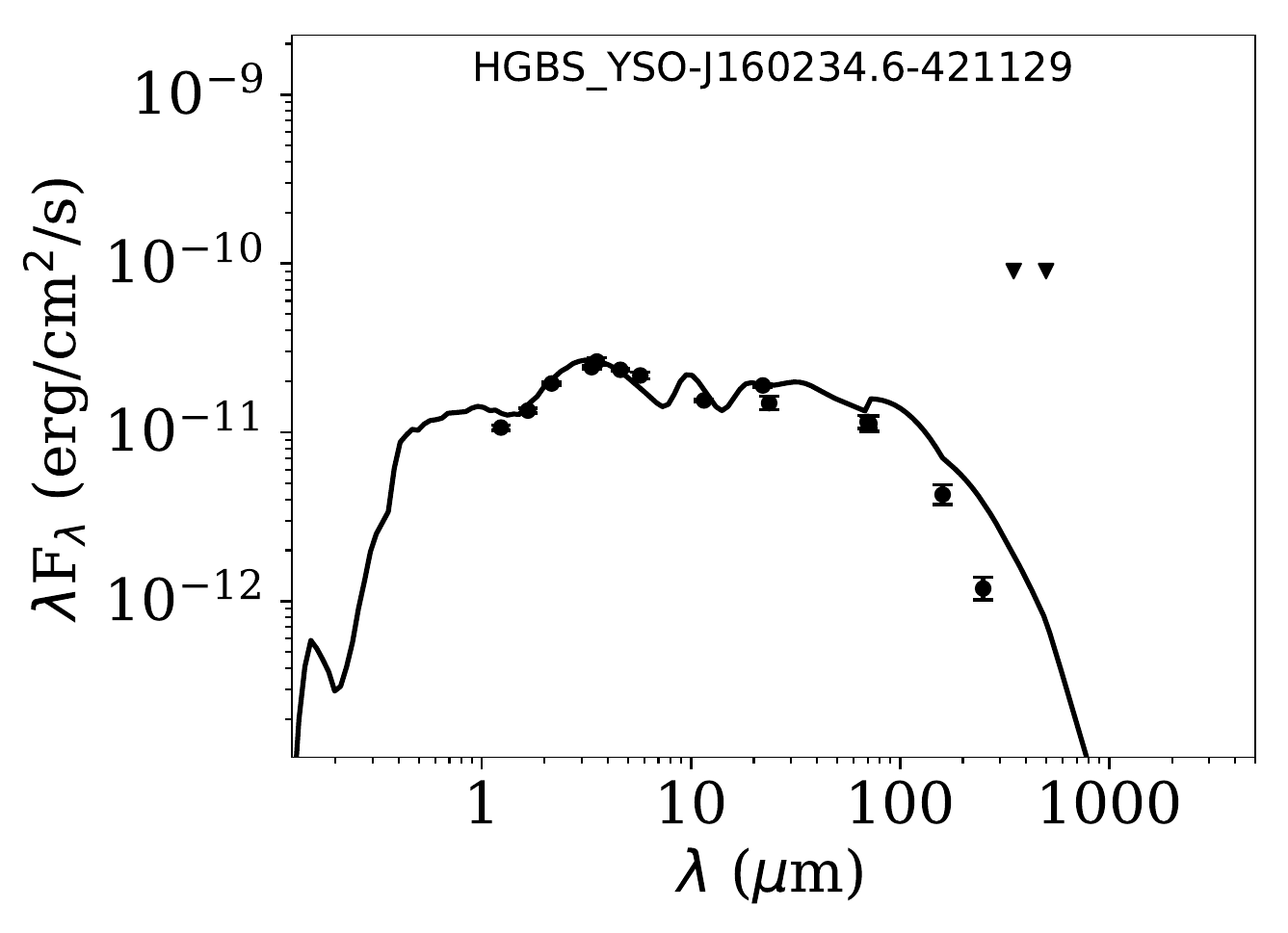}
\caption{}
\renewcommand{\thefigure}{\arabic{figure}}
\end{figure*}

\begin{figure*}[!ht]
\renewcommand{\thefigure}{\arabic{figure} (Cont.)}
\addtocounter{figure}{-1}

\includegraphics[width=0.33\textwidth]{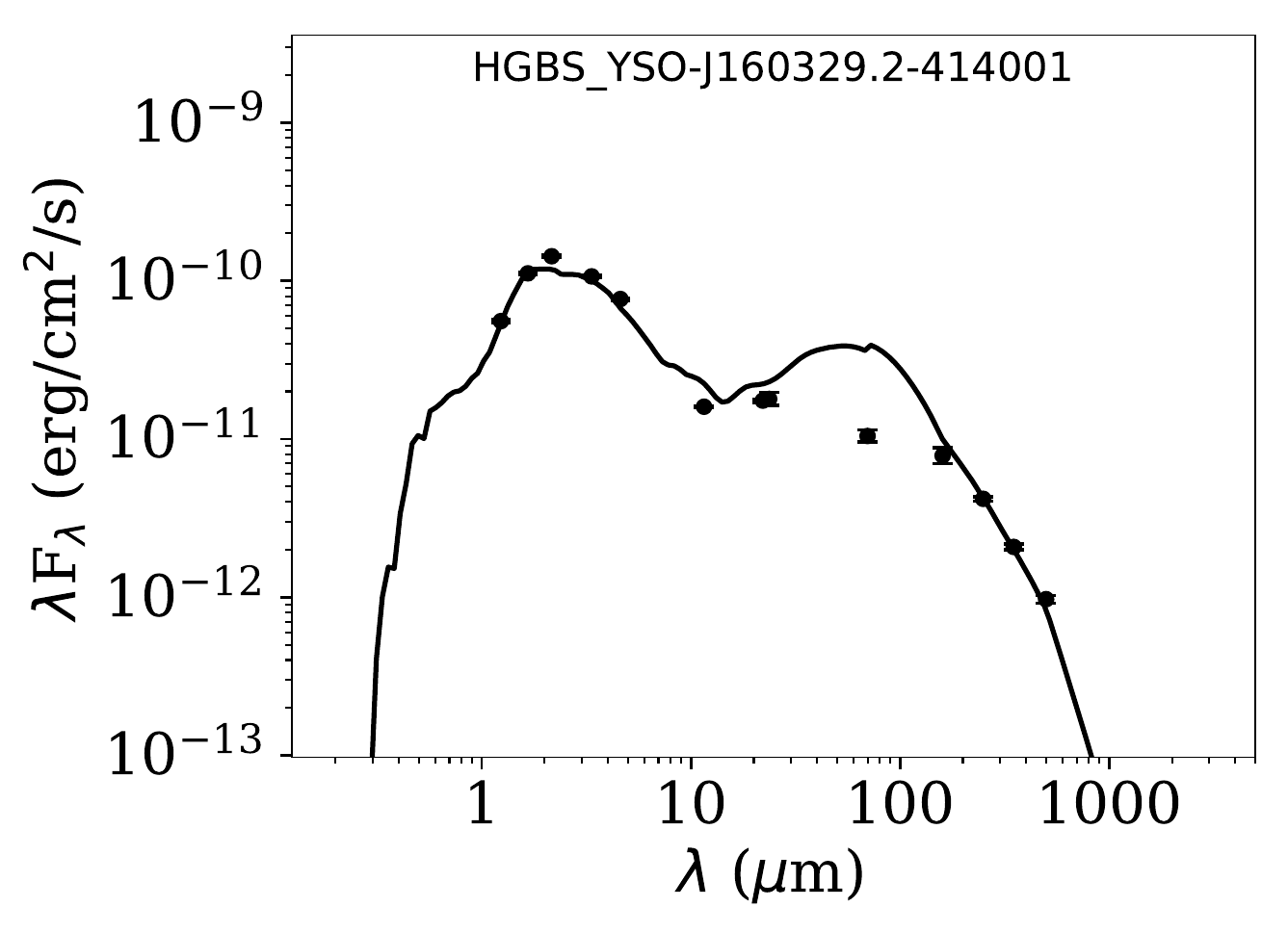}
\includegraphics[width=0.33\textwidth]{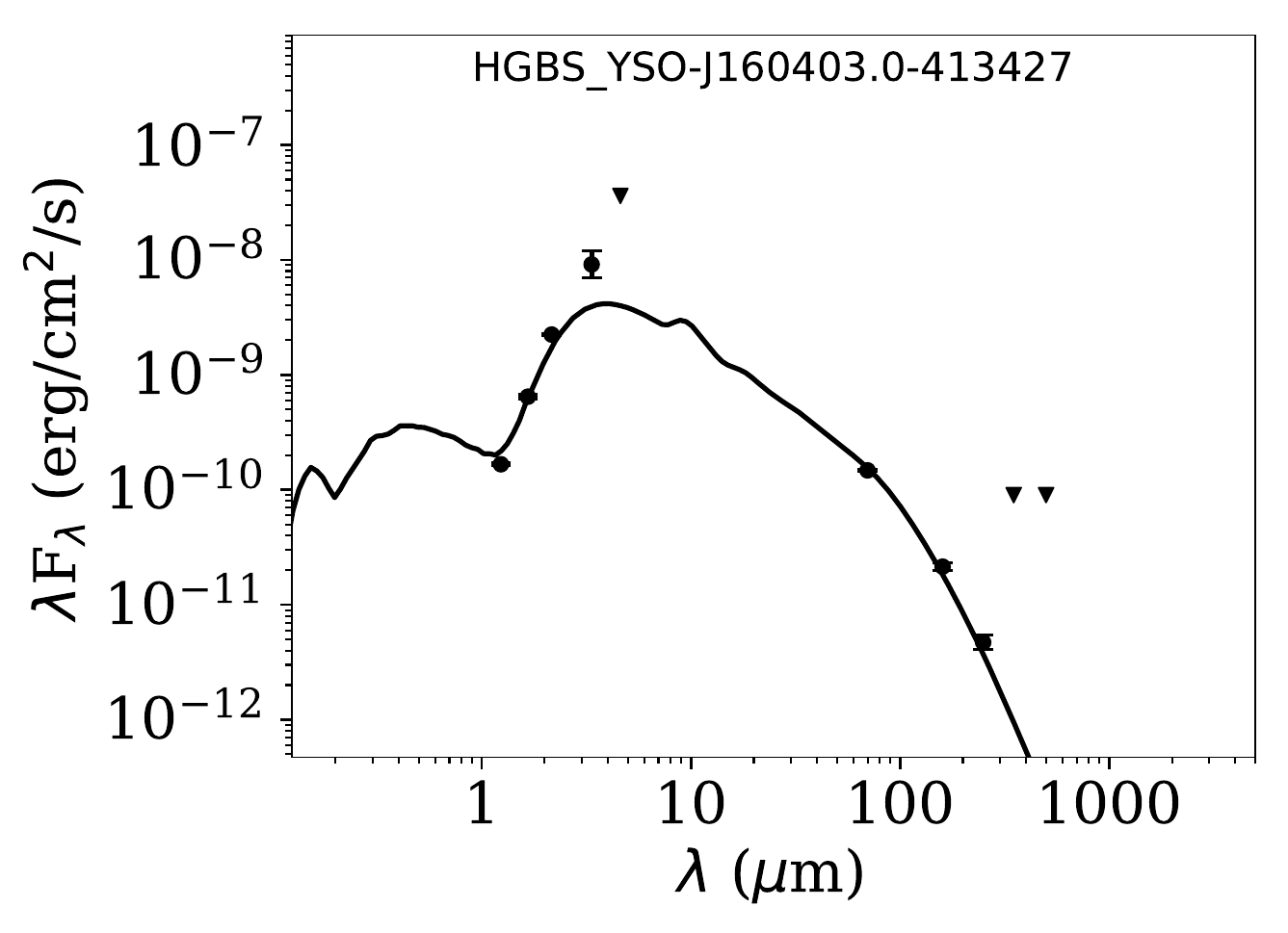} 
\includegraphics[width=0.33\textwidth]{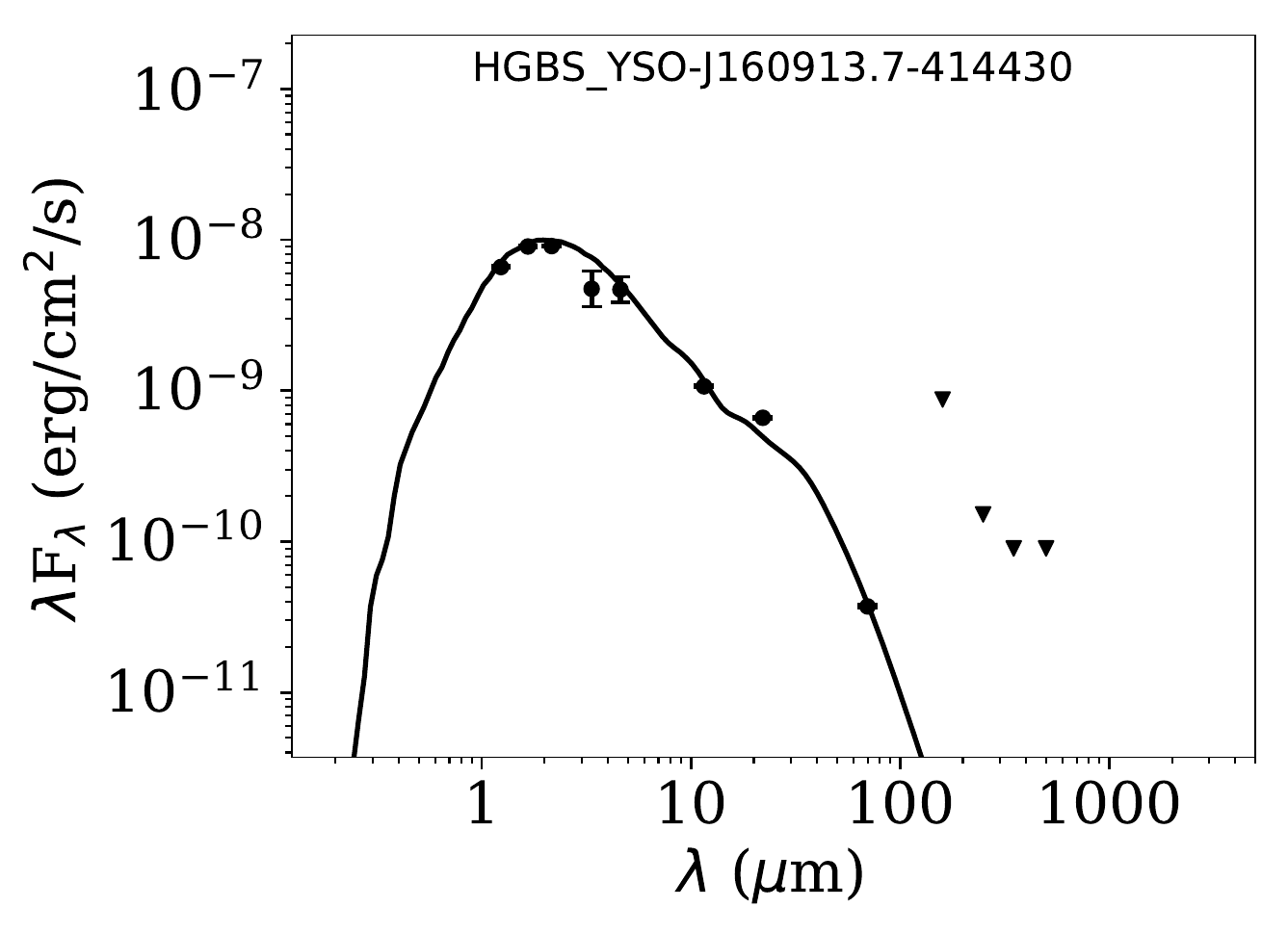}

\includegraphics[width=0.33\textwidth]{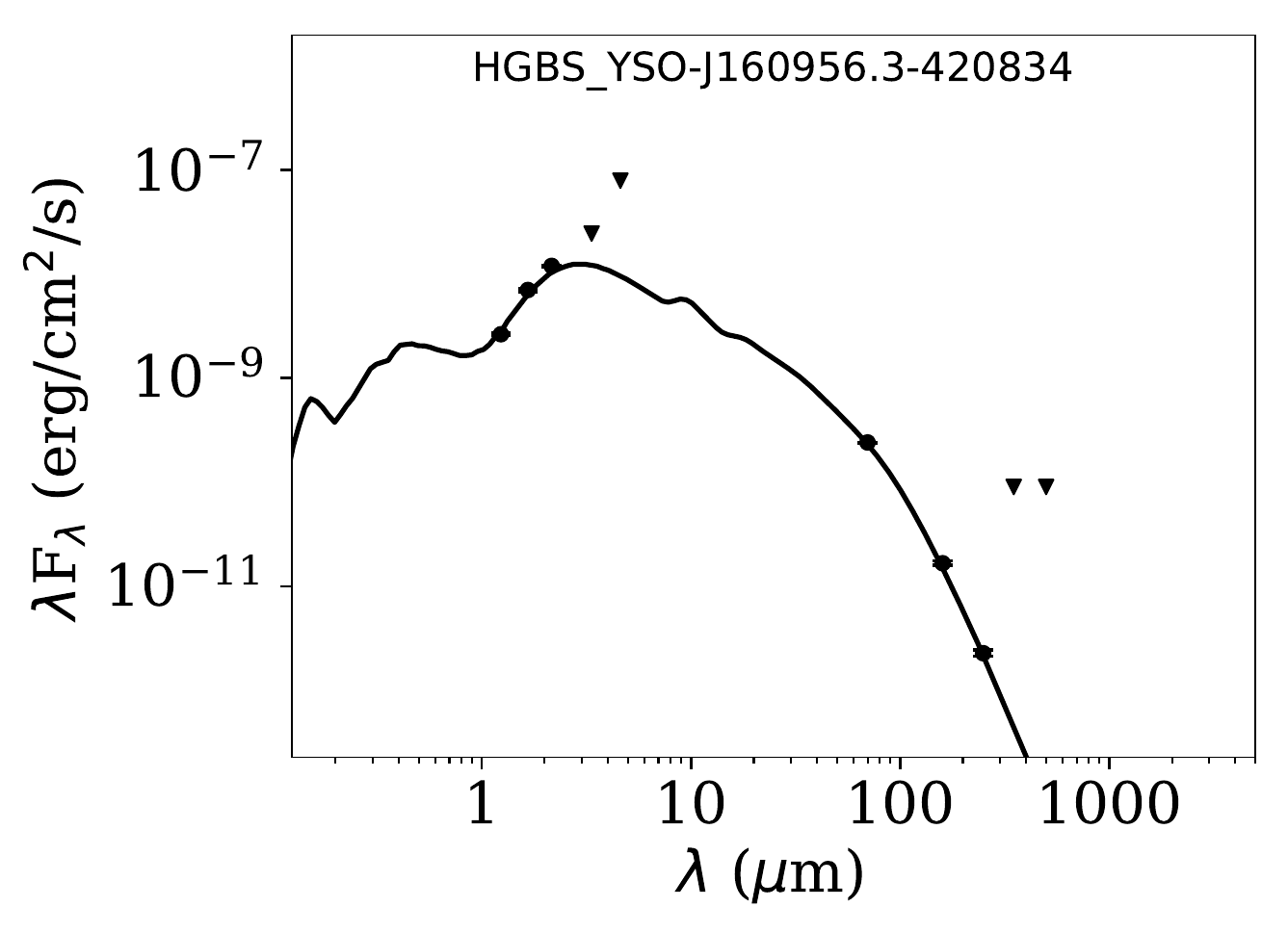}
\includegraphics[width=0.33\textwidth]{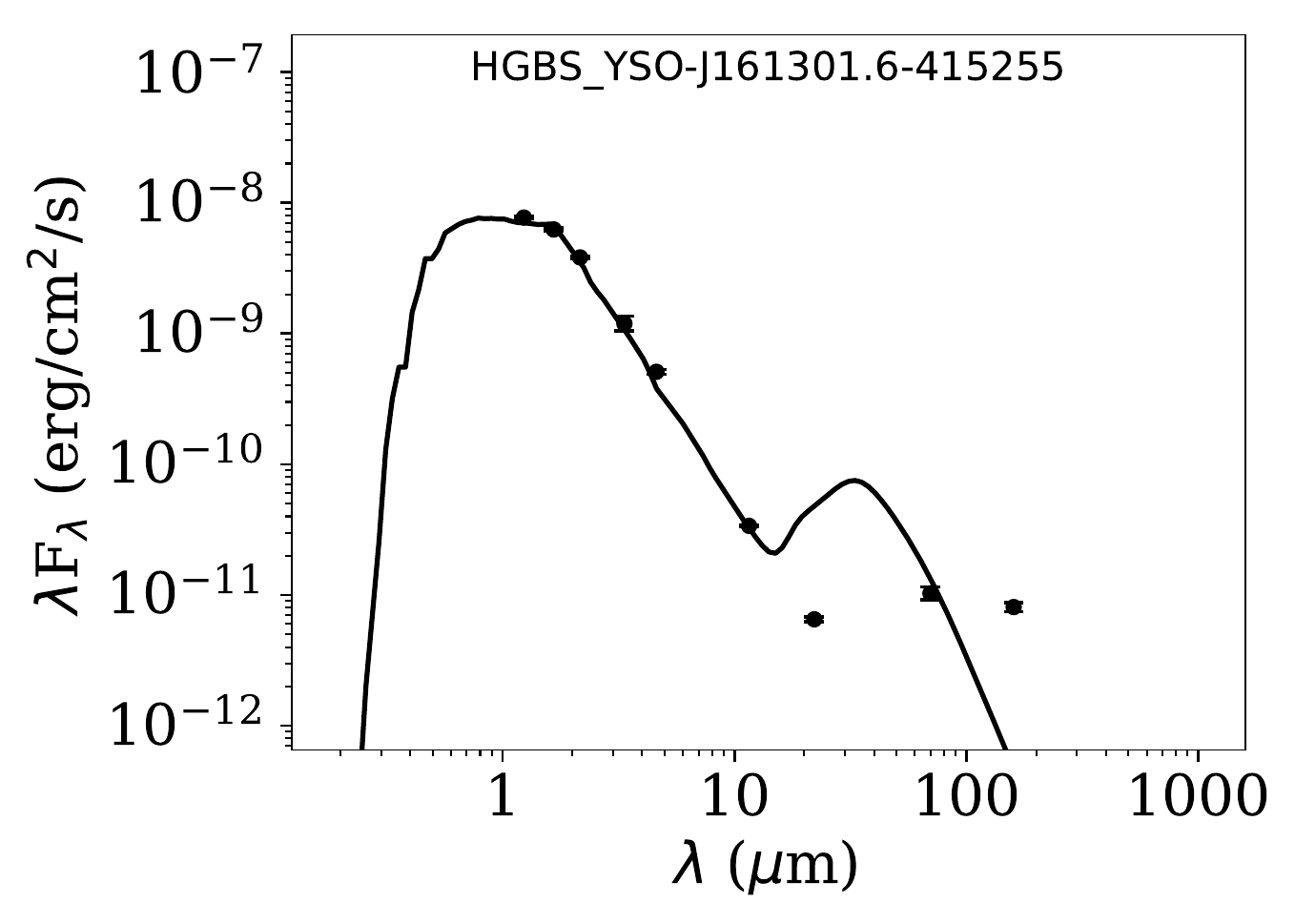}
\caption{}
\renewcommand{\thefigure}{\arabic{figure}}
\end{figure*}

\section{YSOs/protostars catalogue}
\label{sect:yso_cat}

The protostars detected by \her\, in the FIR are objects in different evolutionary stages. 
They span from extremely young objects - that are still embedded in their parental envelopes and have a strong FIR emission - to more evolved YSOs where the envelopes have been mostly dissipated and the continuum emission is stronger at NIR-MIR wavelengths with only a residual FIR contribution. Therefore, a correct classification of the evolutionary stage of these types of YSOs/protostars requires us to sample the SED as much as possible from NIR to FIR wavelengths. For this purpose, we matched our {\it Herschel} catalogue of YSOs and protostars with other infrared catalogues. In particular, we searched for associations within a 6\arcsec\, radius in the \twomass, \spit, and \wise\, catalogues and for most of the sources in our catalogue we were able to build a SED spanning from 1.2 \um\, to 500 \um. These SEDs are shown in Fig. \ref{fig:sed}. In Table \ref{tab:cat} we report the name of the \spit\, and \wise\, association, if present. We took the \twomass, \spit, and \wise\, fluxes of the associated sources from the on-line catalogues by using the Infrared Science Archive (IRSA)\footnote{http://irsa.ipac.caltech.edu}.
It is worth noting that in our catalogue we identified some sources (4, 1, and 9 objects in Lupus I, III, and IV, respectively) that are not present in the \spit-c2d catalogue of candidates YSOs, because of the wider area covered by \her\, maps. Given the spectral range covered by the \her\, instruments, they are particularly sensitive to protostars in the younger evolutionary stages, namely Class 0 and I, still embedded in a robust dusty envelope responsible for the FIR emission. Consequently, they can miss the most tenuous, evolved YSOs where their envelopes have started to dissipate. Therefore, the \her\, YSOs/protostars catalogue is incomplete for Class II and III objects.

\subsection{SED classification}

The shape of the SED is usually used to classify protostars in evolutionary stages. Four evolutionary classes, Class I, Flat, Class II, and Class III, have been defined by \citet{greene94} from the original work of \citet{lada84}, on the basis of a spectral index $\alpha$, where
\begin{equation}
 \alpha=\frac{d(log(\lambda F_{\lambda}))}{d(log \lambda)}
\end{equation}
is the slope, on log-log axes, of the SED between 2.2~\um\, and 24~\um.
To be consistent with the previous YSOs catalogue of \citet{merin08} based on \spit\, data, we used the \spit-c2d limits of $\alpha$ for the definition of the four classes \citep{evans09}. For each source in our catalogue, we calculated $\alpha$ by using a linear least-squares fit to all available data between 2.2 \um\, ($K$ band) and 24 \um\, (\spit-MIPS). We note that our $\alpha$ index values can differ from those given in the c2d catalogue because we include in the fits both \spit\, and \wise\, fluxes whenever available. It is worth noting that \her\, measurements can detect also protostars younger than Class I, the so-called Class 0 defined by \citet{andre00} as objects with $L_{smm}/L_{bol} \ge $ 0.01, where  $L_{\rm smm}$ is the luminosity at $\lambda \ge$ 350 \um. We, therefore, evaluated the $L_{smm}$ of our objects by performing a grey-body fit of the \her\, fluxes at wavelengths $\ge$ 160 \um\, by using the same opacity law used for dense cores (see Sect. 2) and integrating the emission of the best grey-body fit at $\lambda \ge$ 350 \um. Moreover, for all the YSOs/protostars of our \her\, catalogue we calculated the bolometric luminosity by integrating the observed SED in the widest range of available wavelengths (shown in Fig. \ref{fig:sed}).
$L_{\rm bol}$, $L_{\rm smm}$, $\alpha$ index, and evolutionary class are reported in Table~\ref{tab:par_yso}.

\subsection{SED fit}

To define better the evolutionary statuses of the YSOs/protostars of our catalogue, we compared their observed SEDs with the set of synthetic SEDs produced by \citet{robitaille17}. These SEDs are representative of YSOs spanning a wide range of evolutionary stages, from the youngest deeply embedded protostars to pre-main-sequence stars with little or no disk contribution. The full sample of \citet{robitaille17} models is divided into 18 sets where the three major elements that determine the SED of a forming star, namely disk, envelope, and cavity are present or absent and modelled with a different profile.
Given the complexity of the model and the large number of input parameters, there is some degeneracy among the synthetic SEDs so that the model that {\it best} represents the observed SED is not simply the one with the lowest $\chi^2$. In fact, if only one or a small number of models provides a good fit in a specific model set it means that the parameters of such a set need to be fine-tuned to reproduce the data. Hence, that set of models is more unlikely than a different model set where a larger fraction of models can reproduce the data because it requires less fine-tuning. Therefore, instead to simply identify the best fit model as the one with the lowest $\chi^2$, we identify the best model set as the one with the largest number of good models, following the fitting procedure defined \citet{robitaille17} and described in Appendix \ref{ap:sed}.

Given the large numbers of observed fluxes (from 11 to 18, for most of the sources of our catalogue) and the wide wavelength range, usually only a limited number of model sets are able to reproduce the observed SED reasonably.
In Fig. \ref{fig:sed}, we show the observed SEDs of the 35 YSO/protostar candidates of our catalogue for which the fitting procedure converged, together with the model with the lowest $\chi^2$ among the most probable model set. Moreover, in Table \ref{tab:par_yso} we add three columns for three possible input parameters of the SED models, one for the disk, one for the envelope, and one for the cavity. When present, we report the value of the three parameters of the best fit model. In fact, the presence or absence of these three elements in the best fit model and their relative prominence can give a rough indication of the evolutionary stage of the object. In the online material we provide the range of input parameters of all the good models within the best model set (see Appendix \ref{ap:sed}).

\section{Discussion}

\subsection{Comparison with the previous preliminary catalogue}
\label{sect:disc_sfr}

\begin{figure}
\includegraphics[width=0.48\textwidth]{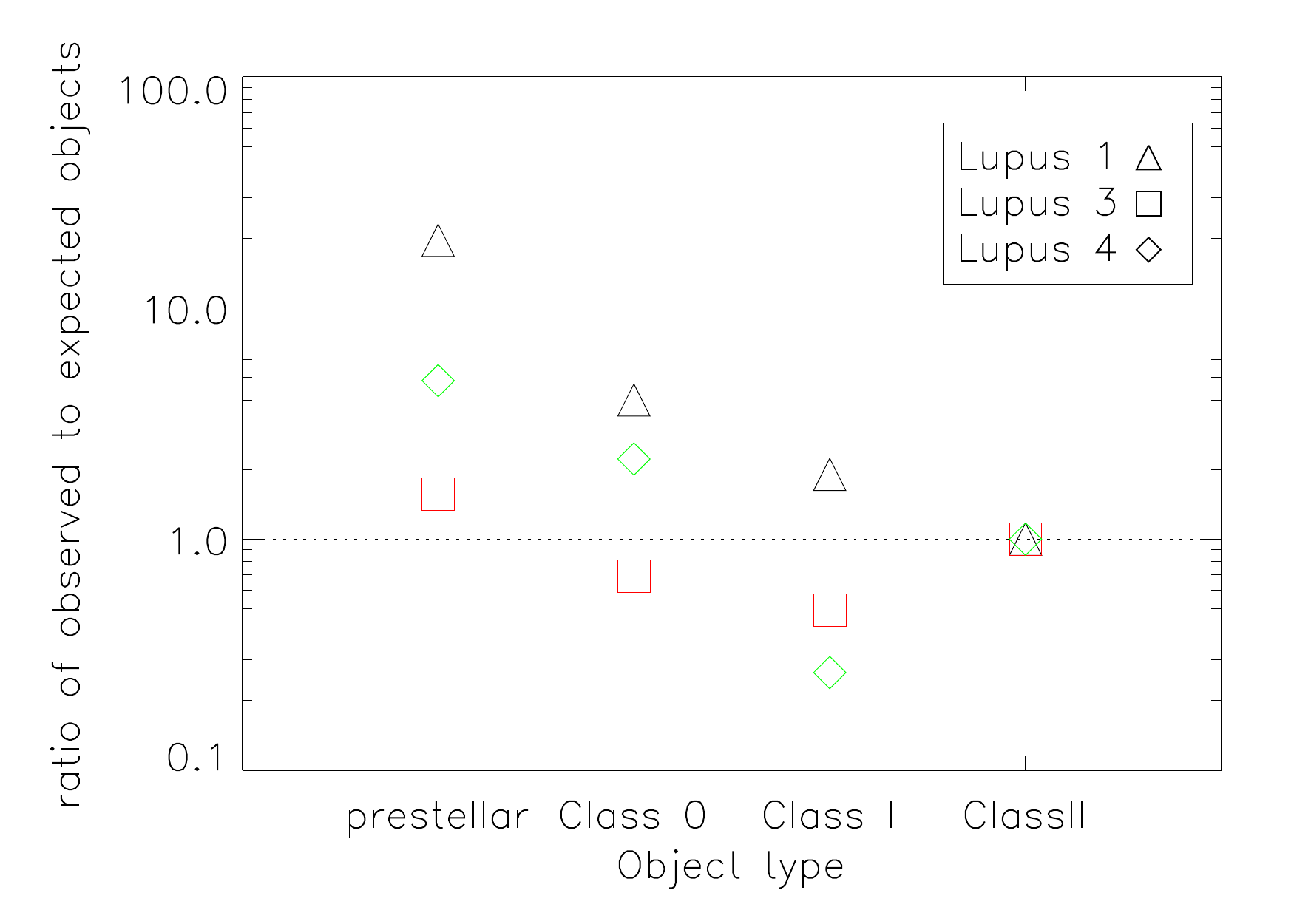}
\caption{Ratio of observed objects number for each evolutionary class with respect to the expected number for a constant SFR. Values above one indicate an accelerating SFR. Values below one indicate a decelerating SFR. Different symbols and colours are used for the ratios of the three Lupus clouds, as in the legend.}
\label{fig:sfr}       
\end{figure}

The total number of detected compact sources (including both starless cores and YSOs/protostars) listed in our catalogue (see Table \ref{tab:cat_sum}) is larger than that presented in the preliminary catalogue of \citet{rygl13}. This increase is not surprising since the \citet{rygl13} preliminary catalogue was compiled on maps produced with a different reduction method and using a different source extraction and flux measurement tool (i.e. {\it CuTEx}) and with more stringent criteria for source selection. Hence, differences in the sources identification and classification are expected (see Sect. \ref{sect:selection}). We performed the same analysis of the star formation history done by \citet{rygl13} using our more extended catalogue. Specifically, we calculated the ratios of the numbers of observed sources in each evolutionary class (prestellar cores, Class 0, and Class I) with respect to the expected numbers in case of a constant star formation rate (SFR), using the Class II as reference class. We assumed the same  lifetimes for each class used by \citet{rygl13}. In Fig. \ref{fig:sfr} we show the ratio of observed-to-expected objects numbers for the prestellar cores, Class 0, and Class I objects in each cloud. Despite the difference in absolute numbers with respect to the values in Fig. 3 of \citet{rygl13}, we find the same trend, confirming their results. For Lupus I we find an increasing SFR with a particularly high ratio for prestellar cores, suggesting that Lupus I is undergoing a major star formation event. Similarly, for Lupus IV the observed number of prestellar cores and Class 0 is higher than what expected for a constant SFR, suggesting that the star formation has accelerated over the past $\sim$ 0.5 Myr.  Conversely, in Lupus III the bulk of the star formation activity has already passed and has decelerated in the last $\sim$ 2 Myr, with a recent small increment of the prestellar cores formation.

\newpage

\subsection{Spatial distribution of sources and cloud clumpiness}

From Figs. \ref{fig:lup1_sou}, \ref{fig:lup3_sou}, and \ref{fig:lup4_sou}, we see that the  different categories of objects identified by \her, namely unbound starless cores, prestellar cores, and YSOs/protostars are differently distributed across the regions, with the prestellar cores clustered in spatially limited parts of the clouds, corresponding to the brighter and higher column density filamentary regions, while the majority of the lower column density filaments are populated by unbound starless cores. Conversely, YSOs are spread across the full cloud, even in regions with very low column density and they can be not associated with any filament. 
To quantify this visual impression, we calculated the percentages of association with filaments for each of the three categories. We used the filaments identification presented in \citet{benedettini15} and considered that a source is associated with a filament if its central position falls within the boundary of the filament. 
We found different percentages of association for the three type of sources. Only 26\% of YSOs/protostars and the 36\% of the unbound cores are associated with filaments while almost all (94\%) prestellar cores are within a filament, in particular close to its central, brightest crest. High percentages of association with filaments for the prestellar cores are found also in other star forming regions within the HGBS project, showing that filaments are the preferred place where the dense condensations - that are the seeds of new stars - are formed (\citealt{andre14} and references therein). On the other hand, YSOs have a lower probability to be associated to filaments because this type of objects during their evolution from the prestellar phase to the YSO phase have had time to migrate from their original position and/or the filament where they were formed is dissipated.

We found that about one third of YSOs/protostars and unbound cores (26\% and 36\%, respectively) are associated with filaments while almost all (94\%) prestellar cores are associated with filaments. This behaviour is in line with one of the main results of the \her\, photometric surveys of star-forming regions that filaments are the preferred place where the dense condensations - that are the seeds of new stars - are formed (\citealt{andre14} and references therein).
The locations of the different types of objects also show that star formation activity is not uniformly distributed across the clouds. As examples, in Lupus I the north-west part of the cloud is populated by only starless cores; in Lupus III the majority of prestellar cores are in the long brightest filament crossing the cloud east-west; and in Lupus IV the prestellar cores are clustered in two locations around R.A. of 16$^h$01$^m$00$^s$ and 16$^h$08$^m$00$^s$.

The total number of starless cores (unbound plus prestellar) in Lupus I is about three times that of the other two clouds, suggesting a possible difference in the level of clumpiness of the three Lupus clouds. To estimate this level, we calculated the number of starless cores per pc$^2$ above a certain column density level for the three regions. We choose a column density contour of 1.8$\times$10$^{21}$ \cmdue, corresponding to a visual extinction of 2 mag, since it roughly corresponds to the border of the filamentary clouds. Within this contour, we obtained 100 starless cores per pc$^2$ in Lupus I, 29 starless cores per pc$^2$ in Lupus III, and 52 starless cores per pc$^2$ in Lupus IV. Therefore, in Lupus I the number of starless cores per pc$^2$ is a factor of two higher than in Lupus IV and a factor 3.4 higher than in Lupus III, confirming a higher level of clumpiness in the Lupus I cloud. To investigate this aspect more deeply, we calculated another estimator of the level of clumpiness, namely the projected distance from the nearest neighbour. For the starless cores sample, we found that the nearest neighbour distance in Lupus I has the lowest median value (0.07 pc), it has a slightly higher value (0.08 pc) in Lupus IV, while in Lupus III we found the highest value (0.1 pc). We then used the Kolmogoroff-Smirnov (K-S) test to compare the distributions of the nearest neighbour distance in the three clouds to see at what level they are different. With this test, we found that it is highly improbable that the nearest neighbour distances of starless cores in Lupus I and III belong to the same kind of distribution (K-S probability $p$ = 0.0002). Similarly, we found very low K-S probabilities when we compare the nearest neighbour distributions of Lupus I and Lupus IV ($p$ = 0.01) and Lupus III and Lupus IV ($p$ = 0.13).

All the previous evidence indicates that Lupus I is the cloud with the highest level of clumpiness. Since the Lupus sub-regions - in general - appear quite similar, we wonder what is the origin of Lupus I's higher level of clumpiness. One possibility could be that it is at a younger evolutionary stage in terms of star formation activity \citep{hughes94,benedettini12,rygl13}. Indeed, a lower number of starless cores should be a natural effect of the evolution of the star formation process. In the absence of continuous infall of gas available for forming new dense cores, the number of dense cores should decrease over a typical time related to the core lifetime, due the core dissipation or evolution into protostars. The typical lifetime of the \her\, starless cores estimated in the Aquila region is $\sim$ 1 Myr \citep{konyves15}.
In Lupus I, star formation activity at the present epoch is proceeding at an increasing rate and the ages of the stellar population are usually lower than 1 Myr \citep{hughes94}.  Therefore, the dense cores created since the onset of the current star formation activity should be, on average, still present. In contrast, in Lupus III we measured a decreasing SFR rate at the present epoch, indicating that the bulk of the star formation activity has passed. On average, the stars in Lupus III are older than 4 Myr \citep{mortier11}, therefore they have an age greater than the fiducial \her\, prestellar cores lifetime ($\sim$ 1 Myr), suggesting that most of the cores formed at the onset of past major star formation events have already been vanished.

Another possibility is that the higher level of clumpiness of Lupus I is a side-effect of the collision of the flows from which it was generated. Indeed, much evidence has been collected \citep{tothill09,benedettini15,gaczkowski17} showing that Lupus I could has been strongly influenced by colliding flows generated by the expanding Upper-Scorpius $\ion{H}{i}$ shell and the Upper Centaurus-Lupus wind bubble. Furthermore, theoretical models have shown that shocked flows are a very efficient mechanism for the formation of filaments and the density perturbations that generate dense cores in molecular clouds (e.g. \citealt{padoan01,heistsch08,inutsuka15}). Conversely, the cloud material in Lupus III and IV, even though they belong to the same molecular complex as Lupus I, have not been affected by the expanding shells, and so have not yet experienced any major collision. 

\subsection{About the existence of a column density threshold for the formation of prestellar cores}

\begin{figure}
\includegraphics[width=0.48\textwidth]{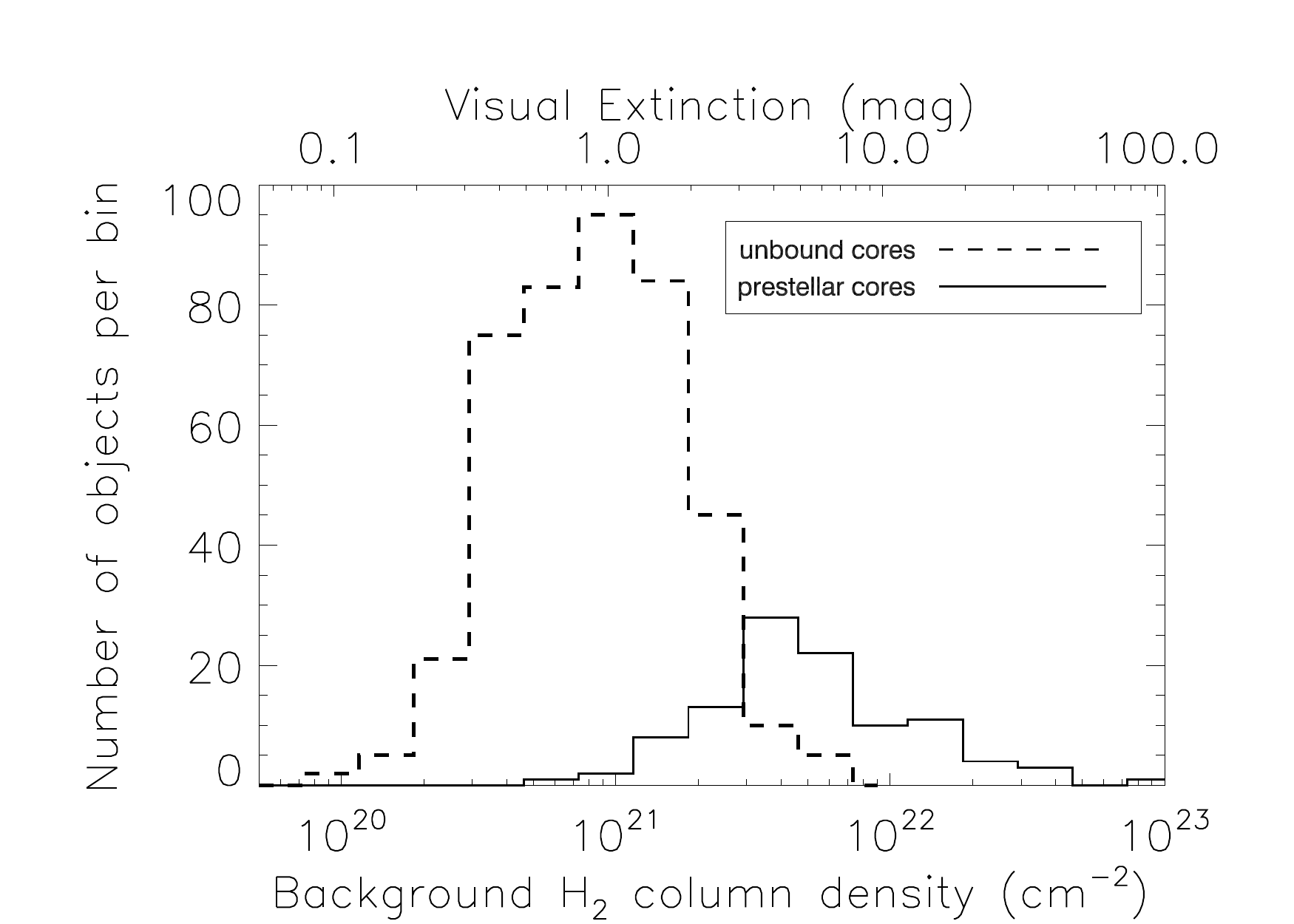}
\caption{Histogram of the background column density for the sample of unbound cores (dashed line) and prestellar cores (solid line). }
\label{fig:backg}       
\end{figure}

In Fig. \ref{fig:backg}, we show the histograms of unbound cores and prestellar cores as function of their local background H$_2$ column density. The distributions of the two categories are well separated. Unbound starless cores are found over lower column density backgrounds (from $\sim$10$^{20}$ \cmdue\, to  $\sim$6$\times$10$^{21}$ \cmdue) while prestellar cores are found above a background of $N$(H$_2$) > 8$\times$10$^{20}$ \cmdue\, with a modal value corresponding to $\sim$ 4 mag of visual extinction, assuming the \citet{bohlin78} conversion factor $N$(H$_2$)/$A_{\rm V}$ = 9.4$\times$10$^{20}$ (\cmdue/mag).

The fragmentation of filaments in roundish cores is a key step of the passage of material from clouds to stars, yet it seems that only cores above a certain background H$_2$ column density threshold become gravitationally bound and eventually evolve into protostars. Theoretically, a column density threshold is expected since in low density environments material is not well shielded from the interstellar radiation field and correspondingly its ionisation fraction is high. In this case, the magnetic pressure becomes efficient in opposing gravitational collapse and limiting the formation of sub-structures \citep{mckee1998}. In the last few decades, some observational evidence for the existence of an H$_2$ column density threshold for star formation has been collected from millimetre and sub-millimetre surveys of dense cores correlated to NIR extinction maps \citep{onishi98,johston04,kirk06}.
These surveys, however, suffer from severe limitations, including their limited capability in probing the full population of dense cores, their limited dynamical range, and the different tracers used to detect the cores and estimate the visual extinction. In this respect, the \her\, data represent a step forwards since they have the advantages of being able to probe prestellar cores and their underlying background simultaneously and of being more sensitive to the detection of dense cores, reaching the level of completeness for core masses above $\sim$ 0.1 \msun\, \citep{konyves15,marsh16,bresnahan18}. 

In earlier analyses of \her\, catalogues, the visual extinction threshold for star formation was defined as the background level above which 90\% of prestellar cores are found. Values of about 7 mag and 6 mag were found for Aquila and Taurus, respectively \citep{konyves15,marsh16}. Ground-based estimates of the threshold are available only for a handful of regions and a large range of values was found for different regions, for example, from $\sim$ 5 mag in Perseus \citep{kirk06} to $\sim$ 9 mag in Taurus \citep{onishi98}. In Lupus, we find a threshold of $\sim$ 2 mag, meaning that 90\% of prestellar cores lie on a background higher than this threshold. Even considering the large uncertainties affecting our H$_2$ column densities, which could be underestimated by up to factor of two, the visual extinction threshold found in Lupus is lower than those measured in other star-forming clouds so far.

It is worth noting that Lupus is a peculiar star-forming cloud complex in term of mass density. For example, it is characterised by very low column densities both for its diffuse medium and for its denser material assembled in filaments. More specifically, the probability distribution function of column density (PDF) in Lupus has a peak equivalent to a visual extinction of less than 1 mag, the lowest value found so far in PDF analyses of star-forming clouds based on {\it Herschel} data, that usually peak at visual extinctions larger than 1.5 mag \citep{benedettini15}. Moreover, the Lupus filaments have average column densities at the lower end of the distribution of filaments column densities found in other clouds \citep{benedettini15,arzoumanian13}.
Despite these lower column densities, star formation is on-going in the Lupus I, III, and IV clouds and we find robust candidate prestellar cores even on backgrounds levels lower than what is usually found in other clouds. The presence of dense cores, some of which have been considered prestellar, in cloud regions of $A_{\rm v}\lesssim$ 5 mag has been observed not only in Lupus but also in Perseus \citep{hatchell05} and Taurus \citep{marsh16}. Therefore, the column density threshold should be interpreted more as a level over which a higher probability exists to find prestellar cores rather than a stringent limit under which star formation is inhibited \citep{konyves15,andre14,lada10,hatchell05}. Moreover, the fact that  different values of the threshold are found in different star-forming clouds is an indication that the star formation column density threshold, if it exists, might depend on the local properties of the host cloud such as the strength of its magnetic field, the local radiation field and its non-thermal velocity dispersion. In fact, these properties control the mechanisms that can provide support against gravitational collapse, such as the magnetic pressure mediated by collisions between neutrals and ions and the turbulent motions that supply non-thermal pressure support.

This conclusion is also supported by the results of an alternative method to look for the possible presence of a star formation threshold in GMCs that is to investigate the relation between the surface densities of the SFR and gas mass, namely the Schmidt conjecture.
By applying the Bayesian method developed by \citet{lombardi13} to four GMCs, \citet{lada13} studied the power-law relation between the protostellar surface density distribution derived from \spit\, catalogues of protostars and the dust surface density distribution measured by the extinction maps derived from 2MASS data. They found that there is no star formation threshold for two clouds, Orion A and Taurus, while for the other two clouds, Orion B and California, data are compatible with the possible presence of a threshold, even if with a large degree of uncertainty.

\subsection{Prestellar cores associated with filaments on average thermally sub-critical}

\begin{figure}
\includegraphics[width=0.49\textwidth]{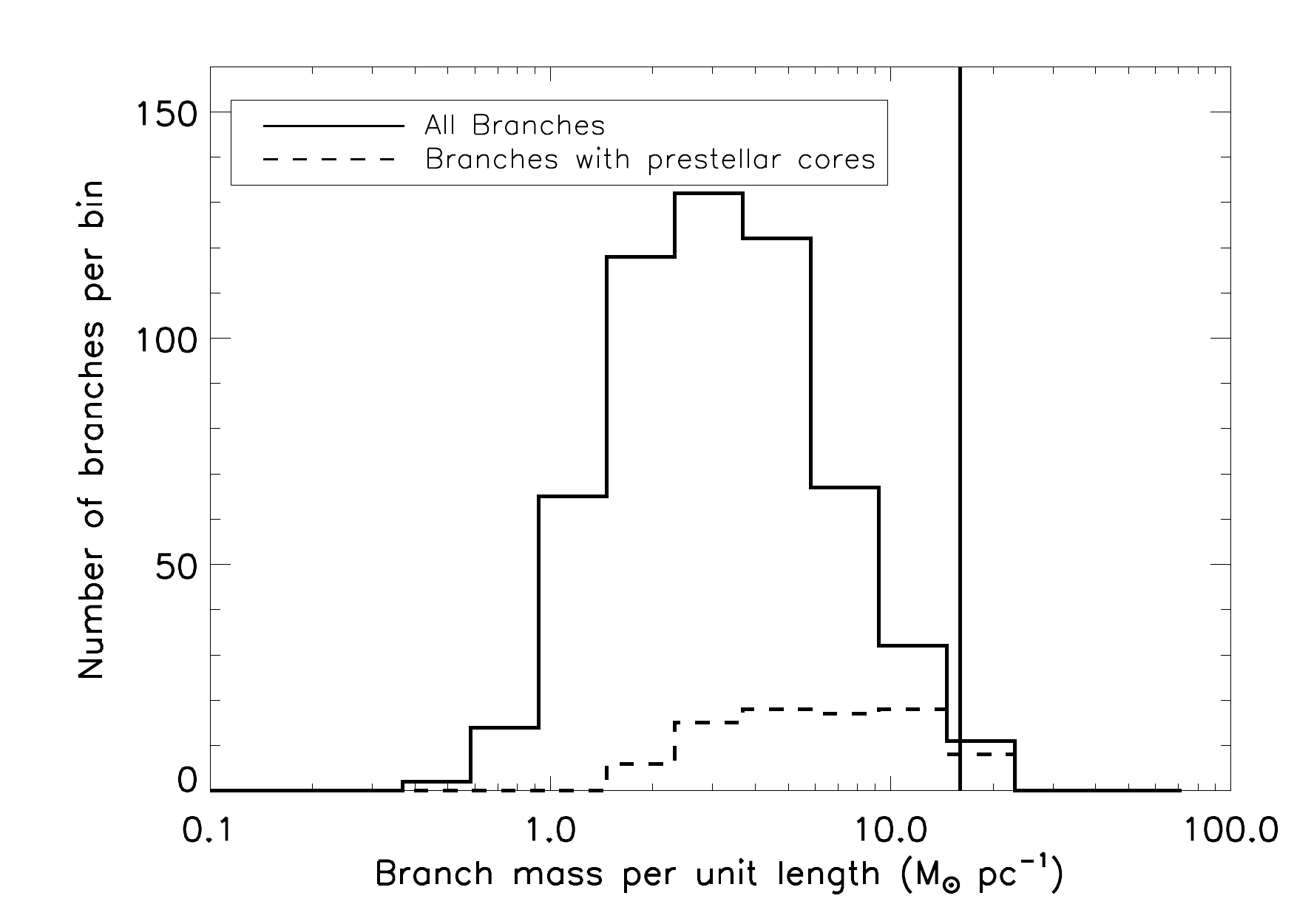}
\caption{Distribution of the average mass per unit length of the filament branches (solid line). Dashed line represents the subset of filament branches associated with prestellar cores. The vertical bar indicates the critical value of 16 \msun\, pc$^{-1}$ for radial gravitational collapse of a cylindrical, isothermal filament with a  temperature of 10 K. }
\label{fig:linemass}       
\end{figure}

In \citet{benedettini15}, we highlighted that the majority of filaments found in Lupus have a mass per unit length lower than the maximum value needed to be thermally supported. This critical value for isothermal infinite cylindrical filaments with a typical temperature of 10 K, confined by the external pressure of the ambient medium, is $\sim$ 16 \msun\, pc$^{-1}$ (see Fig. \ref{fig:linemass}). Previous studies (\citealt{andre10}, 2014; \citealt{arzoumanian11}) have suggested  that only filaments with masses per unit length above this critical value are primarily associated with star formation activity. In Lupus, however, there is evident star formation activity even though the large majority of filaments are sub-critical. More specifically, we find that sub-critical filaments contain prestellar core candidates in which new stars may form (see Fig. \ref{fig:linemass}). 

We stress that this finding is still valid even if we consider the large uncertainty associated with the mass estimates of both filaments and cores. In \citet{benedettini15} we estimated the mass per unit length of a filament by assuming that the total mass measured in the structure was uniformly distributed along the entire structure, meaning that we give the average value along the filament. On the other hand, other studies use a different measure to estimate mass per unit length, considering the average H$_2$ column density of only the denser central part of the filament. The two values differ by a factor of about 1.5, well below the uncertainty associated with the filament column density \citep{schisano14,benedettini15}. 

In Lupus, the fact that prestellar cores are observed in some of its supposedly globally sub-critical filaments is a indication that the condition of overdensity needed to activate  gravitational collapse can be reached only locally and is not necessarily a global property of the filament. Indeed, filaments are highly irregular structures with significant fluctuations of brightness and column density along the principal spine \citep{schisano14,benedettini15}. We conclude that in a low column density regime, such as that of the Lupus clouds, the average mass per unit length is not a good parameter for identifying those filaments where star formation primarily takes place, since a significant fraction of stars may form along filaments that are globally sub-critical.

\begin{figure}
\includegraphics[width=0.5\textwidth]{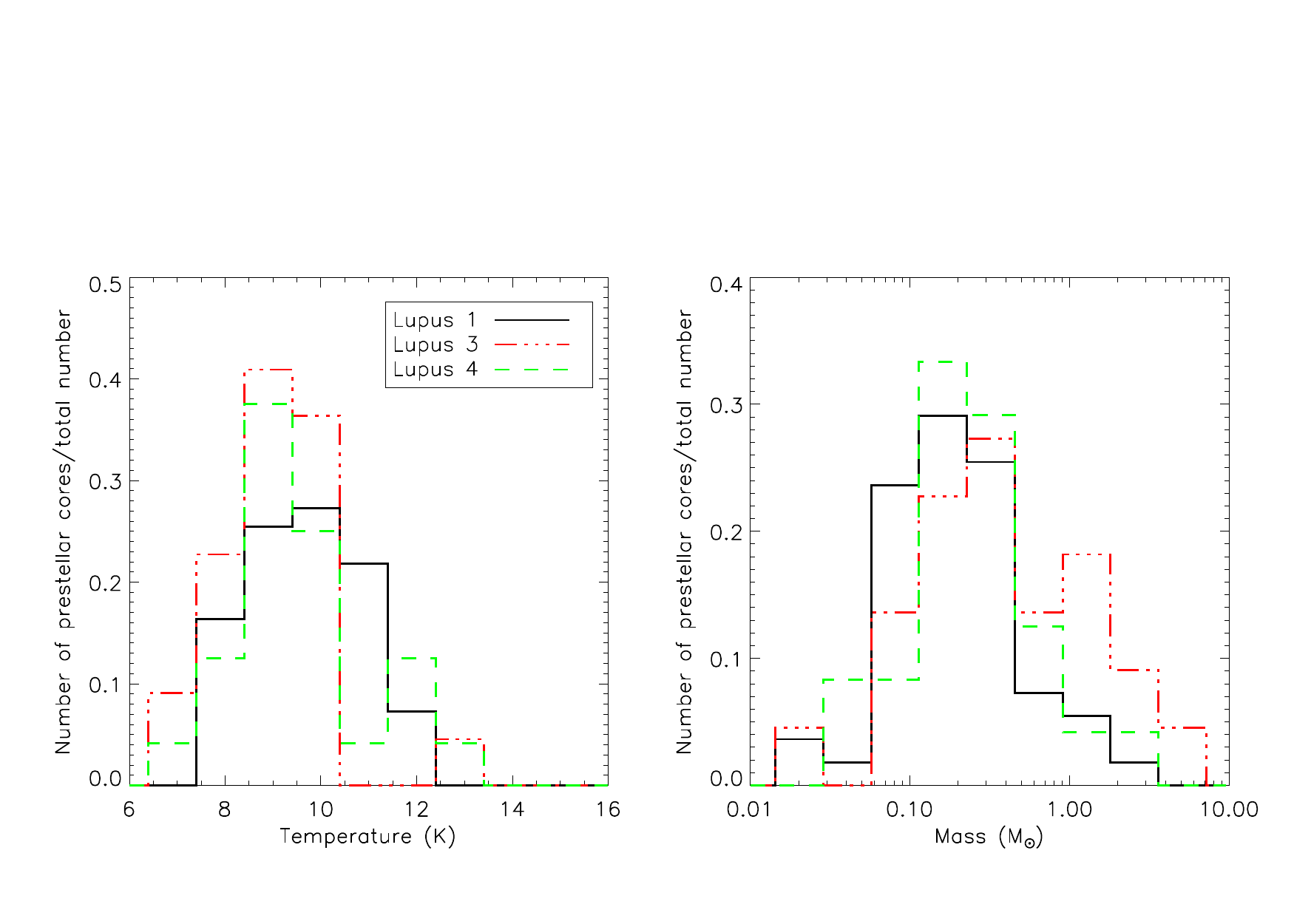}
\caption{Normalised distribution of the temperature (left panel) and mass (right panel) of the population of prestellar cores in the three Lupus clouds. On the y-axis the number of objects per bin relative to the total number of prestellar cores in each cloud is reported.}
\label{fig:t-m}       
\end{figure}

\subsection{Core mass function}

In Fig. \ref{fig:t-m}, we show histograms of the temperatures and masses of the prestellar cores sample for the three Lupus clouds. The temperature distribution is very similar in the three clouds, ranging between 7 K and 13 K with a median value around 10 K. These values are typical values for the prestellar cores identified from \her\, data of both low- and high-mass star-forming regions (e.g. \citealt{konyves15,marsh16,giannini12,rayner17}). Temperatures around 10 K are also the typical values for gas kinetic temperature and dust temperature derived in starless dense cores from single-dish sub-millimetre observations (e.g. \citealt{tafalla02,foster09,keown17}). As noted earlier in Sect. \ref{sect:phys}, however, the temperature derived from SED fitting of \her\, data is an averaged temperature. Indeed, interferometric observations have shown that dense starless cores are not isothermal and have temperatures that drop below 10 K at the core centre (e.g. \citealt{tafalla04,crapsi07}).
Furthermore, in the outer regions of cores, where density and shielding from the external radiation field are smaller, the gas and dust temperatures become uncoupled \citep{galli02}. 

The mass distributions of prestellar cores span a similar range for the three Lupus clouds, with median values of the core masses in the three clouds of 0.17 \msun, 0.38  \msun, and 0.22 \msun\, for Lupus I, III, and IV, respectively. The similar ranges spanned by the temperature and mass distributions of prestellar cores in the three Lupus clouds allow us to merge the three samples, and improve the statistical relevance of subsequent analysis.

\begin{figure}
\includegraphics[width=0.47\textwidth]{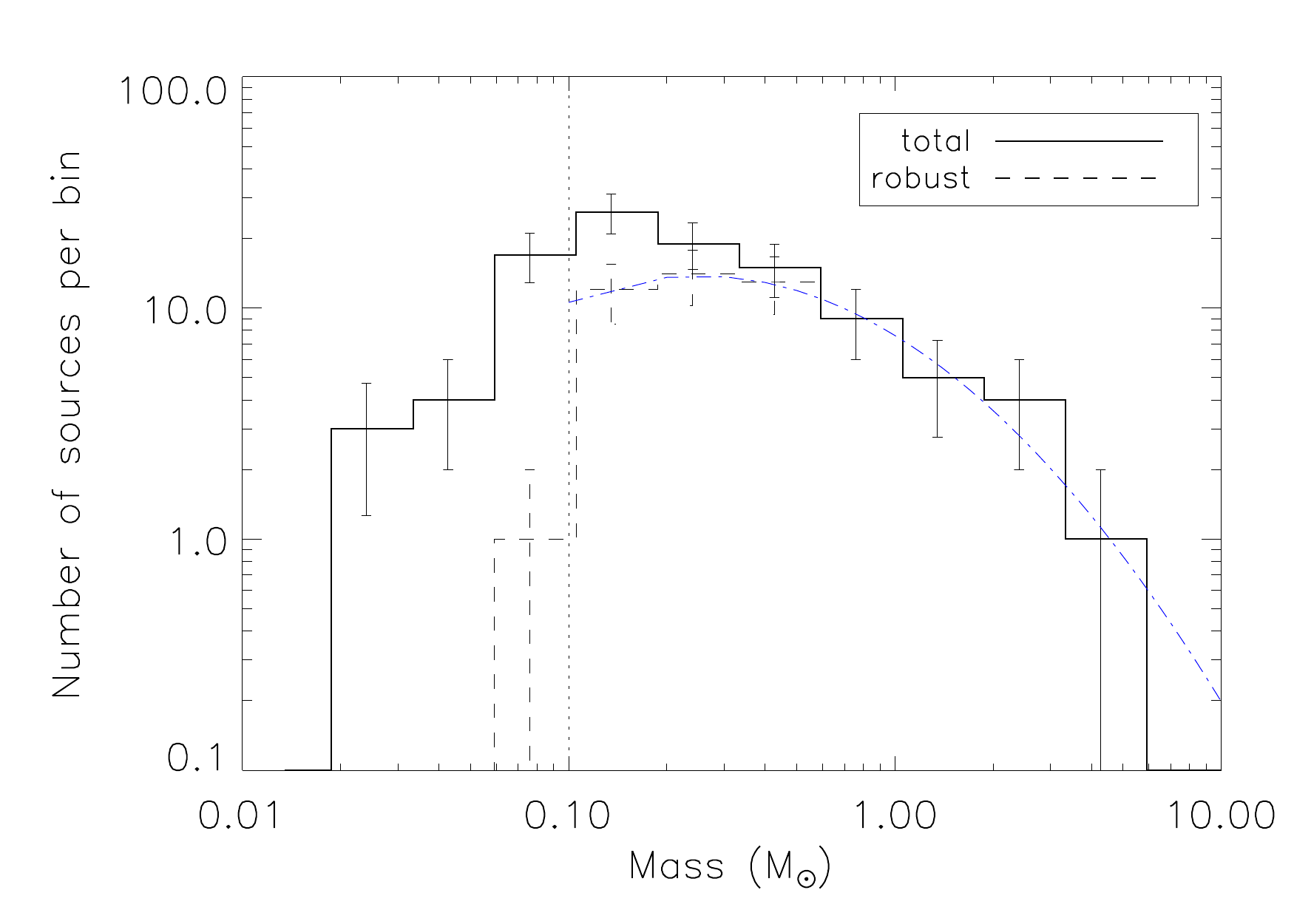}
\caption{Core Mass Function of the total sample of prestellar cores detected in the three Lupus clouds (solid line) and that of the robust sub-sample (dashed line). The dashed-dotted line is the log-normal fit, with central mass of (0.25$\pm$0.04) \msun\, and a standard deviation of 0.55$\pm$0.04. The vertical line indicates the completeness limit of the prestellar cores sample.}
\label{fig:cmf}
\end{figure}

In Fig. \ref{fig:cmf}, we show the core mass function (CMF) of the total sample of Lupus prestellar cores and that of the robust sub-sample. 
A change of the slope of the distribution is present around 0.7 \msun, and, in particular the distribution of the robust prestellar cores, becomes almost flat between 0.1 \msun\, (the completeness limit) and 0.7 \msun, with a small peak around 0.2-0.3 \msun. This shape resembles that of the CMF derived from the \her\, data in Taurus \citep{marsh16}, a cloud quite similar to Lupus both in terms of mass range and total number of prestellar cores. The evidence of a possible flattening or break near a mass of about 0.5 \msun\, is also found in the dense cores mass functions derived from ground-based sub-millimetre surveys (e.g. \citealt{motte98,alves07,enoch08}).
The CMFs derived within the HGBS consortium for Aquila and Taurus are well described by log-normal functions \citep{konyves15,marsh16}. The log-normal fit to the Lupus CMF of the robust sample, for core masses above the estimated completeness limit $\sim$0.1 \msun, is shown in Fig. \ref{fig:cmf}. The best fit function has a central mass of (0.25 $\pm$0.04) \msun\, and a standard deviation of 0.55$\pm$0.04, the same values of the log-normal parameters of the stellar initial mass function (IMF) for multiple systems with masses between 0.12 \msun\, and 1 \msun\, \citep{chabrier05}. The K-S test performed on the cumulative distribution, confirms that the observed CMF has a log-normal form with 99\% of probability. We recall that, for the high mass tail of the distribution, the statistics are quite poor and the uncertainties on the mass estimate are large, therefore the real shape of the CMF for masses larger than 1 \msun\, cannot be robustly constrained.

The similarity of the shape of the CMF and IMF suggests that the prestellar cores identified in the \her\, catalogue may evolve into single stars on a one-to-one basis. Under this hypothesis, the scaling factor in the mass axis between the two distributions is interpreted in terms of star formation efficiency $\epsilon$, where $M_{\rm star}= \epsilon M_{\rm core}$, and it can be estimated from the ratio of the peak of the two distributions. For the Lupus complex the real peak of the CMF is difficult to define since the histogram of the robust prestallar cores sample is almost flat between 0.1 \msun\, to 0.7 \msun. However, if we consider the peak of the log-normal best fit, that has the same value of the \citet{chabrier05} IMF, we found that the star formation efficiency might be extremely high, close to 100\%, much higher than the $\sim$ 40\% found in Aquila \citep{konyves15}. This result is confirmed also if we compare the \her\, CMF with the distribution of the masses of a sample of class II YSOs in Lupus \citep{alcala17}: the range of the stellar masses are similar to the ones of the \her\, prestellar cores and the peak of the stellar masses distribution is between 0.2 \msun\, and 0.3 \msun. A star formation efficiency close to 100\% is, however, difficult to justify from a physical point of view since the accretion phase of a protostar is associated to significant mass loss through jet and outflow.
Probably, we are systematically underestimating the mass of the \her\, prestellar cores. Of all the sources of uncertainty of our prestellar core mass estimate described in Sect. \ref{sect:phys}, the only ones that can produce a systematic shift towards lower masses are the distance and the dust opacity law. If we assume as distance for all the three clouds the upper limit of the  distance estimates, namely $d$ = 200 pc (see Sect. \ref{sect:intro}), the correspondent CMF would peaks at $\sim$0.48 \msun\, with a resulting star formation efficiency of $\sim$50\%, more similar to that found in Aquila. However, the new {\it Gaia} DR 2 distances point towards a distance closer to 150 pc for the three clouds, rather than 200 pc. More likely, the opacity law we adopted (see Sect. \ref{sect:reduction}) might be not fully appropriate for the low mass regime of the Lupus cores. Indeed, in \citet{benedettini15} we already discussed that $\kappa_{\rm 300}$=0.1 cm$^2$g$^{-1}$ - the standard value of the dust opacity at the reference wavelength assumed in the HGBS consortium - could be too high for the H$_2$ column density range of the Lupus clouds and that halving the opacity would reconcile our column density maps to the visual extinction maps. Similarly, reducing the $\kappa_{\rm 300}$ of a factor of two would produce a CMF almost flat between 0.2 \msun\, and 0.6 \msun. In this case, the shape of CMF is still compatible with the log-normal form with 88\% probability and the best fit function has a central mass of (0.34$\pm$0.05) \msun, corresponding to $\epsilon \gtrsim 60\%$, and a standard deviation of 0.41$\pm$0.05.

\subsection{YSOs/protostars}
\label{sect:disc_yso}

\her\, instruments are particularly sensitive to protostars in the younger evolutionary stages, namely Class 0 and I objects, still embedded in substantial dusty envelopes responsible for the FIR emission. In contrast, they can miss the most tenuous, evolved YSOs where the envelope starts to dissipate. Therefore, \her\, data make possible the identification and correct classification of Class 0 and Class I objects more efficiently than previously, but YSOs/protostars catalogues based on only \her\, data can be largely incomplete for Class II and III objects.
Despite this shortcoming, the \her\, catalogue is a vital complement to previous YSOs catalogues of the region, in particular that based on \spit-c2d data \citep{merin08}, because the \her\, maps cover a larger area of sky with respect to the \spit\, maps and because it provides FIR fluxes not available in other catalogues. Indeed, the addition of the \her\, FIR fluxes to the previously known NIR-MIR SEDs is important to estimate the properties of the possible circumstellar envelopes. Such info significantly improves the evolutionary classifications of objects based on their SED shapes, especially for the younger protostars.

Some of the more evolved objects (Class II and III) in our catalogue lack excess emission at MIR wavelengths (typically around 8--24 \um), but present considerable emission at longer wavelengths. This kind of object, originally discovered by \citet{strom89} with IRAS data, are evolved YSOs with transitional disks, that are optically thick and gas-rich protoplanetary disks with astronomical unit-scale inner disk clearings or radial gaps. The union of the MIR and \her\, FIR data is a valid tool for detecting such objects and in our catalogue we find several candidates. Three of them, (HGBS\_YSO-J160711.5-390347, HGBS\_YSO-J160854.5-393743, and HGBS\_YSO-J161051.5-385314) have been previously classified as transitional disks by  \citet{bustamante15}. Eleven other objects in our catalogue (HGBS\_YSO-J153927.9-344616, HGBS\_YSO-J154512.8-341729, HGBS\_YSO-J160709.9-391102, HGBS\_YSO-J160822.4-390445, HGBS\_YSO-J160825.7-390600, HGBS\_YSO-J160829.6-390309, HGBS\_YSO-J160836.1-392300, HGBS\_YSO-J160948.5-391116, HGBS\_YSO-J160002.5-422216, HGBS\_YSO-J160044.6-415530, HGBS\_YSO-J160329.2-414001) were also analysed by \citet{bustamante15} and though not confirmed as transitional disk candidates they were classified simply as objects with infrared excesses higher than the median value. Our SED fitting results (see Table \ref{tab:par_yso}) confirm these sources as Class II objects with a disk where the detected infrared emission can be ascribed to a residual envelope. Finally, three objects in our YSOs/protostars catalogue, namely HGBS\_YSO-J153640.0-3421145, HGBS\_YSO-J160830.7-382826, HGBS\_YSO-J155730.4-421032, from the shape of their SED, are new good candidate YSOs with transitional disks, not present in the list of objects analysed by \citet{bustamante15}.

We classified the evolutionary state of the YSOs/protostars of the \her\, catalogue by using two indicators: the $\alpha$ index, as used in the \spit-c2d catalogue, and the SED fitting. With the latter, we can estimate the presence of three key components of the early phase of the star formation, namely the envelope, the bipolar cavity opened by the outflow, and the disk, whose relevance changes during the evolutionary process. In general, we found that the evolutionary class indicated by the two methods is in agreement (see Table \ref{tab:par_yso}). In Appendix \ref{sect:app_yso}, however, we discuss the few cases where we do not find full agreement.

\section{Summary and conclusions}

The nearby star-forming clouds of the Lupus complex have been mapped in five photometric bands at 70~\um, 160~\um, 250~\um, 350~\um, and 500~\um\, with \her\, photometric instruments within the HGBS key project. In this paper, we presented the catalogues of dense cores and YSOs/protostars in Lupus I, III, and IV, compiled from the extraction of compact sources in the five \her\, maps. Two dedicated procedures for sources extraction and selection have been applied, optimised for dense cores and protostars, respectively. A total of 532 dense cores and 38 YSOs/protostars have been identified in the maps. Catalogues, listing their measured properties, namely position, flux density and size at each band, are supplied. In addition, we also provide catalogues of the physical properties listing mass, temperature, radius, column density, and average volume density for the dense cores and bolometric luminosity, FIR luminosity, $\alpha$ spectral index, and evolutionary class for the protostars.

The physical properties of the starless dense cores were estimated by fitting the observed SEDs with a grey-body function. The comparison of the measured mass with the BE mass was used to select a sample of 103 candidates prestellar cores, complete down to masses of $\sim$ 0.1 \msun. 
Almost all the prestellar cores are associated with the brightest filamentary structures of the clouds, confirming one of the main results of the \her\, photometric surveys of star-forming regions, that filaments are the preferred place for the formation of the dense condensations that will evolve in new stars. Conversely, we found that only about one third of the starless cores and YSOs/protostars are associated with filaments. In Lupus I, we found a higher level of clumpiness of cores, possibly due to its younger evolutionary stage in terms of star formation activity.

In Lupus, we found robust prestellar core candidates even in regions with background column density lower than that measured in other star-forming regions so far. In particular, we found that 90\% of prestellar cores lie on a background higher than a visual extinction threshold of about 2 mag. In other clouds, most of which have typical prestellar core masses higher than the Lupus ones, values of this background threshold from 5 mag to 9 mag were found. The lower limit of the column density background is often interpreted as a column density threshold for the star formation, however the large variability of its value found in several star forming regions indicates that this limit should be interpreted more as a higher probability to find prestellar cores in the parts of the molecular clouds with column densities above it rather than a stringent limit under which the star formation is inhibited. Moreover, its value might depend on the local properties of the single star-forming cloud such as the strength of its magnetic field, its radiation field and its typical non-thermal velocity dispersion.

Overall, the analysis of the prestellar cores catalogue indicates that the physical properties of the Lupus sample are similar to those of other regions studied with \her. One peculiar characteristic is the significant number of prestellar condensations with very low mass ($\lesssim$ 0.2 \msun), in line with the low column density of its ISM and with the low mass of its main sequence stars. We derived the CMF of the prestellar cores total sample. The CMF of the robust prestellar cores has a log-normal form with a peak between 0.2 \msun\, and 0.3 \msun; this implies a very high efficiency in the conversion of the prestallar core mass into the stellar mass that could be an indication that we are underestimating the prestallar core masses of the Lupus clouds. Anyway, the Lupus CMF shape is consistent with previous findings for other star-forming clouds and with the stellar systems IMF.

Noticeably, in the Lupus star-forming regions we found that the majority of filaments that contain prestellar cores have average masses per unit length below the maximum value possible for thermal support. For such low-column density filaments, the mass per unit length averaged along the full filament is not a good parameter for identifying star-forming filaments.

The physical properties of the YSOs/protostars were estimated by building their SEDs over a wide wavelength range, from NIR to FIR wavelengths and fitting the SED with a set of theoretical models. With the SED fitting, we gauged the possible presence around the protostar of three key components of the early phase of star formation: the envelope, the bipolar cavity opened by the outflow, and the disk, whose relevance changes during the evolutionary process. We provide the range of input parameters of the models with  good fit to the observed SEDs.
We estimated the evolutionary status of the YSOs/protostars using two indicators: the $\alpha$ spectral index and the result of the SED fitting.  For about 70\% of the objects, the evolutionary stages derived with the two methods agree. 

The \her\, catalogue of YSOs/protostars, although incomplete for objects in the later evolutionary stages, adds high value to previous catalogues because the addition of FIR fluxes to the previously known NIR-MIR SEDs largely improves the identification and evolutionary classification of the younger protostars based on their SED shape, allowing  estimates to be made of the properties of the possible circumstellar envelope. Moreover, our \her\, YSOs/protostars catalogue of the Lupus I, III, and IV regions includes objects not present in previous catalogues and identifies new candidates of Class 0 objects and YSOs with transitional disks.

\begin{acknowledgements}
KLJR acknowledges financial support by the Italian Ministero dell’Istruzione, Universit\`{a} e Ricerca through the grant Progetti Premiali 2012-iALMA (CUP C52I13000140001). N.S. acknowledges support by the french ANR and the german DFG through the project GENESIS (ANR-16-CE92-0035-01/DFG1591/2-1). P. P. acknowledges support from Funda\c{c}\~ao para a Ci\^encia e a Tecnologia of Portugal (FCT) through national funds (UID/FIS/04434/2013) and by FEDER through COMPETE2020 (POCI-01-0145-FEDER-007672) and also by the fellowship SFRH/BPD/110176/2015 funded by FCT (Portugal) and POPH/FSE (EC). This work has received support from the European Research Council under the European Union's Seventh Framework Programme (ERC Advanced Grant Agreements no. 291294 -- `ORISTARS').
\end{acknowledgements}

\begin{appendix}

\section{Dense cores and YSOs/protostars catalogues identified in Lupus I, III, and IV based on {\it \bf Herschel} data}
\label{ap:cat}

\begin{figure*}
\begin{tabular}{ccc}
\multicolumn{3}{c}{\textbf{\Large Source no. 3}}\\
&&\\
\multicolumn{3}{c}{\textbf{\Large HGBS-J153809.7-34074}}\\
&&\\
\includegraphics[width=0.33\textwidth]{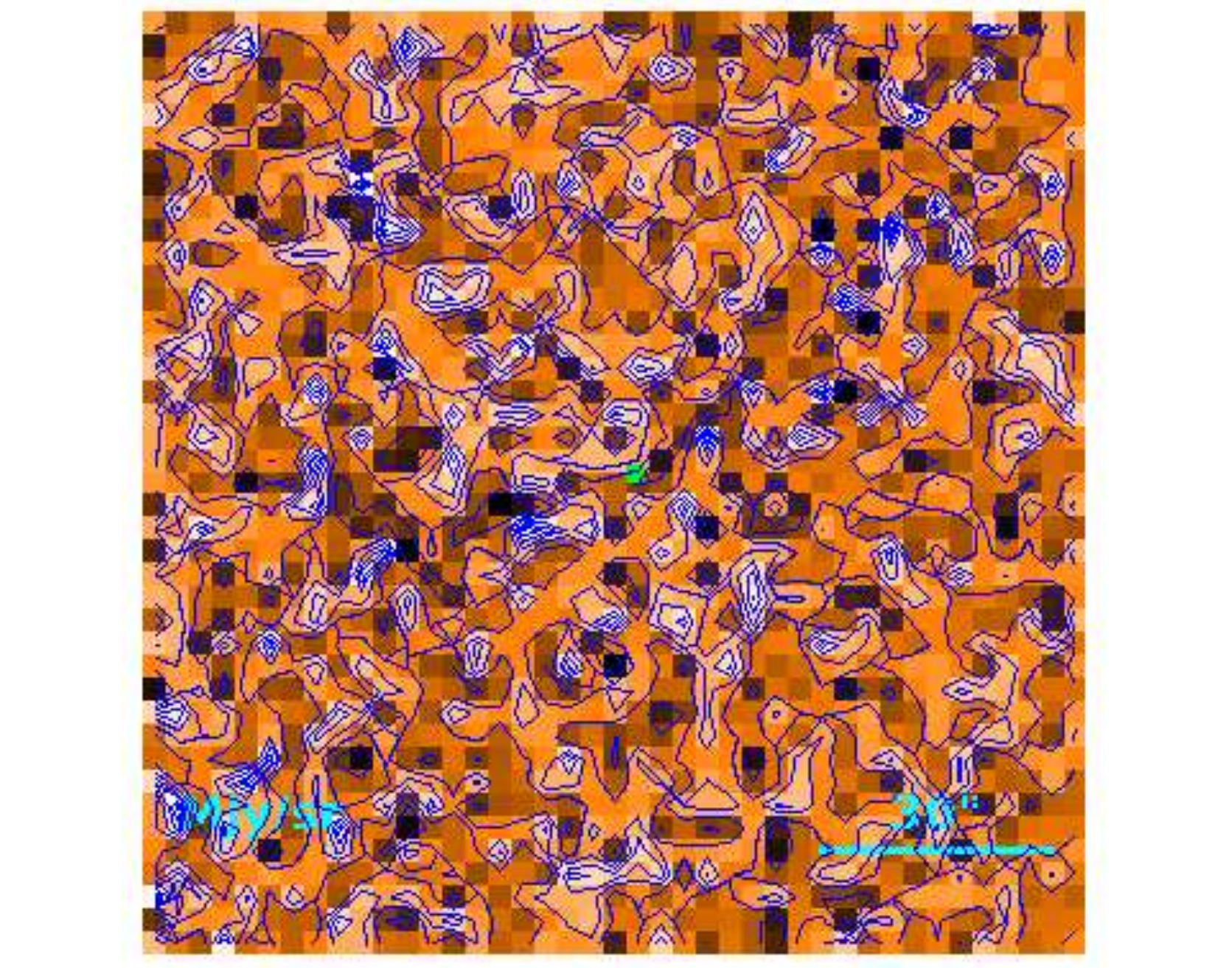}&\includegraphics[width=0.33\textwidth]{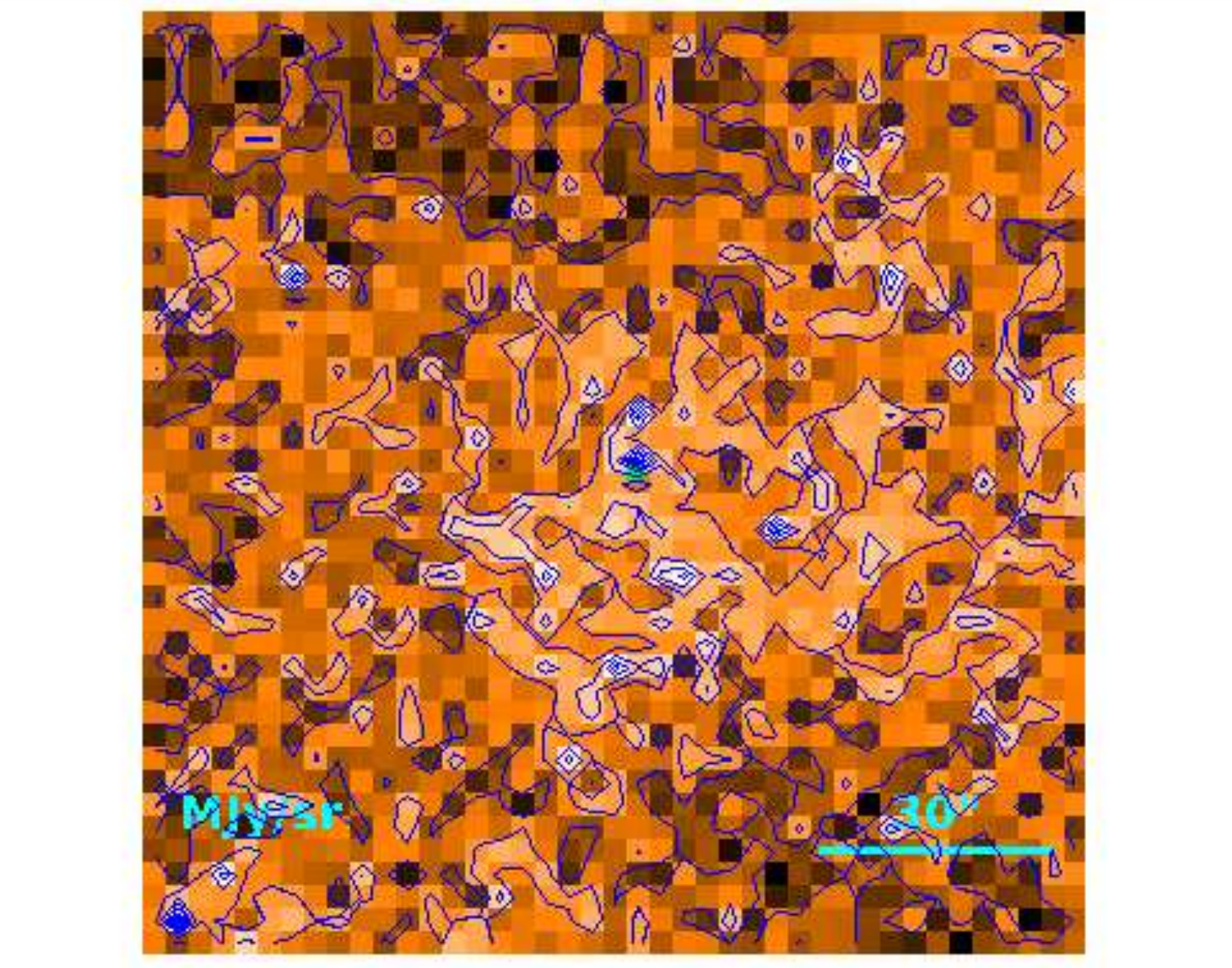}&\includegraphics[width=0.33\textwidth]{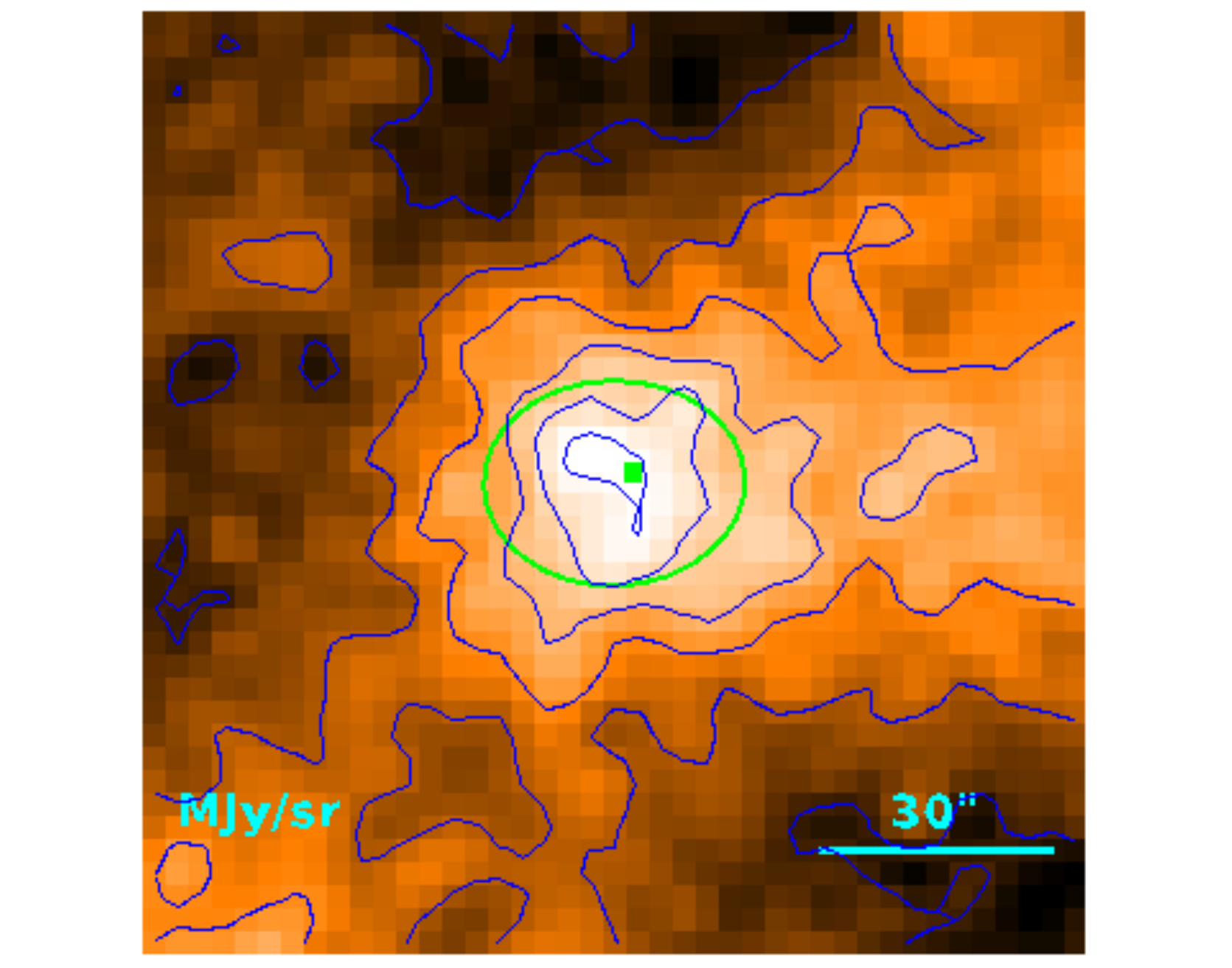}\\
\includegraphics[width=0.33\textwidth]{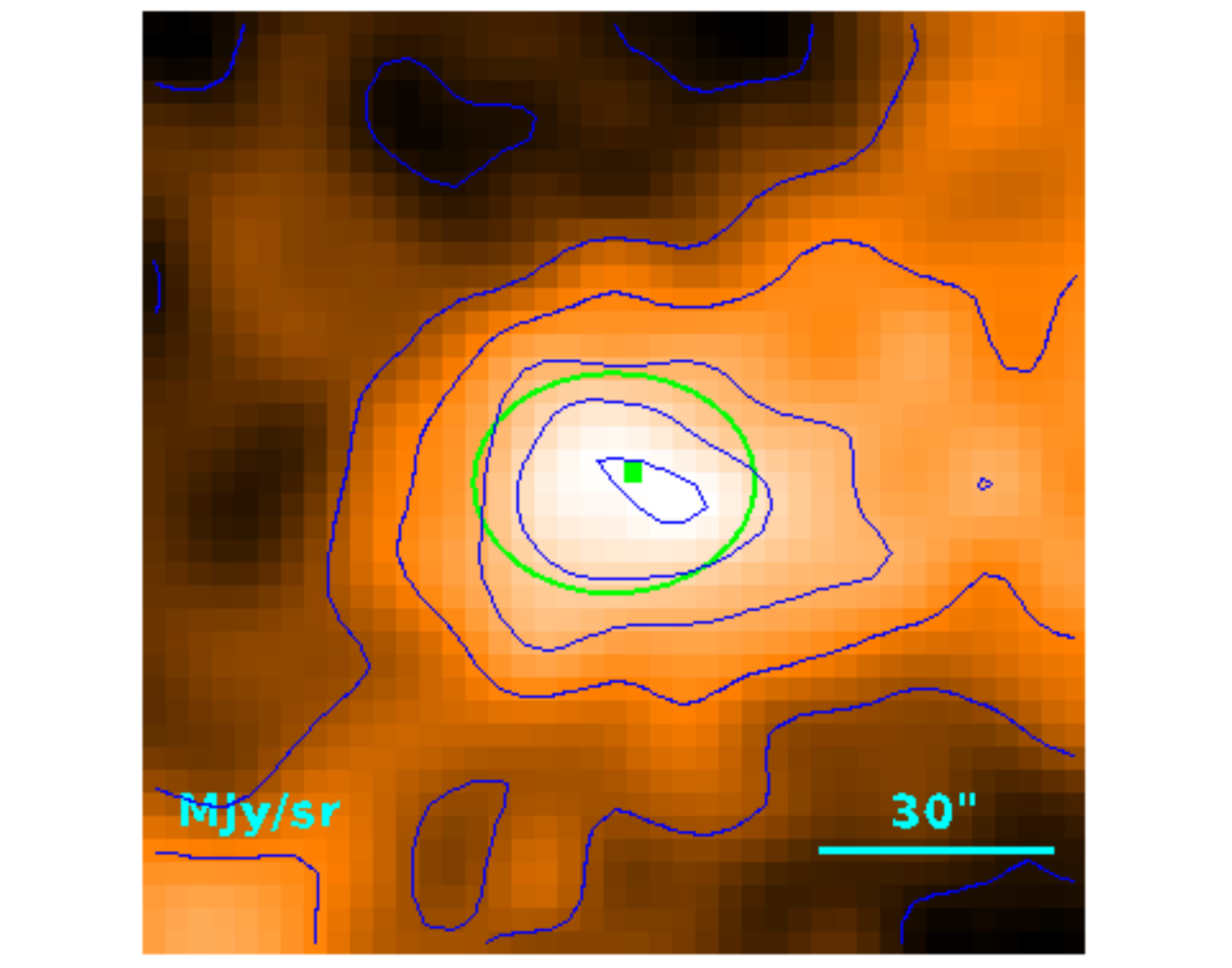}&\includegraphics[width=0.33\textwidth]{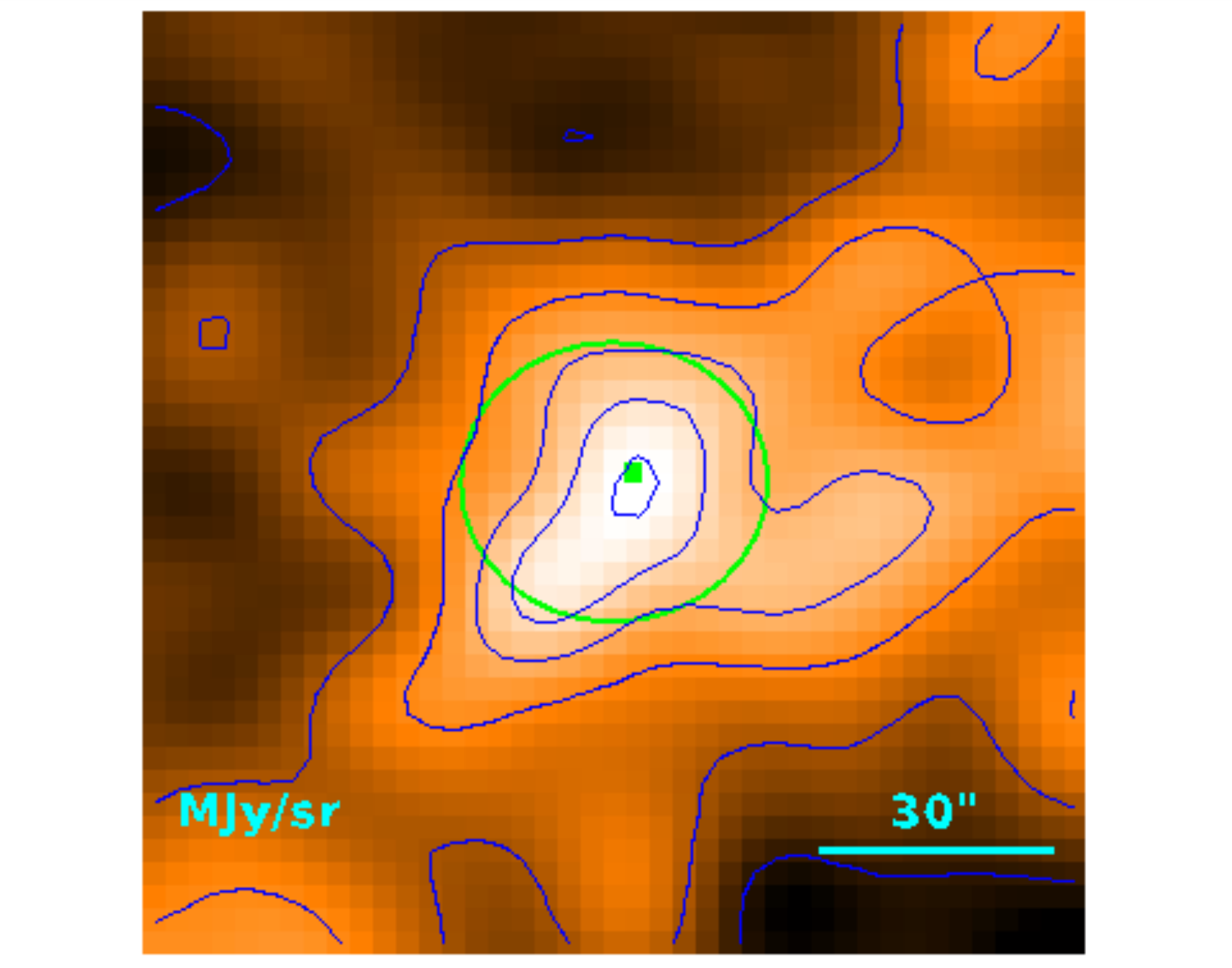}&\includegraphics[width=0.33\textwidth]{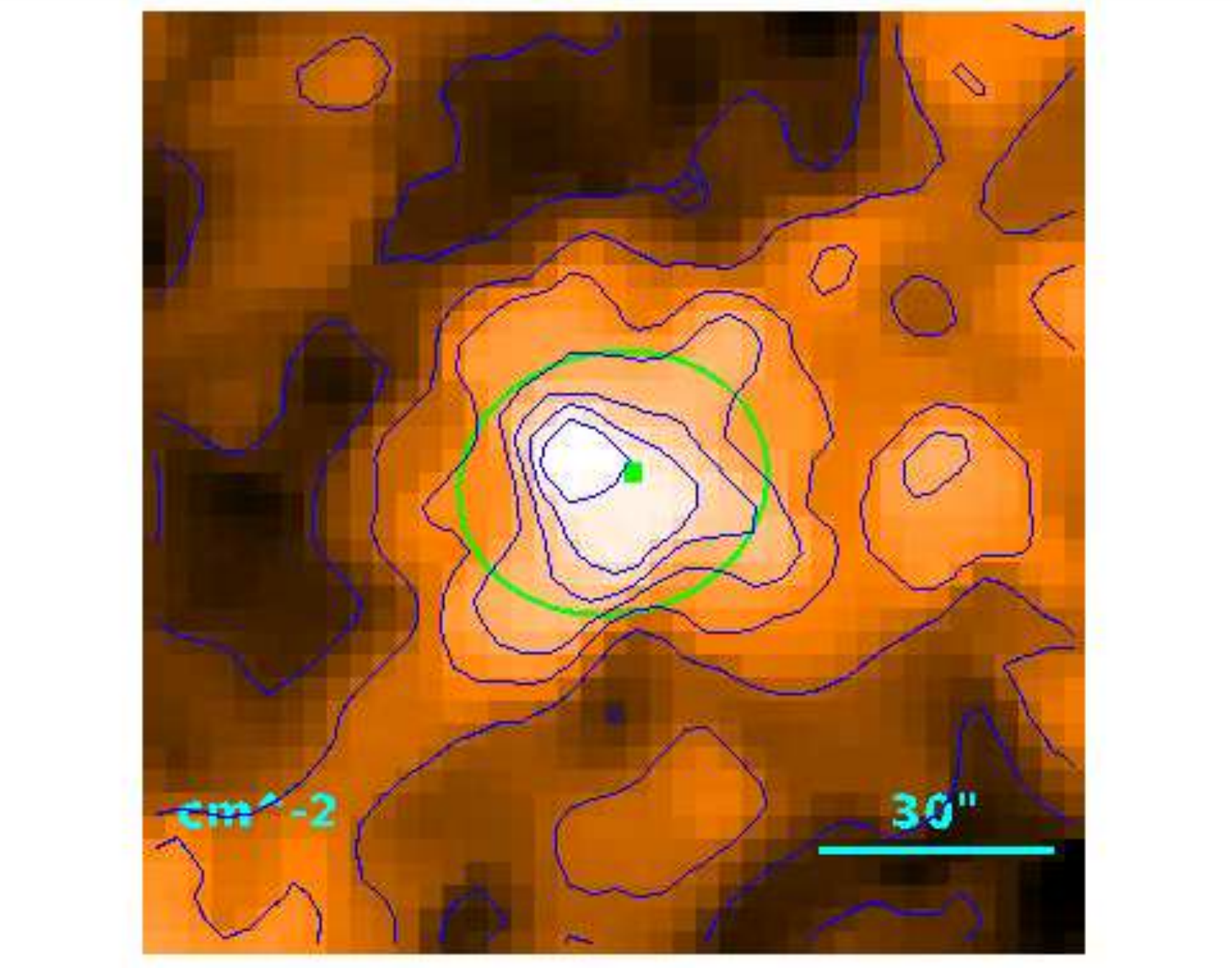}\\
\end{tabular}
\caption{2\arcmin$\times2$\arcmin\, cuts of the \her\,  images at 70 \um, 160 \um, 250 \um, 350 \um, and 500 \um\, and high-resolution column density map (in this order) centred on the HGBS-J153809.7-34074 dense core of our catalogue that has been classified as an unbound starless core. Green ellipses represent the estimated FWHM sizes of the core at each wavelength. Similar images for all the dense cores in our catalogues are provided at http://gouldbelt-herschel.cea.fr/archives.}
\label{fig:small_map_unbound}
\end{figure*}

\begin{figure*}
\begin{tabular}{ccc}
\multicolumn{3}{c}{\textbf{\Large Source no. 123}}\\
&&\\
\multicolumn{3}{c}{\textbf{\Large HGBS-J154024.0-33373}}\\
&&\\
\includegraphics[width=0.33\textwidth]{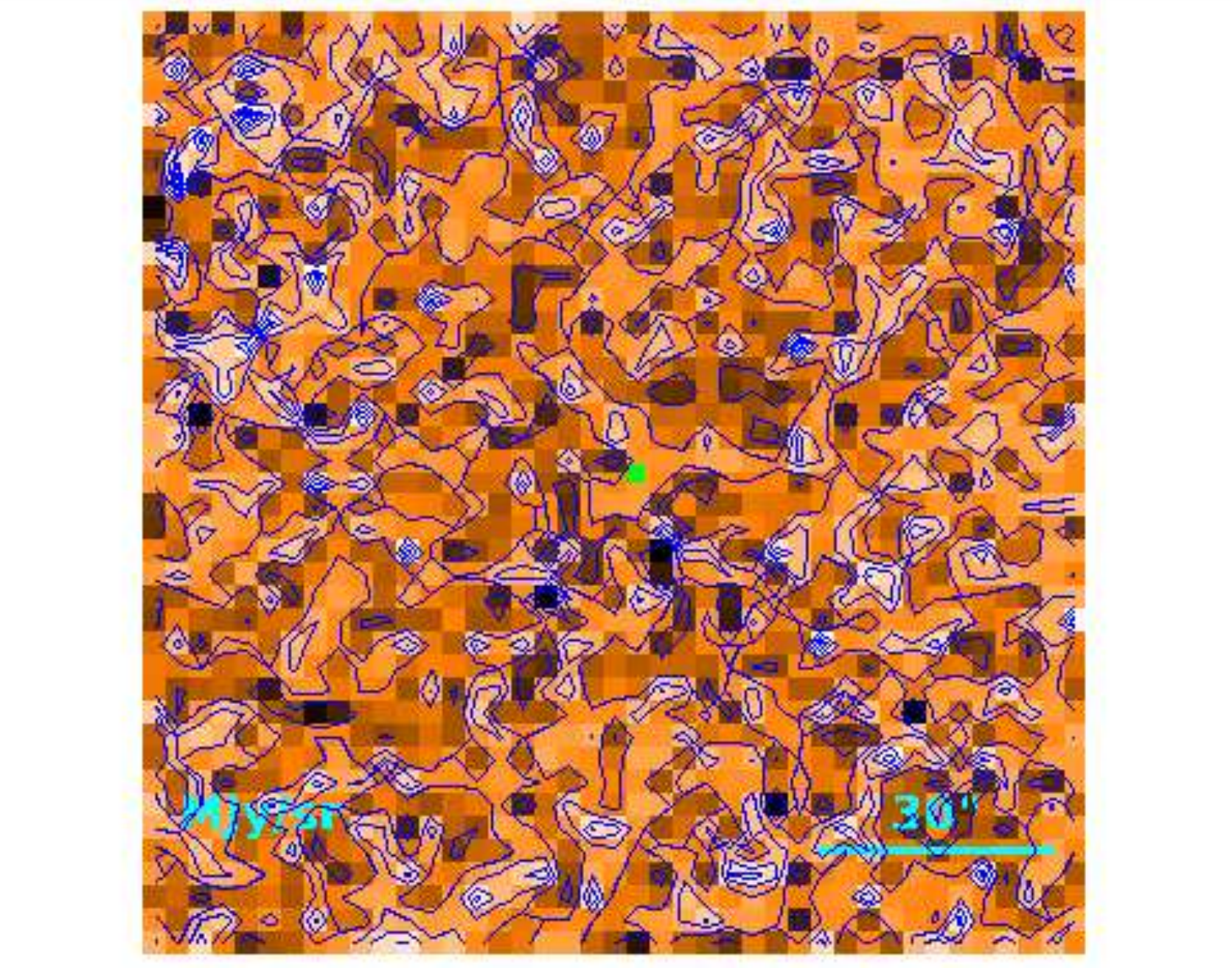}&\includegraphics[width=0.33\textwidth]{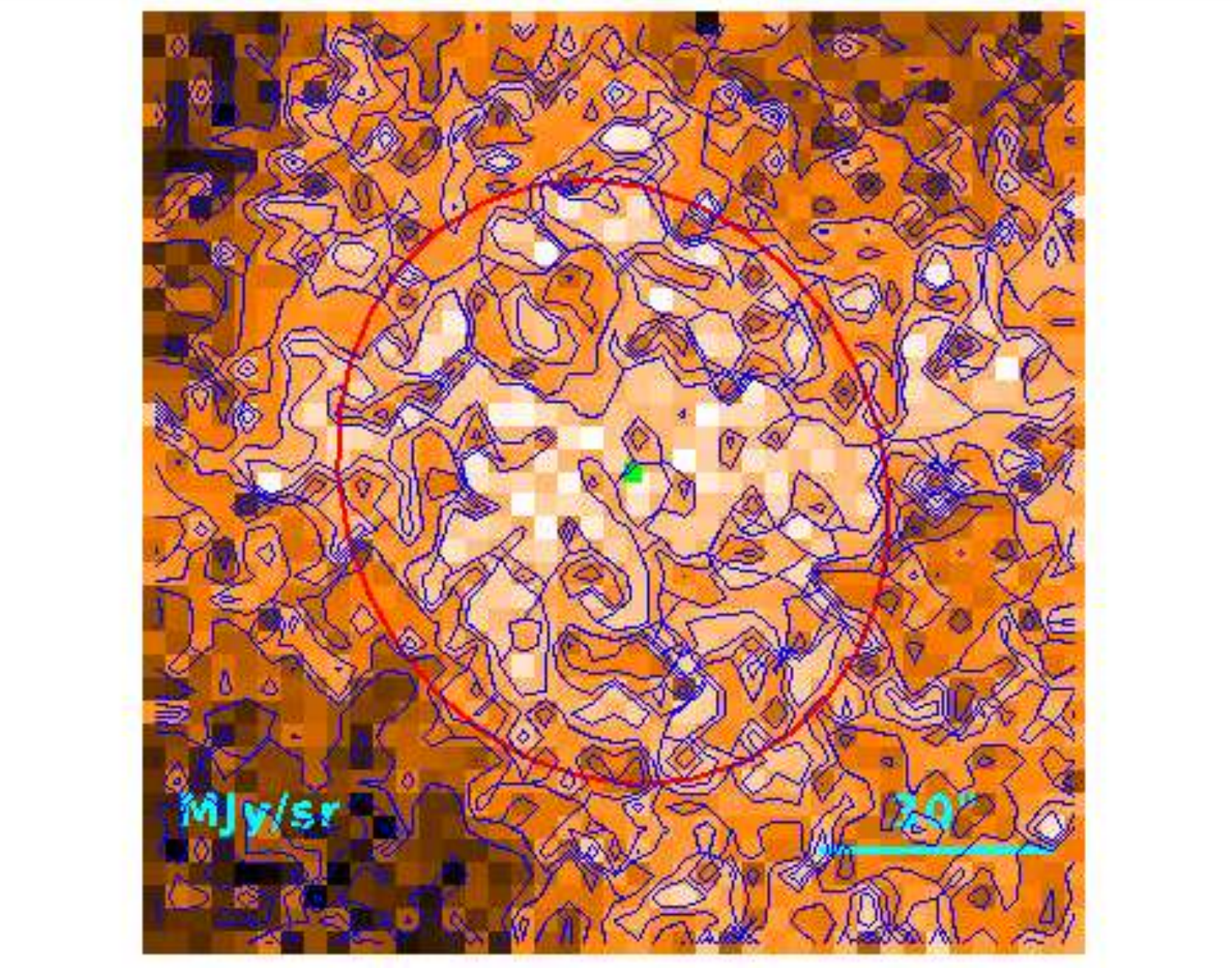}&\includegraphics[width=0.33\textwidth]{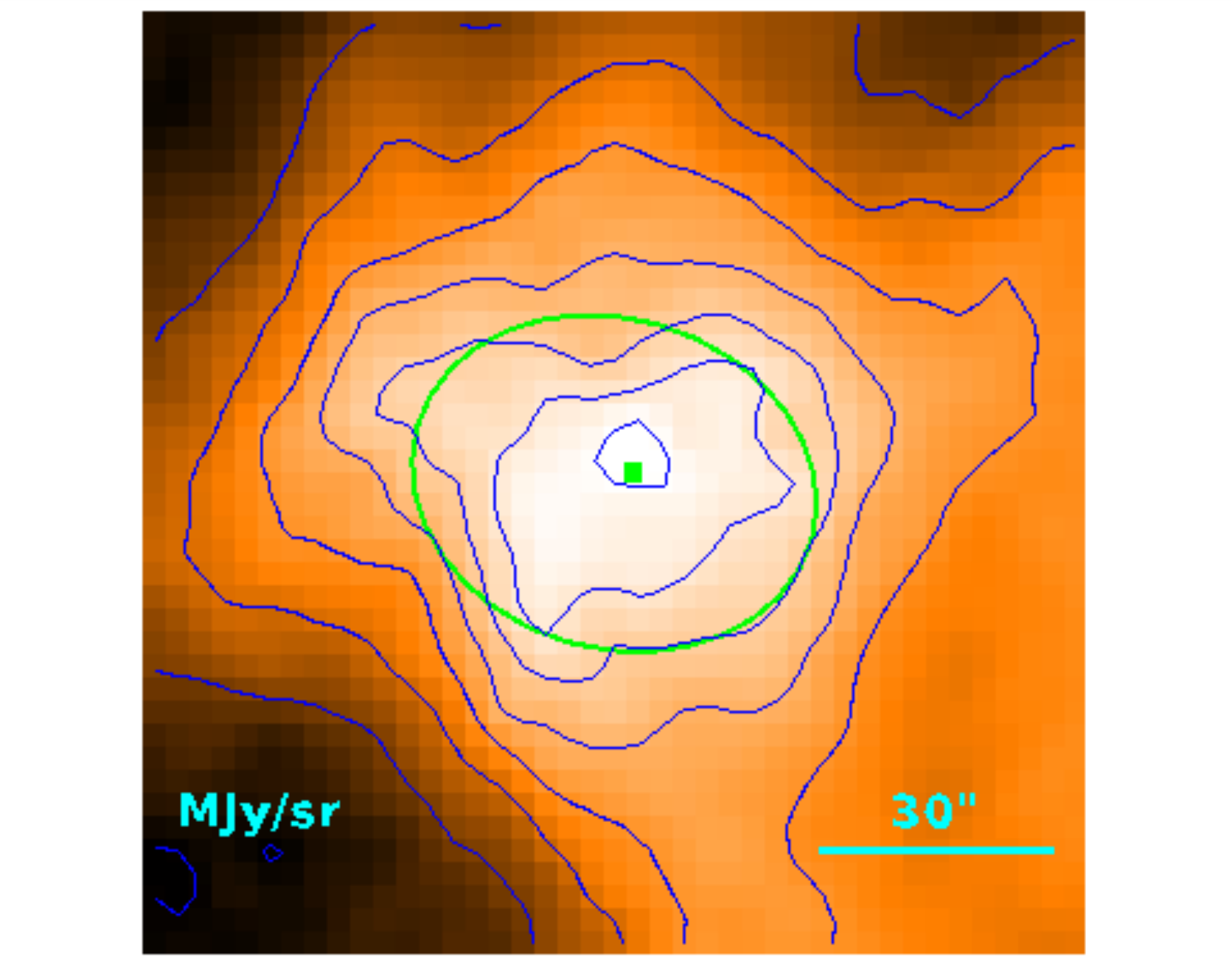}\\
\includegraphics[width=0.33\textwidth]{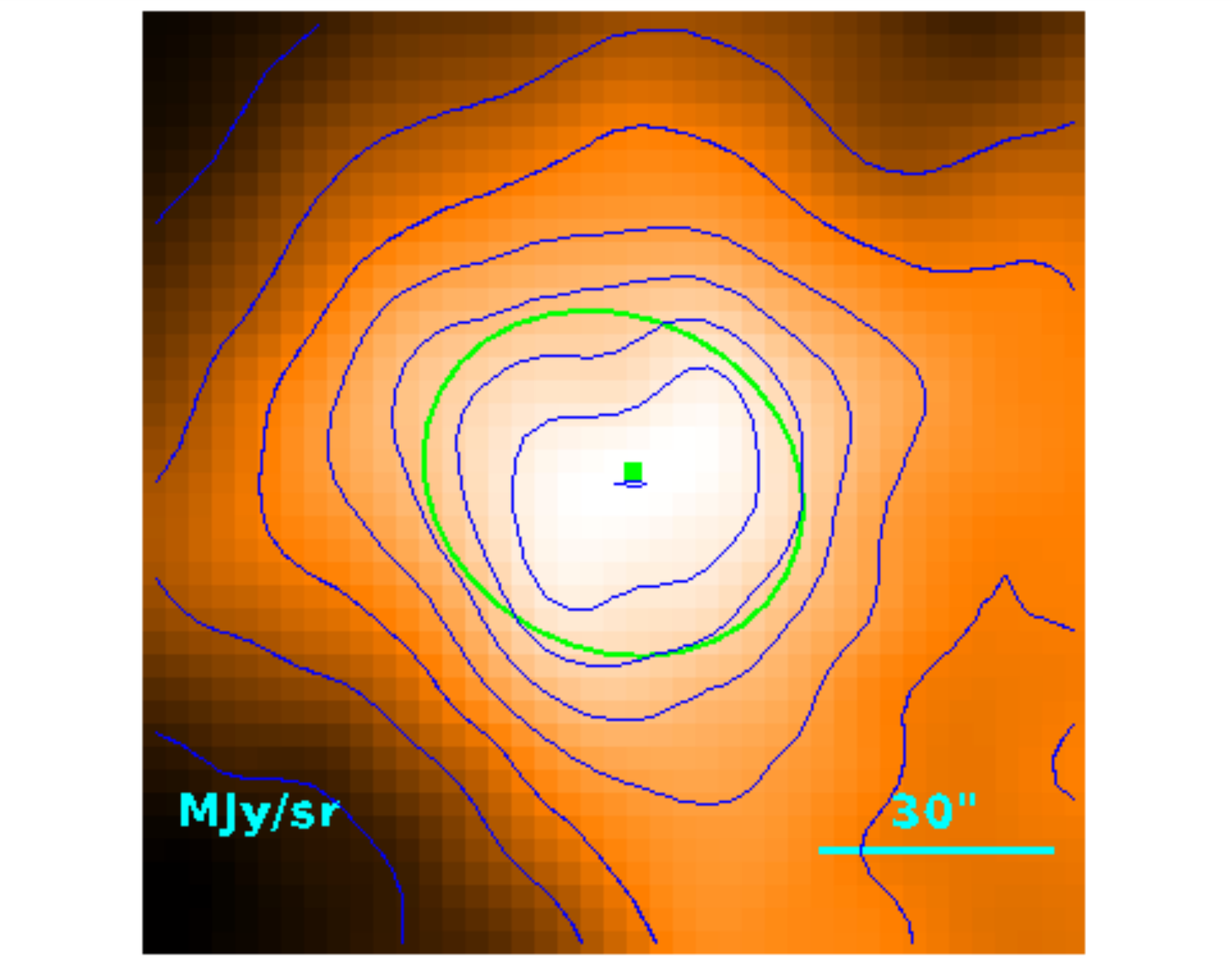}&\includegraphics[width=0.33\textwidth]{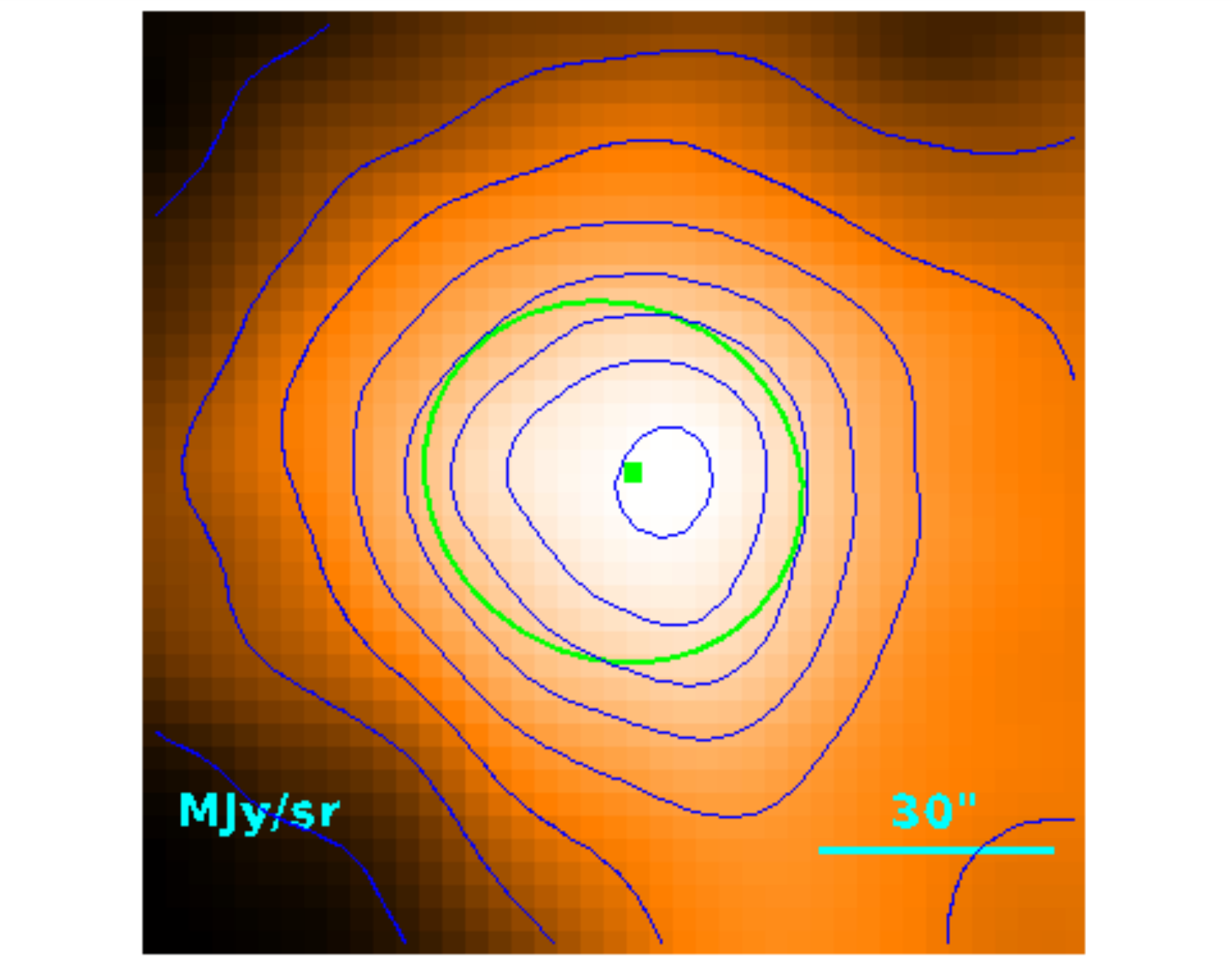}&\includegraphics[width=0.33\textwidth]{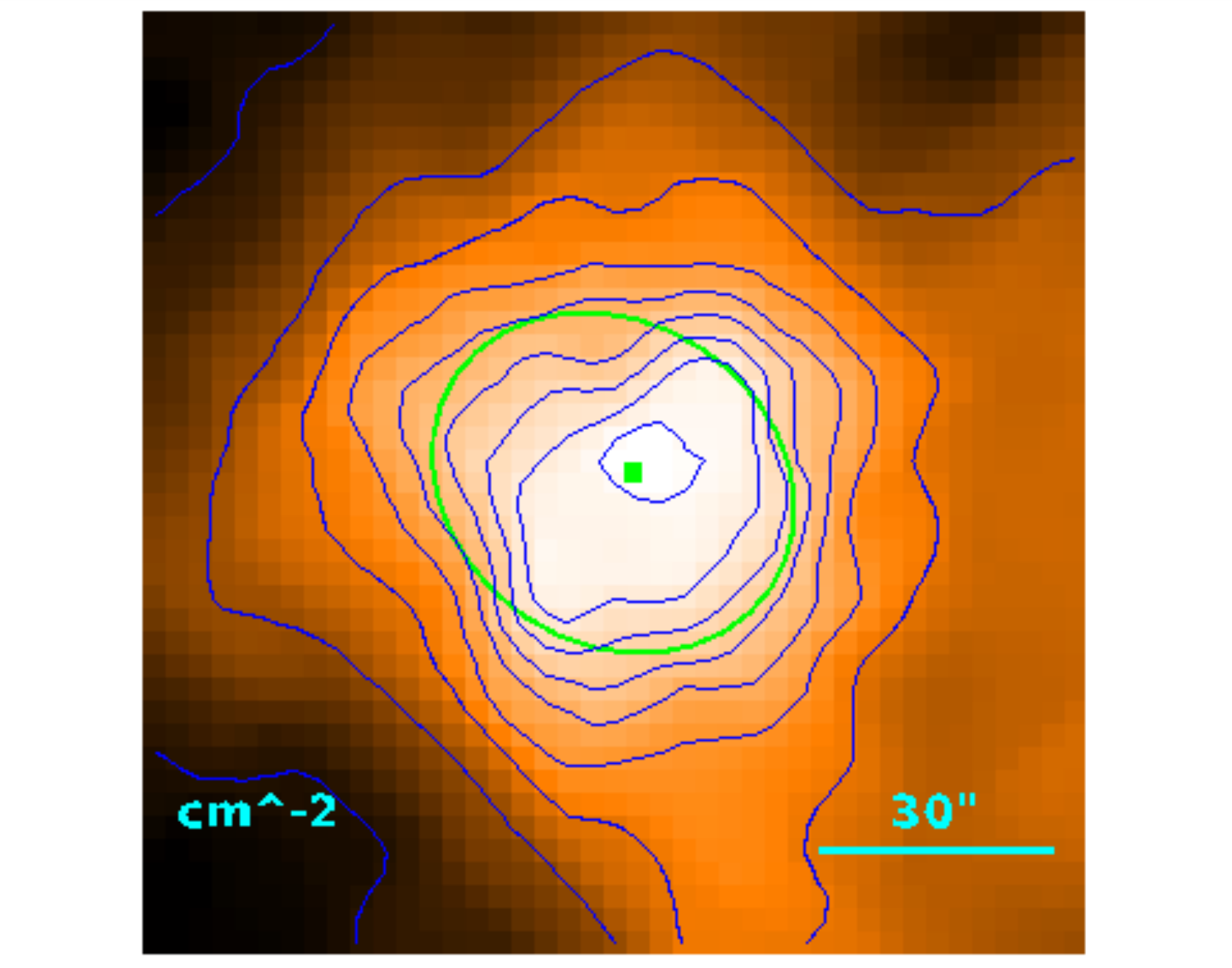}\\
\end{tabular}
\caption{Same as Fig. \ref{fig:small_map_unbound} for the HGBS-J154024.0-33373 dense core of our catalogue that has been classified as a prestellar core. Red ellipse indicates an uncertain flux estimate (see text for details).}
\label{fig:small_map_pre}
\end{figure*}

\begin{figure*}
\begin{tabular}{ccc}
\multicolumn{3}{c}{\textbf{\Large Source no. 6}}\\
&&\\
\multicolumn{3}{c}{\textbf{\Large HGBS\_YSO-J160044.6-415530}}\\
&&\\
\includegraphics[width=0.33\textwidth]{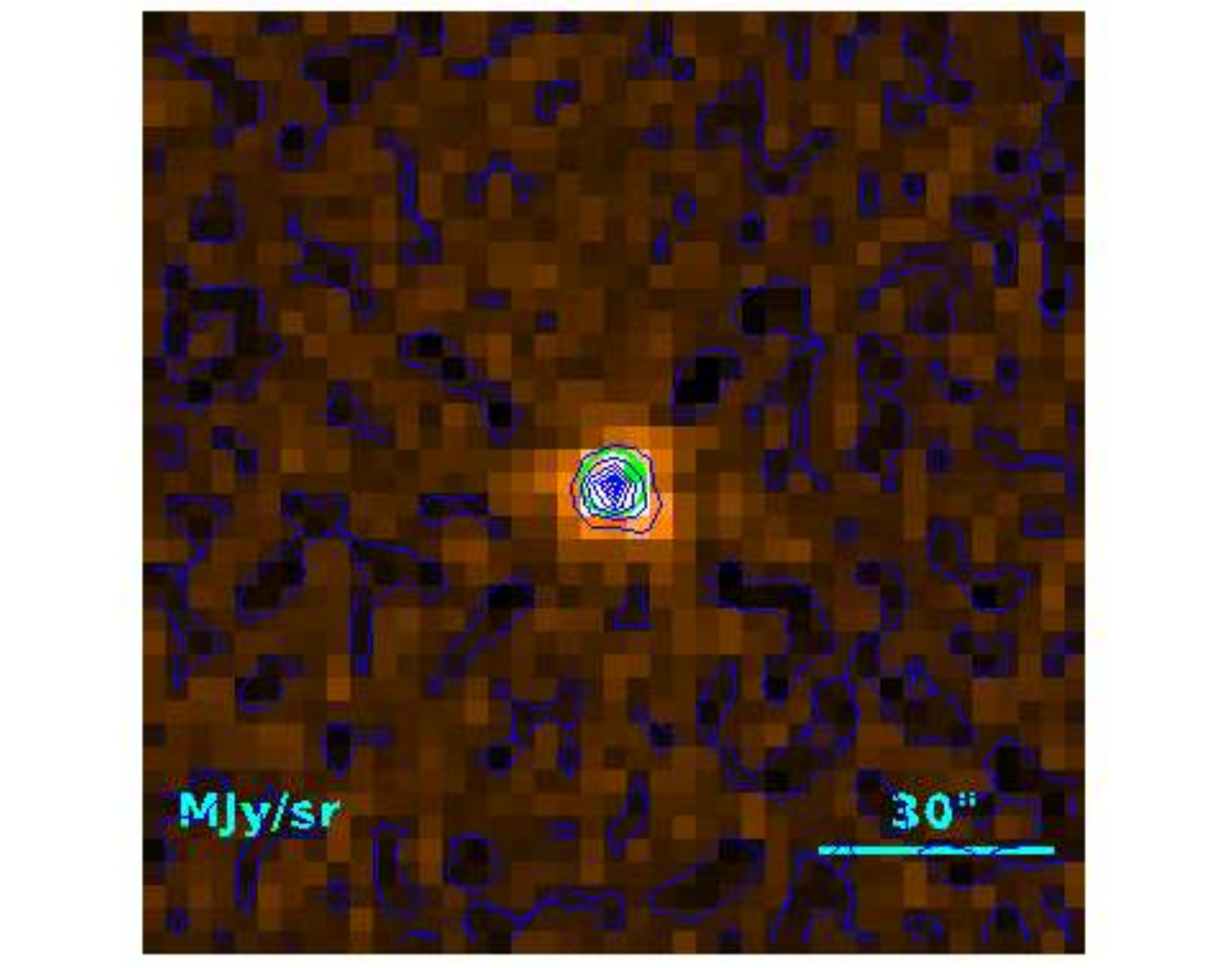}&\includegraphics[width=0.33\textwidth]{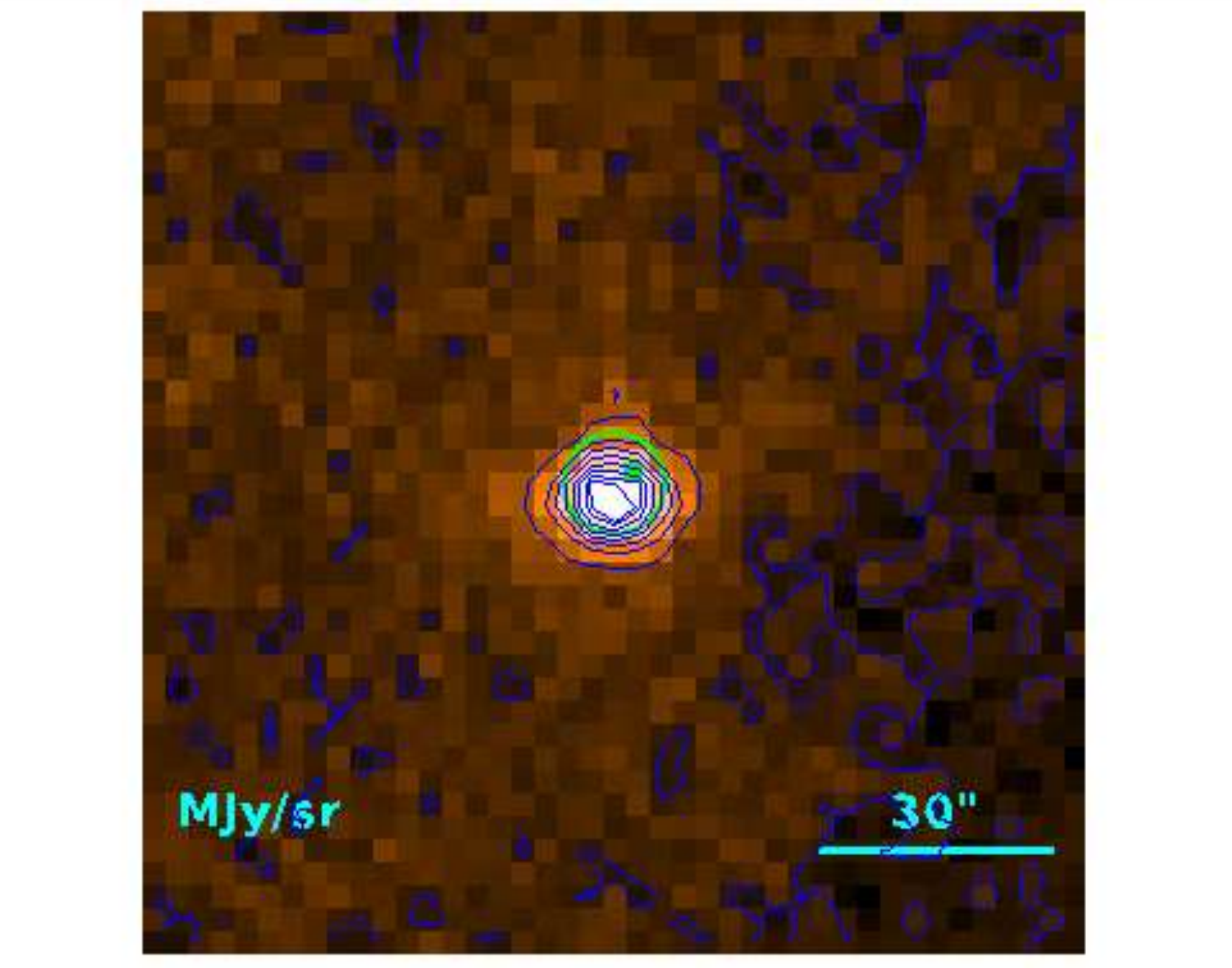}&\includegraphics[width=0.33\textwidth]{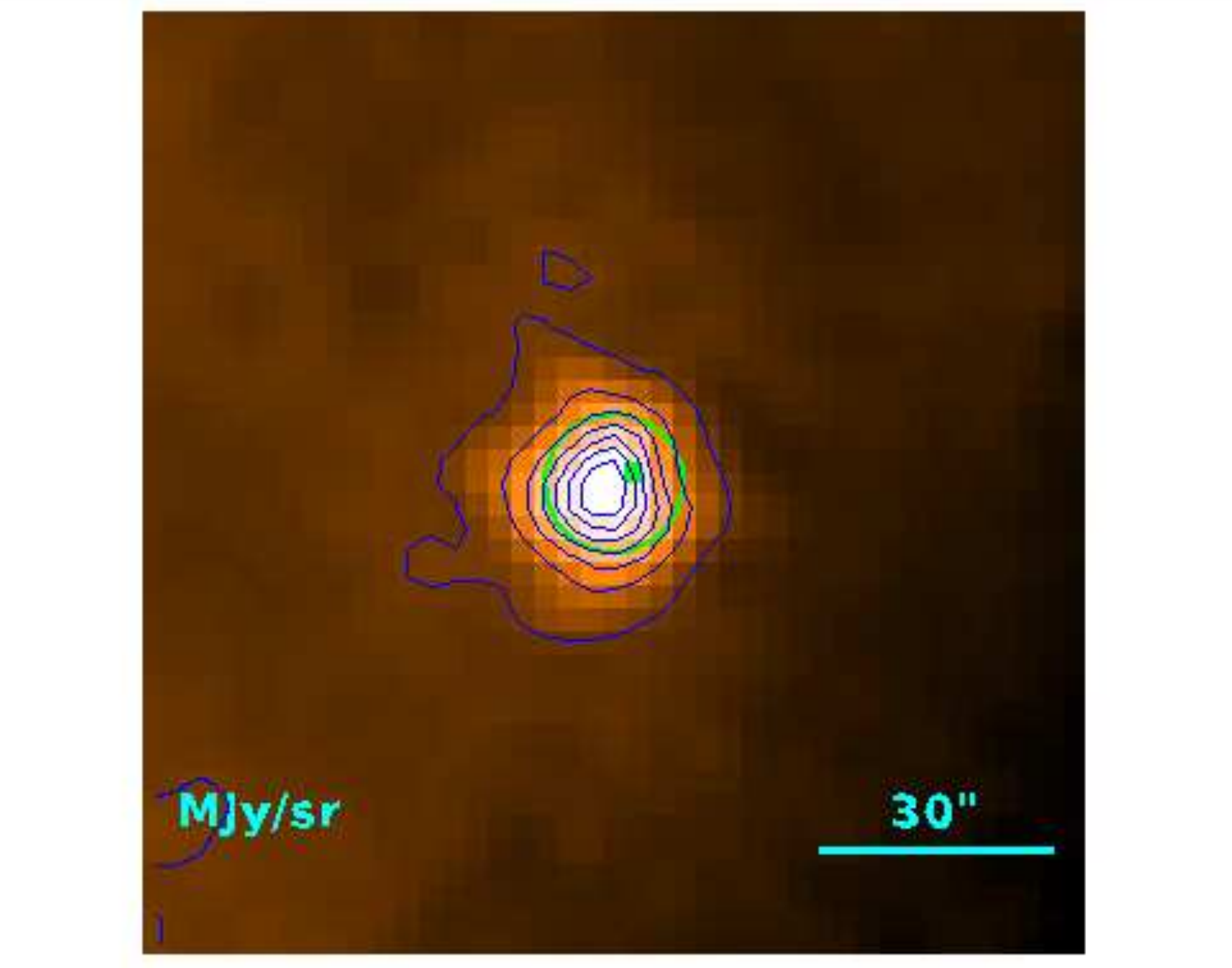}\\
\includegraphics[width=0.33\textwidth]{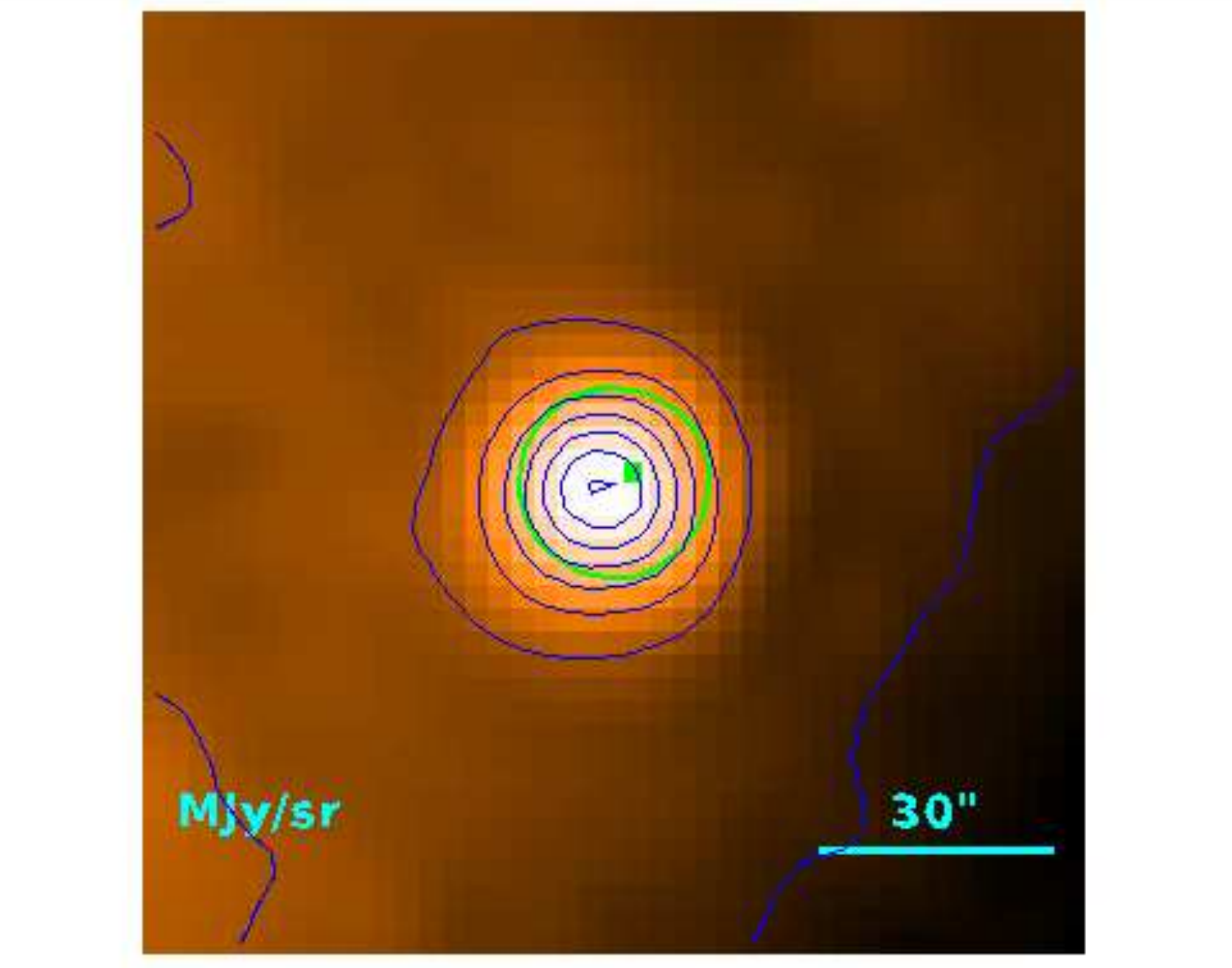}&\includegraphics[width=0.33\textwidth]{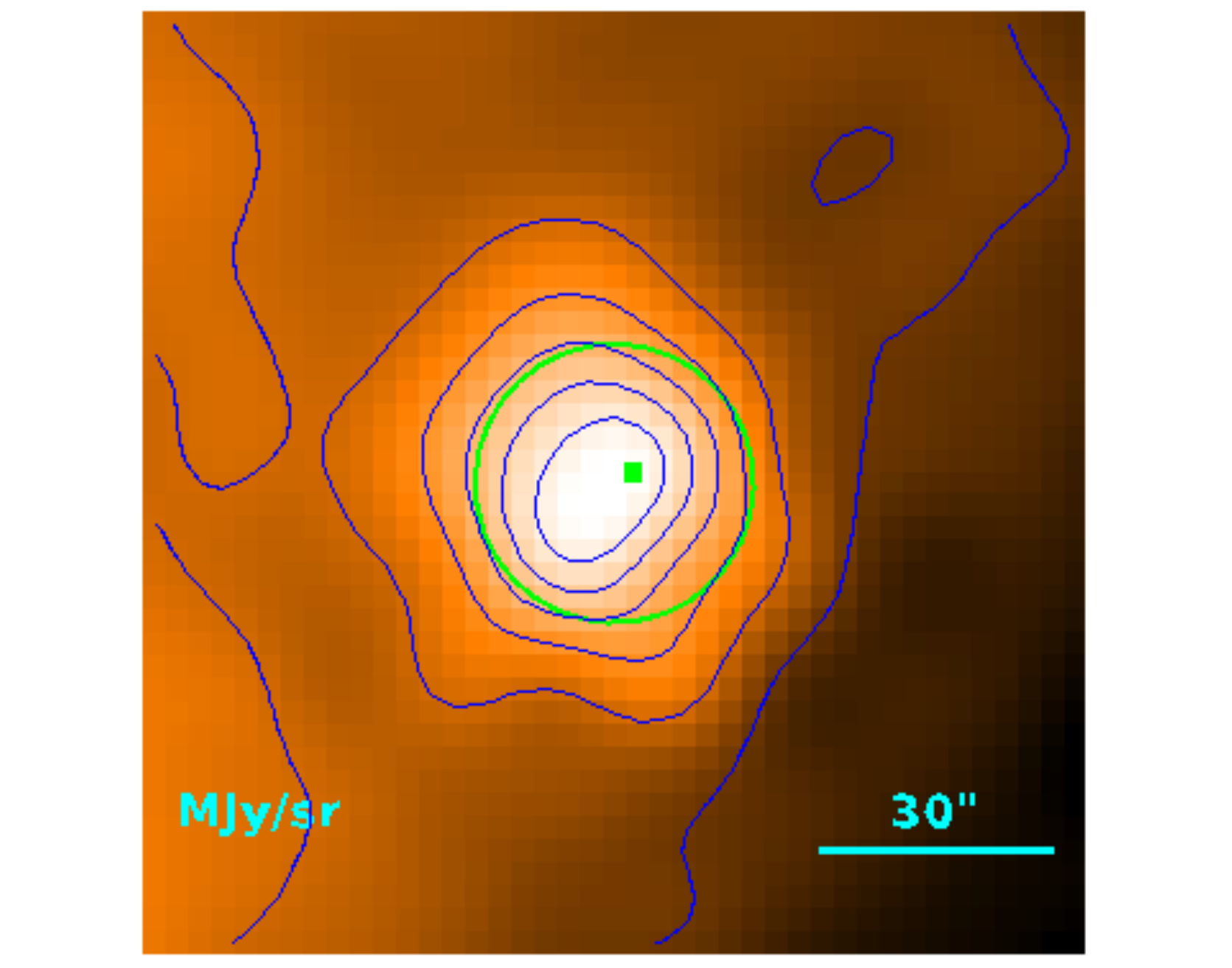}&\includegraphics[width=0.33\textwidth]{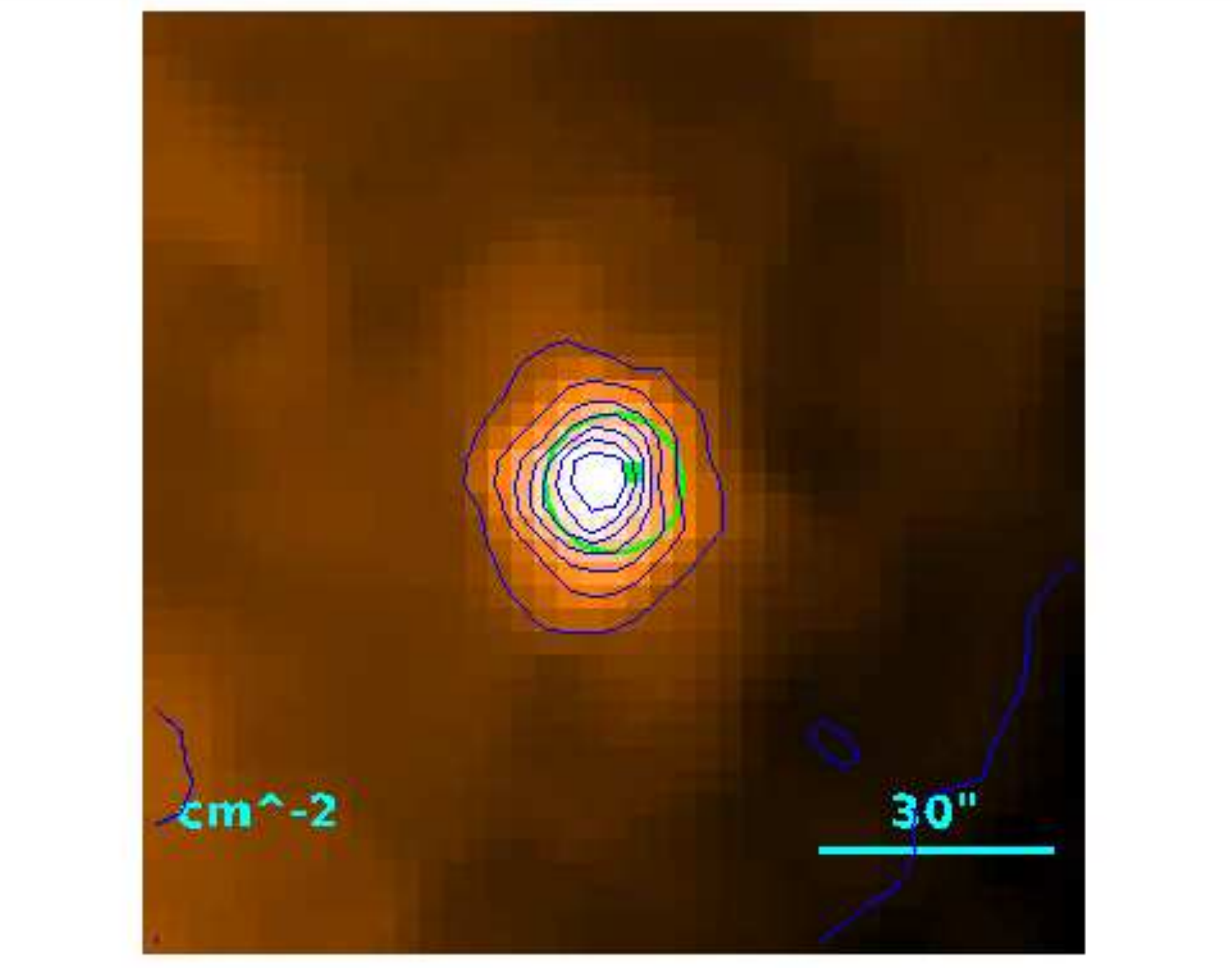}\\
\end{tabular}
\caption{Same as Fig. \ref{fig:small_map_unbound} for the protostar HGBS\_YSO-J160044.6-415530. Similar images for all the YSOs/protostrs in our catalogues are provided at the following URLs: https://owncloud.ia2.inaf.it/index.php/s/TiJtzW5gWo3YwKA, https://owncloud.ia2.inaf.it/index.php/s/LHfkSLF0pI63icB, https://owncloud.ia2.inaf.it/index.php/s/TMnYDhv9uj1mNHg, for Lupus I, III and IV respectively.
. }
\label{fig:small_map_proto}
\end{figure*}

\begin{figure*}
\begin{tabular}{ccc}
 \includegraphics[scale=.25]{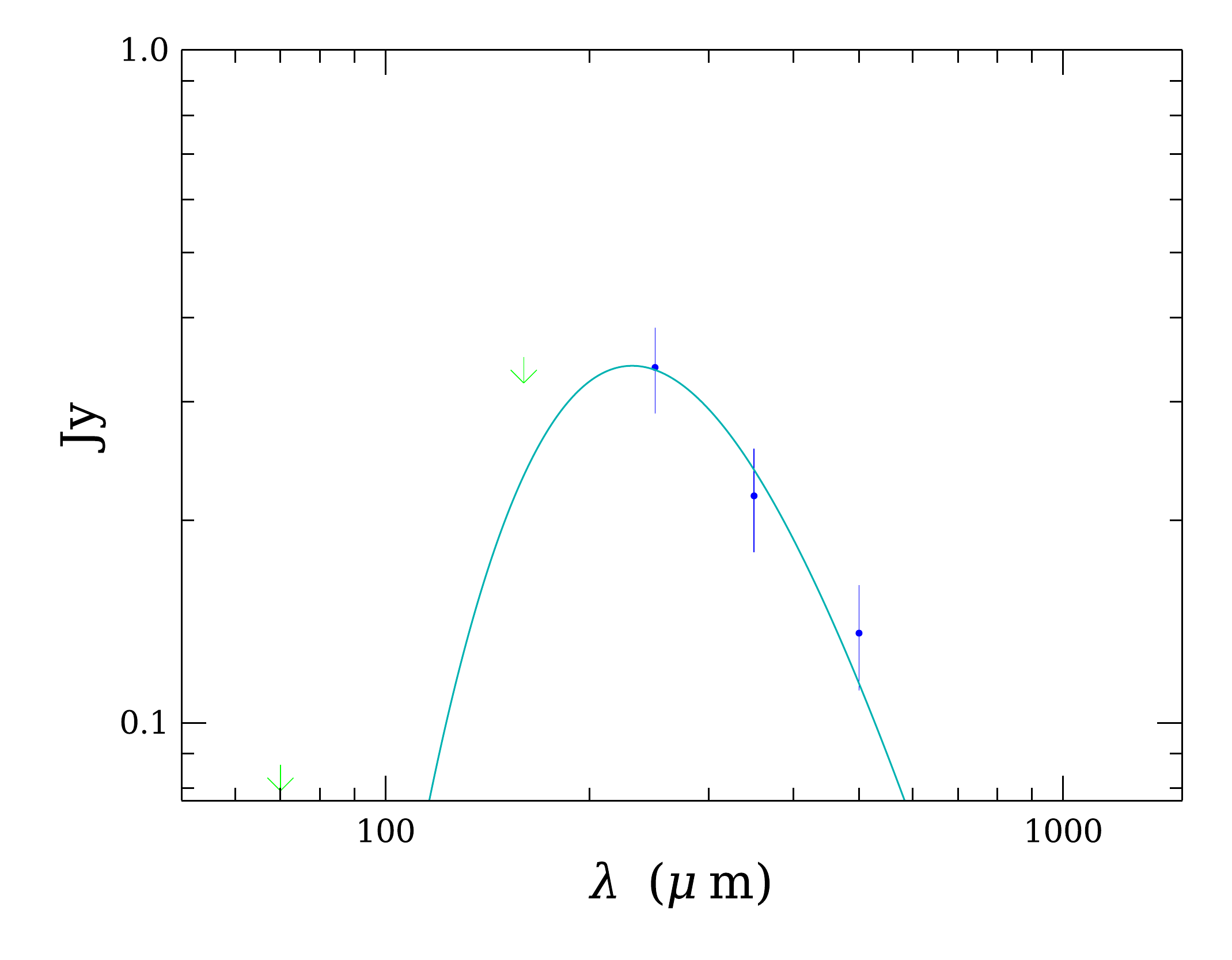} &
 \includegraphics[scale=.25]{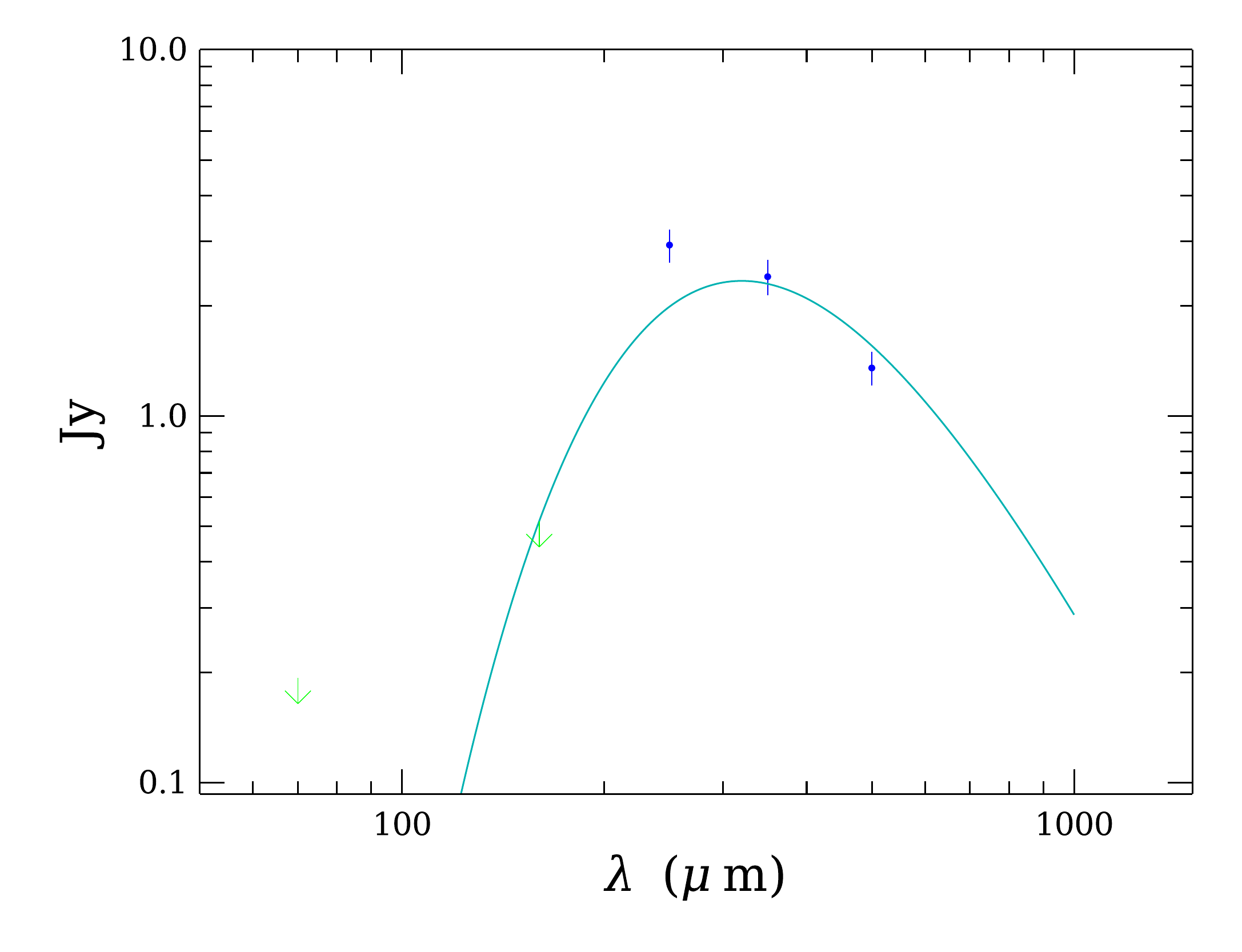} &
 \includegraphics[scale=.25]{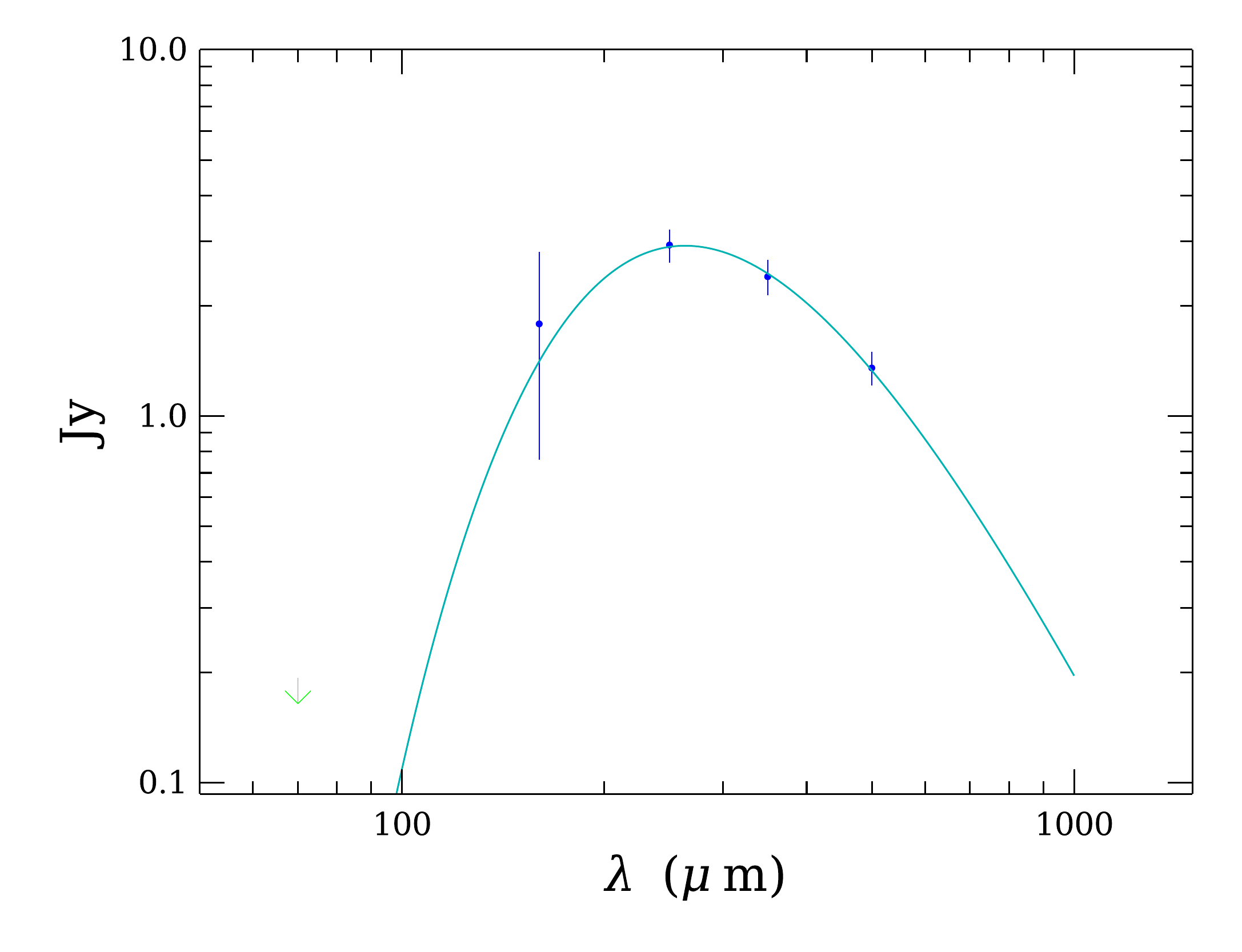} \\
\end{tabular}
\caption{Grey-body fit to the SED at the \her\, bands for sources HGBS-J153809.7-34074 (left panel) and HGBS-J154024.0-33373 (central and right panels) of the dense cores catalogue. Figures of the observed SED and the grey-body fit for all the dense cores in our catalogues are provided at http://gouldbelt-herschel.cea.fr/archives. Upper limits are indicated with green arrows. Source HGBS-J154024.0-33373 was not detected at 160~$\mu$m. In the central panel, we show the best SED fit that respects the upper limit at this wavelength: we cannot find any grey-body model that fits the SPIRE fluxes and the PACS upper limits at the same time. In the right panel, the 160~$\mu$m upper limit is relaxed (see text in Appendix \ref{ap:cat}) and flux with a very low weight is used in the fitting. Now we can find a very good fit to the observed SED.}
\label{fig:sed_exemp}
\end{figure*}

We extracted dense cores and YSOs/protostars in the five \her\, maps of the Lupus I, III, and IV molecular clouds. As examples, in Figs. \ref{fig:small_map_unbound}, \ref{fig:small_map_pre} and \ref{fig:small_map_proto}, we show  2\arcmin $\times$ 2\arcmin\, maps at 70 \um, 160 \um, 250 \um, 350 \um, and 500 \um\, and the high-resolution column density map for an unbound starless core, a prestellar core, and a protostar, respectively. Similar images for all the sources in our catalogues are provided at http://gouldbelt-herschel.cea.fr/archives. 
The complete catalogues of dense cores and YSOs/protostars as well as the physical catalogues are available in electronic form in the online material. Here, we provide a few lines of the catalogues to show the entries we provide.
In Table \ref{tab:cat}, we list the source name and position together with flux density measurements and FWHM at each band as measured in the maps.
The physical properties of dense cores, both unbound and prestellar, are derived from SED fitting at the \her\, bands performed as explained in Sect. \ref{sect:phys}. In  Fig. \ref{fig:sed_exemp}, we show the SED best fit for two sources, a similar image for all the dense cores in our catalogues is provided at http://gouldbelt-herschel.cea.fr/archives. For the unbound starless core HGBS-J153809.7-34074 (left panel of Fig. \ref{fig:sed_exemp}) it is possible to find a theoretical SED that not only fits very well the fluxes in the SPIRE bands, but is also perfectly compatible with the flux upper limits in the PACS bands.
For the prestellar core HGBS-J154024.0-33373, however, the 160 \um\, upper limit is clearly incompatible with the SPIRE fluxes (see the central panel of Fig.~\ref{fig:sed_exemp}). No theoretical SED has a shape that can fit both the observed fluxes and upper limits at the same time. Clearly, it is highly debatable that one upper limit counts more than three fluxes, and in any case the upper limit also has its own uncertainty.

In building the SEDs, we put an upper limit to the flux at a given wavelength when the source is not visible in any clean single scales from \textit{getsources} \citep{menshchikov12}. In this case the flux derived at the measurement stage is very uncertain. When an upper limit is too stringent and does not agree with the fluxes at the other bands, as for source HGBS-J154024.0-33373, we put the \verb+SIG_MON+ parameter of the \textit{getsources} output to a very low value, that is 9.999E-30. When this happen the ellipse that shows the geometrical properties of the source at the given band is shown in red, as shown in Fig.~\ref{fig:small_map_pre}, to warn the reader that the flux is very uncertain. Since each \verb+SIG_MON+ provides the weight of the respective measurement in the SED fitting procedure, and since for valid fluxes this value is always higher than a few units, it is clear that a flux that has a weight 30 or more orders of magnitude smaller than the other fluxes does not play any role in determining the best-fit SED. This change, however, has the important consequence that the upper limit does not longer constrain the shape of the SED.

The result of using a flux with a very low weight instead of as an upper limit is shown in the right panel of Fig.~\ref{fig:sed_exemp}. Now the best-fit SED reproduces very well the fluxes at all wavelengths. We note that the agreement with the 160~$\mu$m datum is by chance, probably because even if \textit{getsources} was not able to detect the source, it could  nonetheless make a valid estimate of its flux. In other cases, however, the datum at 160~$\mu$m is not compatible with the best-fit SED, because its weight in the fitting procedure is so low that it is as if that flux does not count.

The fact that the grey-body model now fits well the data has, of course, important consequences on the physical parameters of the source. In general, we found that the temperature derived when we impose that the fit is consistent with the 160 \um\, upper limit is around 7-9 K, much lower than the average temperature of about 12 K, showing that this upper limit cannot be used. In the particular case of source HGBS-J154024.0-33373, considering the upper limit at 160~$\mu$m results in $T=9.05^{+0.04}_{-0.03}$~K and $M=0.337\pm0.023\,M_\odot$. When the upper limit is not taken into account (again stressing that this does not mean that we are using the 160~$\mu$m flux, but that we are not considering it as a stringent  upper limit), the parameters are $T=10.99^{+0.08}_{-0.07}$~K and $M=0.159\pm0.010\,M_\odot$. The latter temperature is now much more similar to the average value and the mass is less than 50\% the value found when considering the upper limit.

The results of the SED fits for all sources, namely physical radius, mass, SED dust temperature, peak column density at the resolution of the 500 \um\, data, average column density, peak volume density, and average density are given in Table \ref{tab:cores_phys} for the dense cores catalogue.

\begin{sidewaystable*}
\caption{Catalogue of dense cores and YSOs/protostars identified in the HGBS maps of the Lupus I, III, and IV clouds. Only three sources are listed as example here, the complete catalogue is available at CDS. For each of the three clouds, two tables are provided: one for the catalogue of the dense cores and one for the catalogue of YSOs/protostars.}
\label{tab:cat}
\centering
\tiny
\begin{tabular*}{\textheight}{cccccccccccccc} 
\hline\hline  
n & Name & RA(2000) & Dec(2000) & Sig$_{70}$ & \multicolumn{2}{c}{$S^{\rm peak}_{70}$} & $S^{\rm peak}_{70}$/$S_{\rm bg}$ & $S^{\rm conv,500}_{70}$ & \multicolumn{2}{c}{$S^{\rm tot}_{70}$} & $\theta^{\rm a}_{70}$ & $\theta^{\rm b}_{70}$ & $PA_{70}$ \\
  & & (h m s) & (\degr\, \arcmin\, \arcsec) &  &  \multicolumn{2}{c}{(Jy/beam)} &  & (Jy/beam$_{500}$) & \multicolumn{2}{c}{(Jy)} & (\arcsec) & (\arcsec) & (\degr) \\
  (1)& (2)& (3)& (4)& (5)& \multicolumn{2}{c}{(6) $\pm$ (7)}& (8)& (9)& \multicolumn{2}{c}{(10) $\pm$ (11)}& (12)& (13)& (14) \\
\hline
 3 & HGBS-J153809.7-340741 & 15:38:09.77 & -34:07:42.6 & 1.6 & -2.14e-02 & 1.3e-02 & 0.84  & -1.06e-01 & 4.54e-03 & -2.8e-03 & 67 & 10 & 39 \\                                                                            
 123 & HGBS-J154024.0-333734 & 15:40:23.99 & -33:37:33.9 & 0.0 & 7.76e-03 & 9.9e-03 & 0.08 & 1.24e-01 &  2.52e-01 & 3.2e-01 & 72 & 64 & 14 \\  
 6 & HGBS\_YSO-J160044.6-415530 & 16:00:44.62 & -41:55:31.9 & 49.4 & 1.06e+00 & 2.2e-02 & 18.50 & 1.15e+00 & 1.07e+00 & 2.2e-02 & 8 & 8 & -4 \\                                                                    
 \hline\\
 \end{tabular*}
 
\begin{tabular*}{\textheight}{AEDDFEDcccAEDDFEDccc}
\hline\hline  
 
 Sig$_{160}$ & \multicolumn{2}{c}{$S^{\rm peak}_{160}$} & $S^{\rm peak}_{160}$/$S_{\rm bg}$ & $S^{\rm conv,500}_{160}$ & \multicolumn{2}{c}{$S^{\rm tot}_{160}$}  & $\theta^{\rm a}_{160}$ & $\theta^{\rm b}_{160}$ & $PA_{160}$ & Sig$_{250}$ & \multicolumn{2}{c}{$S^{\rm peak}_{250}$} & $S^{\rm peak}_{250}$/$S_{\rm bg}$ &$S^{\rm conv,500}_{250}$ &  \multicolumn{2}{c}{$S^{\rm tot}_{250}$} & $\theta^{\rm a}_{250}$ & $\theta^{\rm b}_{250}$ & $PA_{250}$ \\
 & \multicolumn{2}{c}{(Jy/beam)} &  & (Jy/beam$_{500}$) & \multicolumn{2}{c}{(Jy)} & (\arcsec) & (\arcsec) & (\degr) &  & \multicolumn{2}{c}{(Jy/beam)} &  & (Jy/beam$_{500}$) & \multicolumn{2}{c}{(Jy)} & (\arcsec) & (\arcsec) & (\degr) \\
 (15)& \multicolumn{2}{c}{(16) $\pm$ (17)}& (18)& (19)& \multicolumn{2}{c}{(20) $\pm$ (21)} & (22)& (23)& (24)& (25)& \multicolumn{2}{c}{(26) $\pm$ (27)}& (28)& (29)& \multicolumn{2}{c}{(30) $\pm$ (31)}& (32)& (33)& (34) \\
 \hline
   0.0 & 6.60e-02 & 4.1e-02 & 0.18 & 3.65e-01 & 5.30e-01 & 3.3e-01 & 54 & 50 & 18 & 7.6 & 1.13e-01 & 1.7e-02 & 0.25 & 2.06e-01 & 3.37e-01 & 4.9e-02 & 34 & 27 & -2 \\
   0.0 & 7.55e-02 & 4.3e-02 & 0.10 & 4.94e-01 & 1.78e+00 & 1.0e+00 & 80 & 70 & 66 & 20.0 & 3.85e-01 &  4.0e-02 & 0.35 & 1.03e+00 & 2.93e+00 & 3.0e-01 & 54 & 43 & 16 \\
  53.1 & 1.47e+00 & 4.5e-02 & 3.81 & 1.59e+00 & 1.50e+00 & 4.6e-02 & 14 & 14 & 0 & 86.8 & 1.36e+00 & 2.8e-02 & 1.42 & 1.31e+00 & 1.07e+00 & 2.2e-02 & 18 & 18 & -82 \\ 
  \hline\\
\end{tabular*}

\begin{tabular*}{\textheight}{AEDDFEDcccAEDDEDccc}
\hline\hline  
 Sig$_{350}$ & \multicolumn{2}{c}{$S^{\rm peak}_{350}$} & $S^{\rm peak}_{350}$/$S_{\rm bg}$ & $S^{\rm conv,500}_{350}$ & \multicolumn{2}{c}{$S^{\rm tot}_{350}$}  & $\theta^{\rm a}_{350}$ & $\theta^{\rm b}_{350}$ & $PA_{160}$ & Sig$_{500}$ & \multicolumn{2}{c}{$S^{\rm peak}_{500}$} & $S^{\rm peak}_{500}$/$S_{\rm bg}$  &  \multicolumn{2}{c}{$S^{\rm tot}_{500}$} & $\theta^{\rm a}_{500}$ & $\theta^{\rm b}_{500}$ & $PA_{500}$ \\
 & \multicolumn{2}{c}{(Jy/beam)} &  & (Jy/beam$_{500}$) & \multicolumn{2}{c}{(Jy)} & (\arcsec) & (\arcsec) & (\degr) &  & \multicolumn{2}{c}{(Jy/beam)} &   & \multicolumn{2}{c}{(Jy)} & (\arcsec) & (\arcsec) & (\degr) \\
 (35)& \multicolumn{2}{c}{(36) $\pm$ (37)}& (38)& (39)& \multicolumn{2}{c}{(40) $\pm$ (41)} & (42)& (43)& (44)& (45)& \multicolumn{2}{c}{(46) $\pm$ (47)}& (48)& \multicolumn{2}{c}{(49) $\pm$ (50)}& (51)& (52)& (53) \\
   \hline
  6.4 & 1.06e-01 & 1.9e-02 & 0.26 & 1.33e-01 & 2.17e-01 & 3.8e-02 & 37 & 29 & -1 & 6.9 & 1.17e-01 & 2.1e-02 & 0.33 & 1.36e-01 & 2.4e-02 & 40 & 36 & 0 \\ 
 29.4 & 5.58e-01 & 6.1e-02 & 0.53 & 8.94e-01 & 2.40e+00 & 2.6e-01 & 51 & 43 & 28 &43.7 & 6.06e-01 &  6.4e-02 & 0.65 & 1.35e+00 & 1.4e-01 & 51 & 46 & 32  \\
 52.2 & 8.66e-01 & 2.0e-02 & 0.94 & 8.57e-01 & 7.48e-01 & 1.7e-02 & 25 & 25 & 65 &34.0 & 5.39e-01 & 4.0e-02 & 0.67 & 4.79e-01 & 3.6e-02 & 36 & 36 & 70 \\
   \hline\\
\end{tabular*}

\begin{tabular*}{\textheight}{ccccccccccccccc}
\hline\hline  
 Sig$_{N_{\rm H_2}}$ & $N^{\rm peak}_{\rm H_2}$ &$N^{\rm peak}_{\rm H_2}$/$N_{\rm bg}$ & $N^{\rm conv,500}_{\rm H_2}$ & $N^{\rm bg}_{\rm H_2}$ & $\theta^{\rm a}_{N_{\rm H_2}}$ & $\theta^{\rm b}_{N_{\rm H_2}}$ & $PA_{N_{\rm H_2}}$ & $N_{\rm SED}$ & {\it CuTEx} & Core type & Comments & SIMBAD  & \spit  &  \wise \\
 & (10$^{21}$ cm$^{-2}$) &  & (10$^{21}$ cm$^{-2}$) & (10$^{21}$ cm$^{-2}$) & (\arcsec) & (\arcsec) & (\degr)  &  &  &  &  &   &    SSTc2d \\
 (54)& (55)& (56)& (57) & (58) & (59) & (60) & (61) & (62) & (63)&  (64) & (65)&  (66) & (67) & (68) \\
 \hline
  7.9 & 0.37 & 0.52 & 0.2& 0.7 &  41 & 34 &-12 & 3 & 0 & starless \\          
  65.7 & 2.42 & 0.99 & 1.6 & 2.5 & 50 & 42 & 34 & 3 & 1 & prestellar & tentative bound \\
  65.0 & 3.0 & 1.38 & 0.8 & 2.2 & 18 & 18 & 73 &  5 & 1 & protostellar & & V MY Lup &  J160044.5-415531 & J160044.51-415531.2 \\     
\hline\\
\end{tabular*}

\tablefoot{
Catalogue entries are as follows: (1) source running number; (2) source name = HGBS\_J prefix followed by a tag created from the J2000 sexagesimal coordinates; (3) and (4): Right Ascension and declination of source centre; (5), (15), (25), (35), and (45): detection significance from monochromatic single scales, in the 70 \um, 160 \um, 250 \um, 350 \um, and 500 \um\, maps, respectively (NB: the detection significance has the special value of 0.0 when the source is not visible in clean single scales); (6) $\pm$ (7), (16) $\pm$ (17), (26) $\pm$ (27), (36) $\pm$ (37), (46) $\pm$ (47): peak flux density and its error in Jy/beam as estimated by {\it getsources}; (8), (18), (28), (38), (48): contrast over the local background, defined as the ratio of the background-subtracted peak intensity to the local background intensity; (9), (19), (29), (39): peak flux density measured after smoothing to a 36.3\arcsec\, beam; (10) $\pm$ (11), (20) $\pm$ (21), (30) $\pm$ (31), (40) $\pm$ (41), (49) $\pm$ (50): integrated flux density and its error in Jy as estimated by {\it getsources};
(12), (22), (32), (42), (51): major FWHM diameters of the source as estimated by {\it getsources}; (13), (23), (33), (43), (52):  minor FWHM diameters of the source as estimated by {\it getsources} (NB: the special value of −1 means that no size measurement was possible); (14), (24), (34), (44), (53): position angle of the source major axis, measured east of north; (54) detection significance in the high-resolution column density image; (55) peak H$_2$ column density as estimated by {\it getsources} in the high-resolution column density image; (56) column density contrast over the local background, as estimated by {\it getsources} in the high-resolution column density image; (57) peak column density measured in a 36.3\arcsec\, beam; (58) local background H$_2$ column density as estimated by {\it getsources} in the high-resolution column density image; (59)–(60)–(61): major and minor FWHM diameters of the source, and position angle of the major axis, respectively, as measured in the high-resolution column density image; (62) number of {\it Herschel} bands in which the source is significant (Sig$_{\lambda}$ > 5) and has a positive flux density, excluding the column density plane; (63) {\it CuTEx} flag: 1 if the {\it getsources} source has a counterpart detected with the {\it CuTEx} algorithm, 0 if no {\it CuTEx} counterpart exists; (64) source type: starless, prestellar, or protostellar; (65) comments, (66) closest counterpart found in SIMBAD within 6\arcsec\, from the \her\, position, (67) closest \spit\, counterpart within 6\arcsec\, from the \her\, position.  When present, the Spitzer source name has the form of SSTc2d JHHMMSSs-DDMMSS; (68) closest \wise\, counterpart within 6\arcsec\, from the \her\, position. When present, the WISE source name has the form of JHHMMSS.ss-DDMMSS.s 
}
\end{sidewaystable*}

\begin{sidewaystable*}
\caption{Physical properties of starless cores identified in the HGBS maps of the Lupus I, III, and IV clouds. Only two sources are listed as example, the complete tables, one for each of the three regions, are available at CDS.}
\label{tab:cores_phys}
\centering
\tiny
\begin{tabular*}{\textheight}{ccccccccccccccccccc} 
\hline\hline  
n & Core name & RA(2000) & Dec(2000) &  \multicolumn{2}{c}{$R_{\rm core}$} &  \multicolumn{2}{c}{$M_{\rm core}$} &  \multicolumn{2}{c}{$T_{\rm dust}$} & $N^{\rm peak}_{\rm H_2}$ & \multicolumn{2}{c}{$N^{\rm ave}_{\rm H_2}$} & $n^{\rm peak}_{\rm H_2}$ & \multicolumn{2}{c}{$n^{\rm ave}_{\rm H_2}$} &  $\alpha_{\rm BE}$ & Core type & Comments\\
  &HGBS-J & (h m s) & (\degr\, \arcmin\, \arcsec) &  \multicolumn{2}{c}{(pc)} & \multicolumn{2}{c}{(M$_{\odot}$)} & \multicolumn{2}{c}{(K)} & (10$^{21}$ \cmdue) & \multicolumn{2}{c}{(10$^{21}$ \cmdue)} & (10$^{4}$ \cmtre) & \multicolumn{2}{c}{(10$^{4}$ \cmtre)} & \\
  (1)& (2)& (3)& (4)& (5)& (6)  & \multicolumn{2}{c}{(7) $\pm$ (8)} & \multicolumn{2}{c}{(9) $\pm$ (10)}& (11) & \multicolumn{2}{c}{(12) $\pm$ (13)} &  (14) & (15) & (16)& (17 )& (18))& (19)\\
\hline
  3 & 153809.7-340741 & 15:38:09.77 & -34:07:42.6 & 2.36e-02 & 2.71e-02 & 0.010 & 0.008 & 12.5 & 2.3 &  0.4 & 0.25 & 0.19 & 0.5 & 0.25 & 0.17 & 50.5 & starless \\                        
 123 & 154024.0-333734 & 15:40:23.99 & -33:37:33.9 & 3.04e-02 & 3.31e-02 & 0.159 & 0.005 & 11.0 & 0.1 & 4.0 & 2.45 & 2.06 & 4.6 & 1.96 & 1.51 & 3.4 & prestellar & tentative bound \\

 \hline\\
 \end{tabular*}
\tablefoot{
Table entries are as follows: (1) core running number; (2) core name = HGBS-J prefix followed by a tag created from the J2000 sexagesimal coordinates; (3) and (4): Right Ascension and declination of core centre; (5) and (6): geometrical average between the major and minor FWHM sizes of the core, as measured in the high-resolution column density map after deconvolution from the 18\farcs2 HPBW resolution of the map and before deconvolution, respectively (NB: both values provide estimates of the object’s outer radius when the core can be approximately described by a Gaussian distribution, as is the case for a critical Bonnor-Ebert spheroid); (7) estimated core mass assuming the dust opacity law advocated by \citet{roy14}; (9) SED dust temperature; (8) and (10) statistical errors on the mass and temperature, respectively, including calibration uncertainties, but excluding dust opacity uncertainties; (11) peak H$_2$ column density, at the resolution of the 500 \um\, data, derived from a grey-body SED fit to the core peak flux densities measured in a common 36\farcs3 beam at all wavelengths; (12) average column density, calculated as $N^{\rm ave}_{\rm H_2} = \frac{M_{\rm core}}
{\pi R_{\rm core}^2}\frac{1}{\mu m_{\rm H}}$, where $M_{\rm core}$ is the estimated core mass (Col. 7), $R_{\rm core}$ the estimated core radius prior to deconvolution (Col. 6), and $\mu$ = 2.8; (13) average column density calculated in the same way as for Col. 12 but using the deconvolved core radius (Col. 5) instead of the core radius measured prior to deconvolution; (14) beam-averaged peak volume density at the resolution of the 500 \um\, data, derived from the peak column density (Col. 11) assuming a Gaussian spherical distribution:  $n^{\rm peak}_{\rm H_2} = \sqrt{\frac{4 lm 2}{\pi}}\frac{N^{\rm peak}_{\rm H_2}}{FWHM_{500}}$; (15) average volume density, calculated as $n^{\rm ave}_{\rm H_2} = \frac{M_{\rm core}}{4/3\pi R_{\rm core}^3}\frac{1}{\mu m_{\rm H}}$, using the estimated core radius prior to deconvolution; (16) average volume density, calculated in the same way as for Col. 15 but using the deconvolved core radius (Col. 5); (17) Bonnor-Ebert mass ratio: $\alpha_{\rm BE} = M_{\rm BE,crit}/M_{\rm obs}$ (see text for details); (18) core type: starless, prestellar, or protostellar; (19) comments may be `no SED fit' or “tentative bound” (see text for details).
}
\end{sidewaystable*}

\section{YSOs SED fitting}
\label{ap:sed}

The \citet{robitaille17} synthetic SEDs are divided in 18 sets of models where the central source may or may not be associated with a disk, a circumstellar envelope, a bipolar cavity and an ambient medium. Each set of models represents a different combination of these components and each component is described by several input parameters. The detailed description of the physics of each component and its physical parameters can be found in  \citet{robitaille17}, and here we briefly report the list of the free input parameters of the models. All free parameters are uniformly sampled within an allowed range.
The central star, present in all models, is a spherical source and is defined by a stellar radius $R_{\star}$ and an effective temperature $T_{\star}$.
The passive flared disk has a density distribution that is defined by the disk dust mass $M_{\rm disk}$, the inner and outer radius $R^{\rm disk}_{\rm min}$ and $R^{\rm disk}_{\rm max}$, respectively, the surface density radial exponent $p_{\rm disk}$, the disk flaring exponent $\beta_{\rm disk}$, and the disk scale height $h_{\rm 100AU}$.
Two types of envelopes are included in the models: the first is a spherically-symmetric power-law envelope, and the second is a rotationally flattened envelope as defined by \citet{ulrich76}. For both envelope components, the envelope is truncated at the inner radius $R^{\rm env}_{\rm min}$, and extends all the way to the edge of the grid. The free parameters are the envelope density scaling $\rho_0^{\rm env}$, the radial exponent $\gamma_{\rm env}$ in the case of the power-law envelope, and the centrifugal radius $R_{\rm c}$ in the case of the Ulrich envelope.
The bipolar cavity, present only in models where an envelope component is also present, is defined by the power-law exponent of the cavity opening $c_{\rm cav}$, the opening angle $\theta_{\rm 0}^{\rm cav}$, and the density inside the cavity is set to $\rho_0^{\rm cav}$ or the envelope density, whichever is lowest. 
In several models an ambient medium is added  with dust density of $\rho_{\rm amb}$ = 10$^{-23}$ g~cm$^{-3}$ and temperature $T_{\rm amb}$ = 10 K.
Each model was computed for nine viewing angles randomly sampled between 0\degr\, to 90\degr.


To determine which model {\it best} represents the observed SED, we follow the fitting procedure described in \citet{robitaille17}. We define a `good' model as one with 
\begin{equation}
 \chi^2 - \chi^2_{\rm best} < 9 n_{\rm data} 
\end{equation}
where $\chi^2_{\rm best}$ is the lowest $\chi^2$ among all model sets and $n_{\rm data}$ is the number of data points in the observed SED. We then define the probability of a model set $P$ as the ratio between the number of good models $N_{\rm good}$ and the total number of models with the set $N$
\begin{equation}
 P = \frac{N_{\rm good}}{N}.
\end{equation}
Finally, we assign a score to each set of models as the ratio of $P$ to the mean of the $P$ values for all model sets. A higher relative score indicates a more likely model, therefore we consider as the best model set the one with the highest score.
 
In Table \ref{tab:par_best_fit_yso}, for each object of our \her\, YSOs/protostars catalogue, we report the model set with the highest score, using the same nomenclature of \citet{robitaille17}, and we indicate which of the four model components are included in the best models set. 
In addition, for all the objects we provide, in electronic form, files with the range of physical parameters of all the good models within the best model set and figures of the synthetic SED of the models with the best $\chi^2$ for all 18 sets of models of \citet{robitaille17}. This material is provided at the following URLs:
https://owncloud.ia2.inaf.it/index.php/s/aVkJ07nYKVj9FlD, 
https://owncloud.ia2.inaf.it/index.php/s/SxkNLXMi9iMIFYd, 
https://owncloud.ia2.inaf.it/index.php/s/zUvSRoLLX2CktO8, 
https://owncloud.ia2.inaf.it/index.php/s/CDQFjFPPJeXu2v1, 
https://owncloud.ia2.inaf.it/index.php/s/cSp9bkOZYgx5wuk,
https://owncloud.ia2.inaf.it/index.php/s/opcGd8WGN3NVdoS
.

\begin{table*}
\centering 
\small

\caption{List of the most probable set of models for the YSOs/protostars of our catalogue fitted with the \citet{robitaille17} model. The names of the sets of models are those defined in \citet{robitaille17}. For each model, the presence of disk, envelope, cavity, and ambient medium is indicated. For the envelope, two possible profiles, spherically-symmetric power-law or Ulrich (1976) type, are specified. For all models that contain at least a disk, envelope, or ambient medium, the inner radius is set to the same value for all components, and is either set to $R_{\rm sub}$ (the dust sublimation radius) or is variable in the range $R_{\rm sub}$  to 1000 $R_{\rm sub}$.}             
\label{tab:par_best_fit_yso}      

\begin{tabular} {l c c c c c c c c c} 
\hline\hline                 
source & $n_{\rm data}$  &  best model set & score  & disk & envelope & cavity & ambient & inner radius \\
\hline \\
LUPUS I \\
\hline 
HGBS\_YSO-J153640.0-342145 & 11 & spubhmi &  18.0 & yes &    Ulrich & yes & yes &      variable \\
HGBS\_YSO-J153927.9-344616 & 18 & spubhmi &  14.0 & yes &    Ulrich & yes & yes &      variable \\
HGBS\_YSO-J154011.3-351522 &  8 & sp--h-i &  10.8 & yes &        no &  no &  no &      variable \\
HGBS\_YSO-J154017.6-324649 &  9 & spubhmi &  11.5 & yes &    Ulrich & yes & yes &      variable \\
HGBS\_YSO-J154051.6-342102 & 12 & spubhmi &  18.0 & yes &    Ulrich & yes & yes &      variable \\
HGBS\_YSO-J154302.3-340908 & 13 & s-pbsmi &  18.0 &      no & power-law & yes & yes & $R_{\rm sub}$ \\
HGBS\_YSO-J154512.8-341729 & 15 & sp--s-i &  18.0 & yes &        no &  no &  no & $R_{\rm sub}$ \\
HGBS\_YSO-J154529.8-342339 & 18 & spubsmi &  12.0 & yes &    Ulrich & yes & yes & $R_{\rm sub}$ \\
HGBS\_YSO-J154644.6-343034 & 12 & spubsmi &   8.2 & yes &    Ulrich & yes & yes & $R_{\rm sub}$ \\

\hline \\

LUPUS III \\
\hline 
HGBS\_YSO-J160500.9-391301 &  5 & sp--s-i &  18.0 & passive &        no &  no &  no & $R_{\rm sub}$ \\
HGBS\_YSO-J160708.4-391407 & 18 & spu-smi &   8.0 & passive &    Ulrich &  no & yes & $R_{\rm sub}$ \\
HGBS\_YSO-J160709.9-391102 & 18 & spubsmi &  13.5 & passive &    Ulrich & yes & yes & $R_{\rm sub}$ \\
HGBS\_YSO-J160711.5-390347 & 18 & spubhmi &  12.0 & passive &    Ulrich & yes & yes &      variable \\
HGBS\_YSO-J160822.4-390445 & 18 & spubhmi &  18.0 & passive &    Ulrich & yes & yes &      variable \\
HGBS\_YSO-J160825.7-390600 & 17 & sp--s-i &  10.3 & passive &        no &  no &  no & $R_{\rm sub}$ \\
HGBS\_YSO-J160829.6-390309 & 15 & s-pbhmi &  18.0 &      no & power-law & yes & yes &      variable \\
HGBS\_YSO-J160830.7-382826 & 11 & spubhmi &  18.0 & passive &    Ulrich & yes & yes &      variable \\
HGBS\_YSO-J160836.1-392300 & 18 & sp--smi &  18.0 & passive &        no &  no & yes & $R_{\rm sub}$ \\
HGBS\_YSO-J160854.5-393743 & 13 & s-pbsmi &  18.0 &      no & power-law & yes & yes & $R_{\rm sub}$ \\
HGBS\_YSO-J160901.8-390511 & 15 & spubsmi &   6.0 & passive &    Ulrich & yes & yes & $R_{\rm sub}$ \\
HGBS\_YSO-J160917.9-390453 & 14 & spubhmi &  18.0 & passive &    Ulrich & yes & yes &      variable \\
HGBS\_YSO-J160948.5-391116 & 17 & sp--s-i &  18.0 & passive &        no &  no &  no & $R_{\rm sub}$ \\
HGBS\_YSO-J161051.5-385314 & 18 & spubsmi &   8.4 & passive &    Ulrich & yes & yes & $R_{\rm sub}$ \\

\hline\\

LUPUS IV \\
\hline 
HGBS\_YSO-J155641.9-421925 & 11 & spubsmi &  18.0 & yes &    Ulrich & yes & yes & $R_{\rm sub}$ \\
HGBS\_YSO-J155730.4-421032 & 11 & s-u-hmi &  14.4 &      no &    Ulrich &  no & yes &      variable \\
HGBS\_YSO-J155746.6-423549 & 12 & spubsmi &   9.0 & yes &    Ulrich & yes & yes & $R_{\rm sub}$ \\
HGBS\_YSO-J155916.5-415712 & 12 & sp--smi &   9.0 & yes &        no &  no & yes & $R_{\rm sub}$ \\
HGBS\_YSO-J160044.6-415530 & 18 & spubhmi &  18.0 & yes &    Ulrich & yes & yes &      variable \\
HGBS\_YSO-J160115.5-415233 & 18 & spubhmi &   7.2 & yes &    Ulrich & yes & yes &      variable \\
HGBS\_YSO-J160234.6-421129 & 14 & spubhmi &   6.0 & yes &    Ulrich & yes & yes &      variable \\
HGBS\_YSO-J160329.2-414001 & 13 & spubsmi &   6.0 & yes &    Ulrich & yes & yes & $R_{\rm sub}$ \\
HGBS\_YSO-J160403.0-413427 &  7 & sp--h-i &  18.0 & yes &        no &  no &  no &      variable \\
HGBS\_YSO-J160913.7-414430 &  8 & sp--s-i &  18.0 & yes &        no &  no &  no & $R_{\rm sub}$ \\
HGBS\_YSO-J160956.3-420834 &  6 & sp--s-i &  18.0 & yes &        no &  no &  no & $R_{\rm sub}$ \\
HGBS\_YSO-J161301.6-415255 &  9 & sp--h-i &  18.0 & yes &        no &  no &  no &      variable \\

\end{tabular}
\end{table*}

\section{Notes on particular YSOs}
\label{sect:app_yso}

In this Appendix we discuss the result of the SED fitting for those YSOs of our catalogue that have a poor fit and those where we find some discrepancies with the spectral index classification.

For HGBS\_YSO-J154017.6-324649 we find only a counterpart in \wise\, but not in \spit-IRAC and \twomass. It has an $\alpha$ index of a Class I object but with such significant emission at longer wavelengths that the $L_{\rm smm}/L_{\rm bol}$ ratio is typical of a Class 0 object. The SED fit indicates the presence of a little disk and a robust dusty envelope with a cavity. Its location in a very low column density region and the fact that it is not visible 500 \um\, with \her\, makes it very unlikely that it is a Class 0 objects and favours the interpretation that it is a more evolved object.

For source HGBS\_YSO-J154529.8-342339, the NIR-MIR part of its SED is typical of a Class II object and indeed it is well fitted by a disk. However, it has been associated with a FIR \her\, source, making its $L_{\rm smm}/L_{\rm bol}$ suggestive of a Class 0 object. Its SED model fails to match the FIR part of its SED. In fact, looking at the \her\, maps, it is likely that the quoted fluxes at 350 \um\, and 500 \um, are contaminated by close-by sources.

For source HGBS\_YSO-J160500.9-391301 we have only five photometric points and the best $\chi^2$ SED model is able to reproduce only three of the points, therefore the result is really uncertain.

For source HGBS\_YSO-J160708.4-391407, the $\alpha$ index indicates a Class Flat object but the $L_{\rm smm}/L_{\rm bol}$ ratio is typical of a Class 0 object. However, this source is well detected at NIR wavelengths up to the $J$ band. Moreover, the best fit model of the full SED indicates the presence of a disk with a tenuous envelope, favouring for this object the interpretation of a more evolved Class Flat rather than Class 0.

For source HGBS\_YSO-J160709.9-391102, the $\alpha$ index indicates a Class II object while the SED fitting gives as model with the highest score one with a very low mass disk of 2.7$\times$10$^{-08}$ \msun, seen edge on, with a quite dense circumstellar envelope with a cavity. However, a similar good fit can be also obtained with a model with only a circumstellar disk with mass of 2.8$\times$10$^{-02}$ \msun\, and an inclination angle of 65\degr. This second model is compatible with the Class II classification.

HGBS\_YSO-J160829.6-390309 is a well known T-Tauri star. It is quite luminous at MIR wavelengths up to 160 \um\, that make its $\alpha$ index typical of the Class I objects, however, it is not detected at the \her-SPIRE bands ($\lambda\geq$ 250 \um). It has been classified as higher than the median infrared excess in \citet{bustamante15}, (see Sect. \ref{sect:disc_yso}), and the best set of models is that with e circumstellar envelope and a cavity. However, also models with disk and wihout envelope, compatible with Class II classification, give good fit to the observed SED.

HGBS\_YSO-J155641.9-421925 is the brightest source in Lupus IV and it is a well known Herbig Ae star (HD 142527). The SED best fit reproduces well its MIR-FIR SED but fails to fit the NIR data. 

HGBS\_YSO-J155730.4-421032 is a Class II object. The most probable set of models, however, indicates only the presence of an envelope, though also models with cavity and disk can fit well the observed SED.

For source HGBS\_YSO-J160115.5-415233, the $\alpha$ index indicates a Class Flat object but the $L_{\rm smm}/L_{\rm bol}$ ratio is typical of a Class 0 object. In this case, the SED fit is consistent with the earlier evolutionary stage, indicating the presence of a massive envelope ($\rho_0^{\rm env}$=1.1$\times$10$^{-20}$ g \cmtre) and a meagre disk ($M_{\rm disk}$=4.3$\times$10$^{-8}$ \msun). An additional indication in favour of the youth of this source is the ratio between the column density of HC$_3$N and NH$_3$, reported in \citet{benedettini12}, that is similar to what is observed towards the Class 0 object HGBS\_YSO-J160917.9-390453 in Lupus III and is higher than typical values found in more evolved protostars. The majority of the observational evidence therefore indicates that this source is a good Class 0 protostar candidate.

HGBS\_YSO-J160403.0-413427 and HGBS\_YSO-J160956.3-420834 have strange SEDs lacking the MIR fluxes, indicating a possible accidental association between the NIR and FIR data. Moreover, their $\alpha$ indexes cannot be defined since not enough data are available. In both cases, the SED fits indicate that they are evolved objects, possibly Class II, with a disk and without envelope. HGBS\_YSO-J160956.3-420834 is associated with a known star (HD 142527)

HGBS\_YSO-J161301-415255 is a Class III object. The best fit reproduces well its NIR-MIR SED but fails to reproduce the fluxes between 24 \um\, and 160 \um, indicating the possible presence of another (younger) source close to the YSO. The position of this object is not covered by the \her-SPIRE maps.

\end{appendix}

\end{document}